%% file: main.tex
\documentclass[conference]{IEEEtran}
\IEEEoverridecommandlockouts

\usepackage{amsmath,multicol,amsthm}
\usepackage[ruled]{algorithm}
\usepackage{algpseudocode}
\usepackage[normalem]{ulem}
\usepackage{lineno}  % Add line numbering
\usepackage{ioa_code}
\usepackage{tikz}
\usepackage{pgfplots}
\usepgfplotslibrary{groupplots}
\pgfplotsset{compat=1.18}
\usepackage{amsfonts}
\usepackage{url}

\input{nn_macros}
\input{optimum_macros}

\algrenewcommand\algorithmiccomment[1]{\hspace{2em}{\color{blue} // \textit{#1}}}

\newcommand{\nn}[1]{{\color{red}#1}}

\newcommand{\myemph}[1]{{\it #1}}
\newcommand{\myparagraph}[1]{\smallskip\noindent{\textbf{#1}}}
\newcommand{\WRP}{\par\qquad\(\hookrightarrow\)\enspace}

\title{\optimum{}: Highly Consistent, Scalable, and Secure Multi-Object Memory using 
RLNC}

\author{
\IEEEauthorblockN{
Nicolas Nicolaou,
Kishori M. Konwar,
Moritz Grundei,
Aleksandr Bezobchuk,
Muriel M\'edard
}
\IEEEauthorblockA{
Optimum, MA, USA,
\{nicolas, kkonwar, moritz, bez, mmedard\}@getoptimum.xyz}
\IEEEauthorblockN{Sriram Vishwanath, 
Georgia Tech, USA, sriram@ece.gatech.edu
}
\thanks{The authors would like to thank Professor Nancy Lynch for the valuable discussions and comments that made the contributions of this work stronger. }
}

\begin{document}
%\tracingmacros=0

\maketitle

\input{abstract.tex}


\section{Introduction}
\label{sec:intro}
\input{intro_v2.tex}
%\input{intro_short.tex}

% \section{Related Work}
% \label{sec:background}
% \input{background.tex}

\section{System Setting and Definitions}\label{model}
\label{sec:model}
\input{model.tex}

\section{Erasure Coded Read/Write Protocol with Byzantine Flexnodes}
\label{sec:byz-rw}
\input{byzantine_rw_algorithm.tex}
%\subsection{Algorithm Correctness}
%\input{safety_liveness.tex}

\section{Managing Multiple Objects}
\label{sec:recofi}
\input{multi_object_manipulation}

\section{Dynamic, Multi-Object, Atomic Shared memory}
\label{sec:recofi}
\input{multi_object_v2}

\section{Correctness of \optimum{}}
\label{sec:correct}
\input{sec_optimum_correct}

\section{Experimental Evaluation}
\label{sec:exp}
\input{sec_experiments.tex}

\section{Conclusions}
\label{sec:conclusions}
\input{sec_conclusion.tex}
\bibliographystyle{acm}
\bibliography{biblio}
%\bibliography{biblio,cadambe-refs,evaluation}
%
% \newpage
% \appendix
% \section{Erasure Coded Read/Write Protocol with Byzantine Flexnodes}
% \label{sec:byz-rw-appendix}
% \input{safety_liveness.tex}

% \section{Correctness of \optimum{}}
% \input{sec_optimum_correct}
% \label{sec:correct-appendix}
\end{document}

%% file: nn_macros.tex
\newtheorem{theorem}{Theorem}[section]
\newtheorem{lemma}[theorem]{Lemma}

\newtheorem{definition}[theorem]{Definition}

\newcommand{\Nat}{\ifmmode \mathbb{N} \else $\mathbb{N}$ \fi}

\newcommand{\Real}{\ifmmode \mathbb{R} \else $\mathbb{R}$ \fi}

\newcommand{\Fin}[1]{\ifmmode \mathbb{F}_#1 \else $\mathbb{F}_#1$ \fi}

\newcommand{\case}[1]{
        {\vspace{1em}\noindent{\bf Case #1:}}}

\newcommand{\ceil}[1]{\lceil #1 \rceil}

\newcommand{\tup}[1]{%
    \relax\ifmmode
      \langle #1 \rangle%
    \else
        $\langle$#1$\rangle$%
    \fi
}

\newcommand{\act}[1]{%
    \relax\ifmmode
        \mathord{\mathcode`\-="702D\sf #1\mathcode`\-="2200}%
    \else
        $\mathord{\mathcode`\-="702D\sf #1\mathcode`\-="2200}$%
    \fi
}

\newcommand{\remove}[1]{}

\newcommand{\mtype}[1]{\text{{\sc #1}}}

\makeatletter
\def\mainlistofsymbols{
  \normalsize
  \vspace*{1.5 em}
  \@starttoc{los}
}

\def\partonelistofsymbols{
  \normalsize
  \vspace*{1.5 em}
  \@starttoc{p1los}
}

\def\parttwolistofsymbols{
  \normalsize
  \vspace*{1.5 em}
  \@starttoc{p2los}
}

\def\l@symbol#1#2{\addpenalty{-\@highpenalty} \vskip 4pt plus 2pt
{\@dottedtocline{0}{0em}{8em}{#1}{#2}}}
\makeatother

\newcommand{\newhiddensym}[2]{%
}

\newcommand{\algIOA}[2]{\ifmmode{\text{#1}_{#2}}\else{$\text{#1}_{#2}$}\fi}
\newcommand{\EX}{\ifmmode{\xi}\else{$\xi$}\fi}
\newcommand{\EXF}{\ifmmode{\phi}\else{$\phi$}\fi}
\newcommand{\hist}[1]{H_{#1}}
\newcommand{\state}{\sigma}
\newcommand{\st}{\sigma}
\newcommand{\obj}[1]{o_{#1}}

\newcommand{\qs}{\mathcal{Q}}
\newcommand{\quo}[1]{Q_{#1}}
\newcommand{\inter}[1]{
	\ifmmode{\left(\bigcap_{\mathcal{Q}\in#1}\mathcal{Q}\right)}
	\else{$\left(\bigcap_{\mathcal{Q}\in#1}\mathcal{Q}\right)$}
	\fi
}
\newcommand{\idSet}{\mathcal{I}}
\newcommand{\wSet}{\mathcal{W}}
\newcommand{\srvSet}{\mathcal{S}}
\newcommand{\objSet}{\mathcal{O}}
\newcommand{\confSet}{\mathcal{C}}
\newcommand{\servers}[1]{ #1.Servers}
\newcommand{\quorums}[1]{ #1.Quorums}
\newcommand{\consensus}[1]{ #1.Con}

\newcommand{\op}{\pi}
\newcommand{\rd}{\rho}
\newcommand{\wrt}{\omega}

\mathchardef\mhyphen="2D
\newcommand{\bef}{\rightarrow}

\newcommand{\vid}[1]{\ifmmode{\nu_{#1}}\else{$\nu_{#1}$}\fi}
\newcommand{\seen}{\ifmmode{seen}\else{$seen$}\fi}
\newcommand{\ares}{{\sc Ares}}
\newcommand{\valSet}{{\mathcal V}}
\newcommand{\tsSet}{{\mathcal T}}
\newcommand{\tgSet}{{\mathcal T}}
\newcommand{\tg}[1]{\tau_{#1}}

\newcommand{\maxts}[1]{\ifmmode{maxTS_{#1}}\else{$maxTS_{#1}$}\fi}
\newcommand{\maxtag}[1]{\ifmmode{maxTag_{#1}}\else{$maxTag_{#1}$}\fi}
\newcommand{\maxpair}[1]{\ifmmode{maxMPair_{#1}}\else{$maxMPair_{#1}$}\fi}
\newcommand{\mintag}[1]{\ifmmode{minTag_{#1}}\else{$minTag_{#1}$}\fi}
\newcommand{\maxps}{\ifmmode{maxPS}\else{$maxPS$}\fi}
\newcommand{\conftg}[1]{\ifmmode{confirmed_{#1}}\else{$confirmed_{#1}$}\fi}
\newcommand{\maxconftag}{\ifmmode{\ms{maxCT}}\else{$maxCT$}\fi}

%% file: optimum_macros.tex
\newcommand{\atT}[2]{#1|_{#2}}

\newcommand{\dap}[1]{{DAP(#1)}}
\newcommand{\hash}[1]{\mathcal{H}^{#1}}

\newcommand{\flexSet}{\mathcal{I}}

\newcommand{\optimum}{{\sc Optimum-DeRAM}}
\newcommand{\rambo}{{\sc Rambo}}
\newcommand{\dynastore}{{\sc Dynastore}}
\newcommand{\aresec}{{\sc ares-rlnc}}

\newcommand{\flexpr}[1]{f_{#1}}

\newcommand{\daputdata}[2]{ {\act{put-data}(#1, #2)}}
\newcommand{\dagetdata}[1]{ {\act{get-data}(#1)}}
\newcommand{\dagettag}[1]{ {\act{get-tag}(#1)}}

\newcommand{\distance}[1]{ d(#1) }

\newcommand{\suc}[1]{\mathfrak{D}_{suc}(#1)}

\algblockdefx[Variables]{Variables}{EndVariables}%
{{\bf State Variables:}}{}

\algblockdefx[Operations]{Operations}{EndOperations}%
{{\bf Operations:}}{}

\algblockdefx[Operation]{Operation}{EndOperation}%
[2]{{\bf operation} $\act{#1}$(#2)}%
{{\bf end operation}}

\algblockdefx[Procedure]{Procedure}{EndProcedure}%
[2]{{\bf procedure} $\act{#1}$(#2)}%
{{\bf end procedure}}

\algblockdefx[Receive]{Receive}{EndReceive}%
[2]{{\bf Upon receive} (#1)$_{\text{ #2 }}${\bf from} $\flexpr{j}$}%
{{\bf end receive}}

\newcommand{\setEC}{\mathcal{E}}
\newcommand{\ecSet}{\setEC}
\newcommand{\ec}[2]{\tup{#1, #2}}
\newcommand{\coeff}[1]{\mathbf{A}_{#1}}
\newcounter{ALC@line}
\newcommand{\members}[1]{ #1.Flexnodes}
\newcommand{\objects}[1]{ #1.Objects}

%% file: abstract.tex
\begin{abstract}
This paper introduces \optimum{}, a highly consistent, scalable, secure, and decentralized shared memory solution. Traditional distributed shared 
memory implementations offer multi-object support by multi-threading a single object memory instance over the same set of data hosts.  While theoretically sound, the amount of 
resources required made such solutions prohibitively expensive in practical systems. \optimum{} proposes a decentralized, reconfigurable, atomic read/write shared memory (DeRAM) that: (i) achieves \textit{improved performance} and 
\textit{storage scalability} by leveraging Random Linear Network Codes (RLNC); (ii) \textit{scales} in the number of supported atomic objects by introducing a new \textit{object placement} and \textit{discovery} approach based on a \textit{consistent hashing ring}; (iii) \textit{scales} in the number of participants by allowing dynamic joins and 
departures leveraging a blockchain oracle to serve as a registry service; and (iv) is \textit{secure} against malicious behavior by tolerating Byzantine failures. Experimental results over a globally distributed set of nodes, help us realize the performance and scalability gains of \optimum{} over previous distributed shared memory solutions (i.e., the ABD algorithm \cite{ABD96}).

\end{abstract}

%% file: intro_v2.tex
Since the inception of the von Neumann architecture computing relied on a central processing unit (CPU) and a memory 
space to store both data and applications. A variance of the same architecture was later adopted in multi-core
systems where multiple central processing units (the CPU-cores) perform parallel computations by communicating data
through a \textit{shared memory} space. With the advent of mobile computing devices and Web3.0 applications (e.g., Ethereum \cite{ethereum2}), the distributed computing paradigm emerged where geographically dispersed nodes collaborate through message passing to accomplish a common goal (e.g., the implementation of a distributed ledger \cite{ethereum2}). 

%Due to the absence of a shared memory, 
{\bf State Machine Replication} (SMR) \cite{lamport1978time} appeared as an attractive solution for newly developed 
distributed applications. SMR requires that all devices execute the same operations on the same order on a replicated system state. Therefore, to achieve SMR, devices need to reach \emph{consensus} \cite{Lamport86} on the operation order, a problem proven difficult to solve in asynchornous, fail-prone environments \cite{FLP85} (like the internet). 

In shared-memory systems, read/write registers are the simplest and most fundamental object type, forming the basis for solving numerous distributed computing problems~\cite{Raynal2018}. However, in asynchronous message-passing environments, such as those building Web3.0, there is no inherent shared memory. To bridge this gap, researchers have explored techniques for emulating an \textit{atomic} shared memory over a distributed network of nodes (ADSM). Atomicity, or linearizability~\cite{Lynch1996, HW90}, ensures that operations on a shared read/write (R/W) object appear to execute sequentially, even when they accesses the object concurrently at different nodes. This is the strongest and most intuitive consistency guarantee, which makes it highly desirable for Web3.0 application developers. 
% This challenge is particularly relevant for blockchain and Web3 platforms, which must maintain a globally consistent state across untrusted nodes while optimizing performance and minimizing the cost of consensus.

For nearly three decades, researchers have developed a vast body of work~\cite{ABD96, CDGL04, LS02, AKMS09, GNS09, CadambeLMM17} 
%\cite{ABD96, CDGL04, FL03, LS02, GLS03, AKMS09, FHN16, GNS09, CadambeLMM17} 
implementing ADSM emulation over message-passing networks, both for static and dynamic environments. 
Ensuring atomicity in a decentralized, asynchronous, and fail-prone setting however, is inherently challenging. Initial solutions adopted 
replication-based strategies to offer data availability and survivability, at the expense of high storage and 
communication costs to maintain consistency, despite 
communication optimization attempts \cite{CDGL04, GNS06, GNS08}.
% . Some attempts were made to reduce the communication overhead 
% \cite{CDGL04, GNS06, GNS08} with the storage costs remaining induct. 

More recently, to address both communication and storage overheads, researchers have explored the use of erasure coding, particularly Reed-Solomon codes~\cite{CT06, CadambeLMM17, DGL08, SODA2016, radon, GIZA2017, KNML19}, to reduce redundancy while preserving fault tolerance. ARES~\cite{ARES19}, one of the most recent such algorithm, managed to bring erasure-coded solutions to dynamic environments, where nodes may join or leave unpredictably. With a modular design ARES introduced the ability to reform set of replicas and modify the replica access algorithm on the fly.

Despite significant advancements in ADSM
%shared atomic memory 
algorithms, existing approaches share a common limitation: they primarily focus on implementing a \textit{single atomic object} and rely on composing multiple instances to form a uniform, multi-object memory. While theoretically sound, deploying multiple instances of ADSM in real-world networks presents practical challenges, not yet addressed by existing solutions:

\begin{itemize}
% \item {\bf Node Selection:} How do we determine which nodes participate in each instance?
\item {\bf Object Placement:} How should objects be assigned to ADSM instances and network nodes?
\item {\bf Dynamic Discovery:} In a continuously evolving network, how can nodes (re)locate objects and their replicas?
\end{itemize}

%Furthermore, 
In the case of erasure coded solutions, the usage of block codes (e.g Reed Solomon), requires prior 
synchronization on the code rate hindering the dynamic nature of the system.

Essentially, a monolithic solution over a cluster of distributed nodes, does not scale in
storage, number of supported objects, and operation concurrency. Storage is limited to the capacity of the 
smallest node in the cluster, resource demands increase with the number of objects supported, 
and eventually the cluster becomes a bottleneck for concurrent operations.

% 

% Implementing distributed algorithms in Web3 environments presents unique challenges due to the decentralized, trustless nature of the network. These challenges include handling adversarial actors, network latency, and ensuring security and privacy in public and permission-less settings. Web3 algorithms must balance decentralization and performance, manage state immutability, and deal with issues like forking and finality delays. Additionally, designing incentive structures to encourage honest participation and overcoming the resource inefficiencies of consensus protocols are key difficulties. 
% These factors make the development of distributed algorithms in Web3 more complex than in traditional systems.

\myparagraph{Our Contributions:}
%
%To address these challenges, w
We introduce \optimum{}, a \textit{scalable}, dynamic, read/write ADSM, that tolerates 
Byzantine failures 
%shared atomic memory algorithm designed for asynchronous message-passing environments that 
and efficiently manages multiple objects. \optimum{} uses composition to infer an atomic memory from multiple objects. However, unlike traditional approaches, \optimum{} has a clear strategy on \textit{how to manipulate} multiple objects, favoring operation performance and service scalability. 
At its core, \optimum{} leverages Random Linear Network Coding (RLNC) ~\cite{RLNC2006} 
to enhance, fault tolerance, storage and latency efficiency, all at once. 
%In contrast, Random Linear Network Coding (RLNC)~\cite{RLNC2006} offers a more flexible and efficient alternative. 
%By allowing nodes to generate coded pieces dynamically, RLNC minimizes the need for strict synchronization.
Storage 
%Additionally, the new algorithm 
\textit{scalability} is further boosted
%in terms of storage, 
by utilizing an innovative \textit{hash ring} approach to 
distribute multiple objects among different node clusters, avoiding this way the bottlenecks of singleton clusters. Such distibution
also allows the service to \textit{scale} in terms of concurrent operations, as 
%in \optimum{} 
each operation may contact, a different cluster. 
Lastly, \optimum{} \textit{scales} with the number of participants by introducing join and depart protocols that handle (re)clustering and 
object migration, without affecting concurrent read and write operations. 
%while maintaining robustness. 
%
%
% This flexibility makes it an attractive approach to implement cost-efficient and fault-tolerant atomic memory in a message-passing environment, particularly for Web3 systems that require scalable and decentralized storage solutions.
%
Enumerated, our key contributions include:

\begin{enumerate}
    \item {\bf Erasure-Coded, Byzantine-Tolerant DeRAM:}
We develop an ADSM or decentralized RAM (DeRAM) algorithm, extending the modular approach proposed in \ares{}~\cite{ARES19}, by 
supporting RLNC-based coding for enhanced efficiency and 
%to provide fault tolerance and efficient retrieval under 
tolerating Byzantine failures.  

\item {\bf Scalable Multi-Object Distribution \& Discovery:}
We introduce a consistent hashing approach for establishing object placement over a subset of nodes in a static cluster. 
We show how both nodes and objects can be placed over a common logical \textit{ring} and we define a \textit{distance} metric
that measures the relative distance of each object to each node on the ring. 
Distance is then used to decide object placement to the nodes.
%An algorithm decides which nodes should hold each object based on that distance. 

% , where a subset of nodes stores shared objects, ensuring balanced storage utilization and scalable service %discovery 
% in dynamic environments.

%\item {\bf Efficient Reconfiguration via Blockchain Oracle \& RLNC:}
\item {\bf Scalable Node Participation:}
We then introduce join and depart protocols, that allows new nodes to join the service and existing nodes to depart 
gracefully. Both protocols leverage a consistent oracle (e.g, a consensus service \cite{L98}, or a blockchain ~\cite{CosmosSDK}) as a node registry, and they handle 
asynchronously the re-allocation of objects as nodes join and depart. RLNC plays a key role in the transmission 
of data, as re-coding allows the distribution of data without decoding. 

\item {\bf Evaluation Results:}
We implement the protocol and we perform preliminary experiments showcasing the scalability of \optimum{} compared to
the classic ABD~\cite{ABD96} algorithm. Our evaluation focuses on four scalability dimensions: (i) object size, (ii) number 
of objects, and (iii) number of nodes, and (iv) operation concurrency. 

% We design a reconfiguration algorithm that leverages a Cosmos blockchain oracle~\cite{CosmosSDK} to track the node participation order and efficiently manage incremental joins and departures.
% Our approach enables a newly joined node to retrieve and store objects efficiently using RLNC, without requiring full decoding and re-encoding. Instead, it constructs a new coded element directly from a subset of existing coded elements, reducing both communication overhead and computational costs.
\end{enumerate}

\input{background}

%% file: model.tex
\optimum{} implements a multi-object, atomic, read/write (R/W), 
distributed shared memory $\mathcal{M}$, over an asynchronous message-passing system in a decentralized setting with certain fraction of the nodes possibly acting maliciously.
%with processes communicating through reliable point-to-point channels.

\myparagraph{Memory Objects.} A shared atomic memory can be emulated
by composing individual atomic memory objects with identifiers from a 
set $\objSet$. 
%Therefore, herein we focus in implementing a single atomic \textit{read/write} memory object. 
A read/write object $\obj{}\in\objSet$ stores a value from a set $\valSet$ and 
supports two operations: (i) a $write(v)$ operation that modifies the value of $\obj{}$ to a value $v\in\valSet$, and (ii) a $read()$ operation that returns the value of $\obj{}$. Due to the composable nature of atomic objects, we can focus on the implementation of one such object, and therefore, a collection of such objects emulates a shared atomic memory.

\myparagraph{Flexnodes.} Our system consists of a collection of fail-prone processes,
%non constant (i.e., \emph{dynamic}) set of processes, 
so called \emph{flexnodes}, with unique identifiers from a totally-ordered set $\flexSet$. 
%\nn{[NN: in case of permissionless we may want to avoid this set.]}. 
Each flexnode can perform read/write operations, and store data for  objects in $\objSet$. %We assume that 
Each pair of flexnodes in $\idSet$ communicates via \emph{messages} in a peer-to-peer fashion through \emph{asynchronous} and \emph{reliable} channels.

\myparagraph{Executions.} 
% We model processes as I/O automata (IOA). A distributed algorithm $A$ is a collection of IOA, where $A_p$
% is the automaton assigned to {process} $p\in\idSet$. The \textit{state}, of a process $A_\pr$ is determined over a
% set of state variables, and the state $\state$ of $A$ is a vector containing the state of
% each process. Each automaton $A_\pr$ implements a set of actions. When an action $\acts{}$ occurs 
% it causes the state of $A_\pr$ to change, say from 
% some state $\state_p$ to some different state $\state_p'$. We call the triple $\tup{\state_p, \acts{}, \state_p'}$
% a \textit{step} of $A_\pr$. Algorithm $A$ performs a step, when some automaton $A_\pr$ performs a step.
% %: (i) receives a
% %message, (ii) performs local computation, (iii) sends a message. 
% %Each such action
% %causes the state at $p$ to change. 
% An action $\acts{}$ is \textit{enabled} in a state $\state$ if $\exists$ a step $\tup{\state, \acts{}, \state'}$ to some state $\state'$.
%
An \textit{execution} $\EX$ of a distributed algorithm $A$ is a sequence (finite or infinite) of alternating states and actions, starting with an initial state. 
The history $\hist{\EX}$ is the subsequence of actions in $\EX$. An 
operation $\op$ is invoked in $\EX$ when its invocation action appears in $\hist{\EX}$, and responds to an operation when the response action appears (sometime after the invocation) in $\hist{\EX}$. An operation is \emph{complete} in $\EX$ when both the invocation and the matching response appear in $\hist{\EX}$ in order. $\hist{\EX}$ is complete if each operation invoked in $\EX$ is complete. An execution $\EX$ is \textit{well-formed} if no process invokes an action 
before all previously-invoked operations have completed,
%operation before the completion of a previously invoked operation, 
and it is 
\textit{fair} if enabled actions appear in $\EX$ infinitely often. In the rest of the paper 
we consider executions that are fair and well-formed. Operation $\op$ \textit{precedes} $\op'$ in real-time in $\EX$, 
denoted by $\op\bef\op'$, if $\op$'s response appears before $\op'$'s 
invocation in $\hist{\EX}$. Two operations invoked in $\EX$ are \textit{concurrent} if neither precedes the other. 
% A process
% $\pr$ \textit{crashes} in an execution if it stops taking steps; otherwise $p$ is \textit{correct} or \textit{non-faulty}.
% We assume a function $c.\mathcal{F}$ to describe the failure model of a configuration $c$.

\myparagraph{{\bf Adversarial models.}}
We assume that up to $b$ processes in $\idSet$ may be \textit{Byzantine} \cite{LSP82} in an execution $\EX$, 
that is, if it crashes or if it behaves 
%they may stop participating 
%in the service or they may behave 
maliciously (i.e., diverge from their algorithm in an arbitrary manner).
A process $p\in\flexSet$ is \emph{faulty} in an 
execution $\EX$ if it crashes or if it is malicious in \EX; otherwise $p$
is \emph{nonfaulty}. We assume that every process that hosts data may be faulty. 
Any process that invokes a read or a write operation follows the protocol\footnote{As writes and reads may 
be invoked from external entities, we left the study of byzantine behavior during 
reads and writes as a future work to preserve protocol simplicity and clarity.}.
We do assume however that a process invoking a read or a write may \textit{crash} 
at any stage during an operation.
 
 % Processes that follow the algorithm are called \em{correct}, otherwise \em{faulty or Byzantine}. 
 We assume an adversary that can coordinate the Byzantine processes to compromise the system.
We further assume that the adversary controls the network and as such controls the scheduling of all transmitted messages in the network, resulting in asynchronous communication. However, we assume that the adversary cannot prevent the eventual delivery of messages between nonfaulty processes and messages cannot be altered in transit across the network.

% \myparagraph{{\bf Failure Model}.}
% %
% % \nn{[NN: Here we should define the failure model in terms of flexnodes and/or equations. It appears that the distribution of the equations to the flexnodes will be detrimental to also specify how many flexnode failures we may tolerate. Note that we have a tradeoff here of latency vs storage efficiency: the more equations we keep in each node for an object, the less flexnodes we should communicate, and vice-versa.]}
% %

\myparagraph{{\bf Hash Functions}.} We assume that there is an authentication scheme in place that supports two operations: \act{sign}() and \act{verify}(). A sender flexnode $\flexpr{}\in\flexSet$ may use \act{sign}($\flexpr{}, m$), given 
a message $m$ and his identifier $\flexpr{}$, to generate a signature $s$ for $m$. 
Given $s$, a receiver node may use \act{verify}($\flexpr{},m,s$) that evaluates 
to true iff $\flexpr{}$ executed \act{sign}($\flexpr{},m$) in some previous step. 
We assume that signatures are \emph{unforgeable}, i.e. no process (including the Byzantine ones) other than $\flexpr{}$ may invoke \act{sign}($\flexpr{}, m$). 

We also assume the use of \emph{collision-resistant} cryptographic hash function $\hash{}: \{0, 1\}^* \rightarrow \{0, 1\}^n$, for which on sufficiently large output size $n$, and inputs $x$, $y$ the following properties hold: 
%\begin{itemize}
%\item[(i)] 
(i) {\em Deterministic}: if $x=y$, then $\hash{}(x)=\hash{}(y)$; 
%The function produces the same output for the same input every time.
%\item[(ii)] 
(ii) {\em Fixed Output Length}: The output has a fixed size $n$, regardless of input size; and 
%\item[(iii)] 
(iii) {\em Collision Resistance}: if $x\neq y$, then $\hash{}(x)\neq \hash{}(y)$ with very high probability.
 \myparagraph{{\bf Encoding/Decoding with RLNC}.} %In algorithm \treas{}, 
 %
 % \nn{[NN: give the intuition and the properties of the LRNC here. Coded elements, or the equations we will store on the flexnodes could be denoted by the tuple $\tup{\Vec{\alpha}, X'}$, where $\Vec{\alpha}$ is the coefficient vector of size $k$ and $X'$ the outcome of applying the vector on the original object fragments.]}
 %\nn{[NN: below is an attempt to describe RLNC.]}
 %
 Flexnodes use Random Linear Network Codes (RLNC)~\cite{RLNC2003} over a finite field $\mathbb{F}_q$ to encode data objects. In particular, for a given parameter $k$, a flexnode encodes a value $val \in\valSet$, using RLNC, to $n$ ($n$ at least as large as $k$) coded elements, such that any $k$ of the coded elements are sufficient to decode the value $v$. For encoding we do the following: $(i)$ split $val$ into a vector $\mathbf{v} = (v_1, v_2, \ldots, v_k)$ of $k$ elements of $\mathbb{F}_q$; $(ii)$  select a invertible matrix $\mathbf{A}$ of coefficients selected uniformly at random from $\mathbb{F}_q$ such that $\mathbf{A}$ consists of $n$ rows and $k$ columns, where $n \geq k$; and $(iii)$ multiply $\mathbf{A}$ with $\mathbf{v}$ to 
generate a vector of $\mathbf{c} = (c_1, c_2, \cdots, c_n)$ of $n$ elements as  $\mathbf{A}\mathbf{v} = \mathbf{c}$.
%The multiplication 
%is formally illustrated below: 
\remove{KMK
\begin{equation} 
    \begin{pmatrix}
        a_{1,1} & a_{1,2} & \ldots & a_{1,k} \\
        a_{2,1} & a_{2,2} & \ldots & a_{2,k} \\
        \vdots & \vdots & \ldots & \vdots \\
        a_{n-1,1} & a_{n-1,2} & \ldots & a_{n-1,k} \\
        a_{n,1} & a_{n,2} & \ldots & a_{n,k}
    \end{pmatrix}
    \begin{pmatrix}
        v_{1} \\
        v_{2} \\
        \vdots \\
        v_{k}
    \end{pmatrix}
    =
    \begin{pmatrix}
        c_1 \\
        c_2 \\
        \vdots \\
        c_{n-1} \\
        c_{n}
    \end{pmatrix}
\end{equation}
}
We then construct the coded elements as the pair $\ec{\coeff{i}}{c_{i}}$
to represent the coefficients on the $i_{th}$ row in $\mathbf{A}$ and the corresponding generated element $c_{i}$. Given that the generated coefficients are not co-linear, then any $k$ coded elements can be used to recover the elements $(v_1, v_2, \ldots, v_k)$ and thus value $val$. The size of each coded element is 
$\frac{v}{k}$. The key property of RLNC is that data can initially be encoded into the $n$ scalars $c_{i}$, but nodes that possess several $c_i$'s can also create new scalars by taking random linear combinations of the existing $\ec{\coeff{i}}{c_{i}}$. When these are transmitted, they (very likely) provide a new independent vector as long as the receiving node has not already received all the inputs for the linear combination.

%\nn{[NN: probably we should define colinearity here.]}

\remove{
\begin{figure}[h]
    \centering
    \includegraphics[width=0.8\textwidth]{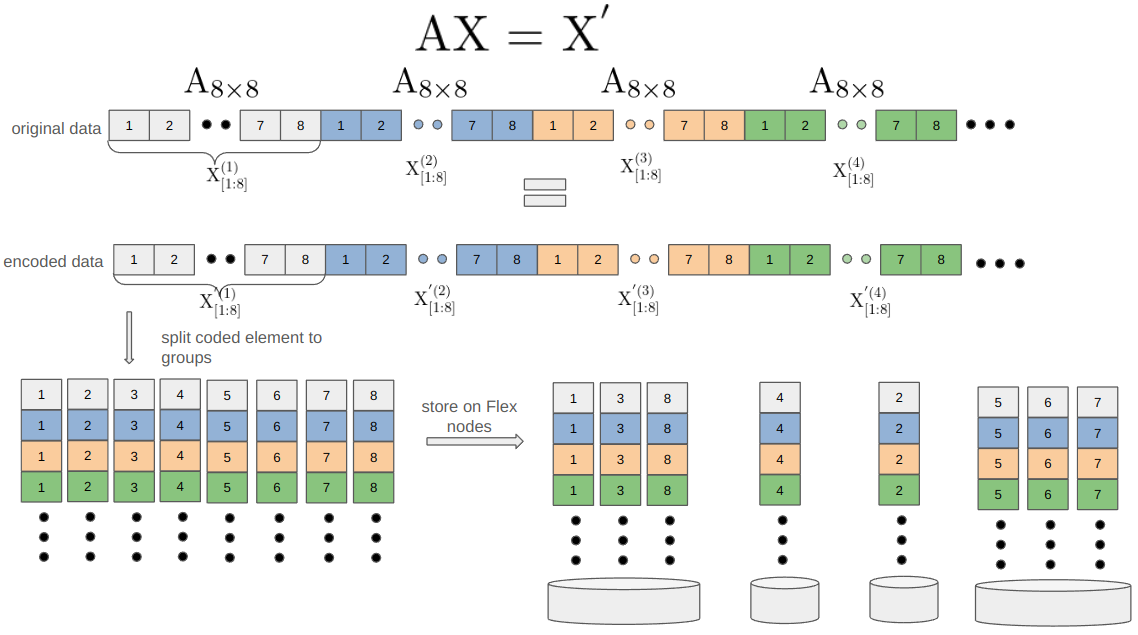}
    \caption{Original data is encoded to an extension field elements, which are encoded by multiplying with a random matrix, here $A$ of dimension $8 \times 8$. The encoded data elements are split into sequences into separate files, and finally, these files are partitioned across a set of Flex nodes, possibly unevenly.}
    \label{fig:my_label}
\end{figure}

\begin{figure}[h]
    \centering
    \includegraphics[width=0.8\textwidth]{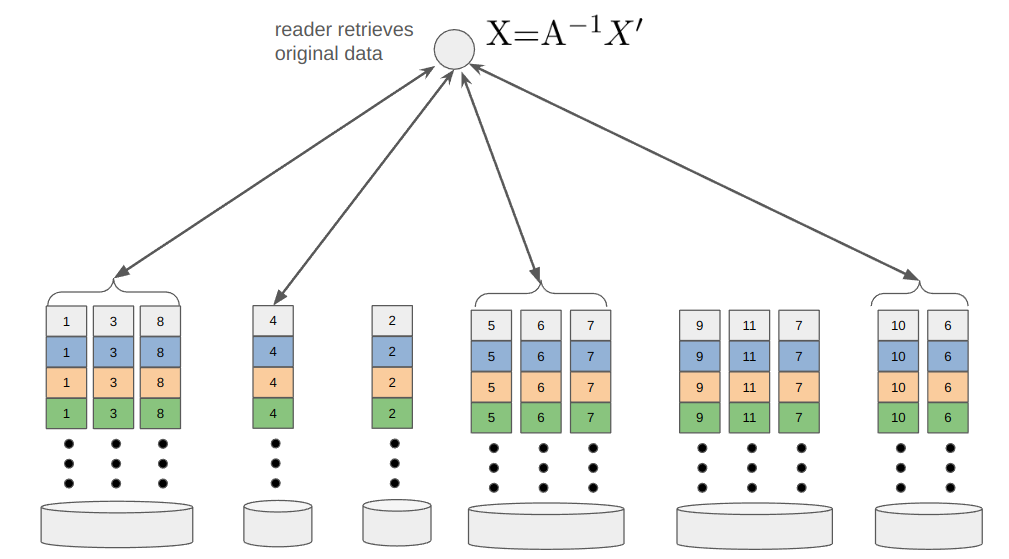}
    \caption{Original data is decoded by a reader by reading the coded elements from a set of Flexnodes.}
    \label{fig:coded-storage-decoding}
\end{figure}
}

\begin{figure}[h]
    \centering
    \begin{minipage}{0.49\textwidth}
        \centering
        \includegraphics[width=\linewidth]{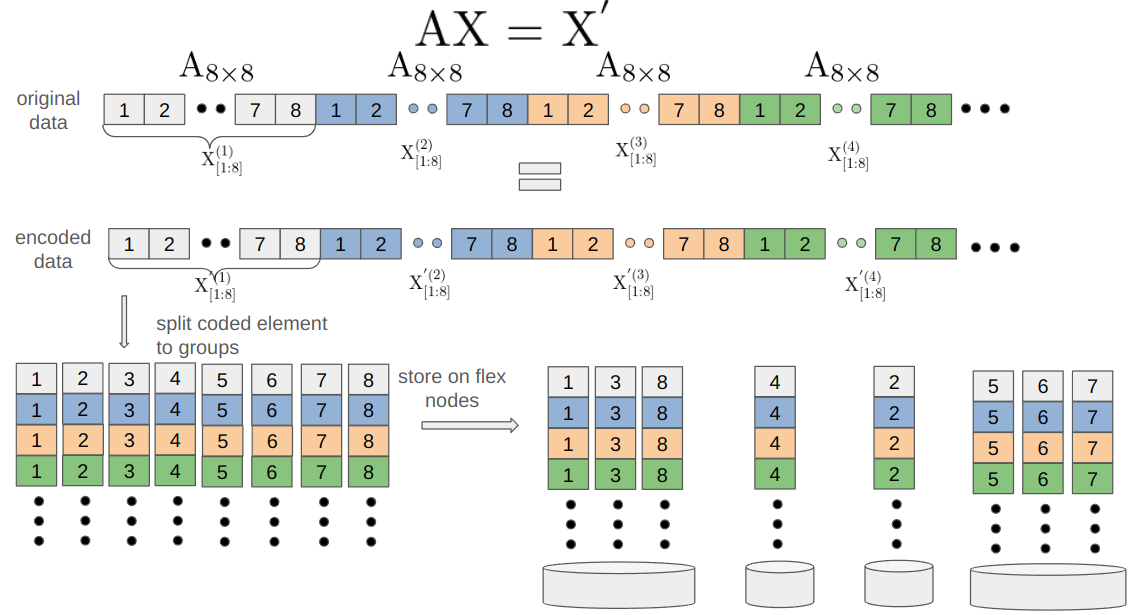}
        \caption{Original data is encoded  by multiplying with a random matrix, here $A$ of dimension $8 \times 8$.}
        \label{fig:coded-storage}
    \end{minipage}\hfill
    % \begin{minipage}{0.49\textwidth}
    %     \centering
    %     \includegraphics[width=\linewidth]{2024-OPTIMUM/ICDCS26/images/coded-storage-decoding.png}
    %     \caption{Original data is decoded by a reader by reading the coded elements from a set of Flexnodes.}
    %     \label{fig:coded-storage-decoding}
    % \end{minipage}
\end{figure}
%\nn{[NN: add discussion on DHT!]}

%\nn{[NN:replace the following with discussion on quorums.]}
\myparagraph{Clusters and Quorums.} A \emph{cluster} is a subset of $\flexSet$, with some additional attributes.
%, i.e. $\confSet\subseteq \idSet$. 
% of %\emph{logical} or \emph{physical} 
% flexnodes. 
% Flexnodes are grouped into logical sets called 
% \emph{clusters}. 
Each cluster is assigned 
% A cluster, with 
a unique identifier $c$ from a set of cluster identifiers $\mathcal{C}$,
and can be used to implement a set of atomic memory objects in $\objSet$. 
%The composition of atomic objects yields an atomic memory space. Thus, the composition of the atomic objects implemented by all the 
% clusters yield the atomic memory $\mathcal{M}$. 
%
%More formally, a 
A cluster $c\in \mathcal{C}$ consists of: (i) $\members{c}\subseteq\flexSet$ the set of flexnodes in $c$, (ii) $\objects{c}\subseteq \objSet$ the set of memory objects
implemented by $c$, and (iii) $\quorums{c}$ a quorum system over the members of $c$.
%, and (iii) a set of operations implementated in $c$. 
%Clusters are \emph{dynamic}, meaning that flexnodes that are not in $\members{c}$ may \emph{join} or existing flexnodes may \emph{depart} the cluster. 
% membership of the cluster, $\members{c}$, and the 
%\emph{Reconfiguration} is the process responsible for modifying the 
%membership of a cluster. 
%
The quorum system $\quorums{c}$ is a set of subsets of the 
members of $c$, or \emph{quorums}, such that every two subsets have a non-empty intersection.
More formally, $\quorums{c}\subseteq 2^{\members{c}}$ such that $\forall \quo{1},\quo{2}\in\quorums{c}$ we have 
$\quo{1}\cap\quo{2}\neq\emptyset$. We assume that $\exists \quo{}\in\quorums{c}$ that is \textit{nonfaulty}, i.e., 
 $\forall \flexpr{}\in \quo{}, \flexpr{}$ is $nonfaulty$.

% The utilization of signatures enables us to employ the class of \textit{b-dissemination} Byzantine quorum 
% systems, as presented in \cite{MR98}. A \emph{b-dissemination} quorum system guarantees that the 
% intersection of any two quorums contains at least one nonfaulty element, even in the presence of $b$ faults. 
% More precisely, $\qs = \quorums{c}$ is a \emph{b-dissemination} quorum system iff $\forall 
% \quo{1},\quo{2} \in \qs$ and $|\quo{1}\cap\quo{2}| \geq b+1$. 

Using signatures we assume that $\qs = \quorums{c}$ is 
a class of \textit{b-dissemination} Byzantine quorum 
system \cite{MR98} where for every pair $\quo{1},\quo{2} \in \qs$, $|\quo{1}\cap\quo{2}| \geq b+1$. 
% A \emph{b-dissemination} quorum system guarantees that the 
% intersection of any two quorums contains at least one nonfaulty element, even in the presence of $b$ faults. 
% More precisely, $\qs = \quorums{c}$ is a \emph{b-dissemination} quorum system iff $\forall 
% \quo{1},\quo{2} \in \qs$ and $|\quo{1}\cap\quo{2}| \geq b+1$. 
%
Additionally, assuming 
%If we assume 
%the use of erasure codes, where 
that each process stores a single RLNC coded element, 
then it is sufficient to guarantee fault tolerance if for every pair $\quo{1}$ and $\quo{2}$ of quorums in $\qs$, 
$|\quo{1}\cap\quo{2}| \geq b+k$: every operation will receive at least $k$ coded 
elements from nonfaulty processes. 
% it is sufficient 
% for the intersection of two quorums 
% %must 
% to contain at least $k$ nonfaulty elements. Consequently, it must be the case that $\forall \quo{1},\quo{2} \in \qs$ and $
% |\quo{1}\cap\quo{2}| \geq b+k$.  
% % Furthermore, we assume that at any given time in the execution, there 
% % exists a single quorum in $\qs$ that does not contain any failed flexnodes, i.e. $\exists \quo{}\in \qs, 
% % s.t. \forall \flexpr{}\in Q, \flexpr{}  $ is $correct$. 

Moreover, given that $b$ processes may crash, then a process that invokes an operation cannot wait for more than $|C|-b$
replies, for $C= \members{c}$. Therefore, every quorum $Q$ in $\qs$ can have size $|Q|\leq|C|-b$.
From this it follows that two quorums $\quo{1}$ and $\quo{2}$ may differ in $b$ elements, i.e. $|\quo{1}\setminus\quo{2}|\leq b$.
Hence, it is necessary for the size of each quorum 
to be $|\quo{}|\geq 2b+k$ to guarantee that their intersection has at least $b+k$ elements.
%
% Given that $|\quo{1}\cap\quo{2}| \geq b+k$ and that $b$ of the flexnodes in $\quo{1}$ may not be included in 
% %reply to an operation that obtained 
% the quorum $\quo{2}$, then we can guarantee that the two quorums overlap
% on $b+k$ elements only if $|\quo{1,2}|\geq 2b+k$. Moreover, if $C= \members{c}$, any quorum should have size at most $|\quo{}|\leq |C|-b$,
% otherwise an operation may not receive enough responses to terminate. 
From the two inequalities it follows:
\[
|C|-b\geq 2b+k ~\act{implies}~b\leq \frac{|C|-k}{3}
\]
From this it follows that the size of each quorum $\quo{}\in\qs$ is:
\[
|\quo{}|\leq|C|-b ~\act{implies}~|\quo{}|\leq \frac{2|C|+k}{3}
\]
To achieve the highest fault tolerance, for the sequel we assume that $b<\frac{|C|-k}{3}$, and we assume a \textit{majority quorum} system with the size for each quorum to be $\quo{}\in\qs$, $|\quo{}|=\ceil{\frac{2|C|+k}{3}}$. To this end, the size of each quorum depends on our choice of the cluster size $|C|$,
and the replication factor $k$ of the coding scheme we use (for $k=1$ being replication). 

\myparagraph{Consistency.} 
We consider \textit{atomic/linearizable} \cite{Lynch1996, HW90} R/W objects that support two types of operations:  
$\act{read}()$ and $\act{write}(v)$. 
%and $\act{reconfig}(c)$. 
A \act{read}() returns the value of the atomic object, while 
a $\act{write}(v)$ attempts to modify the value of the object to $v\in\valSet$.
%, and the $\act{reconfig}(c)$ that attempts to install a new configuration $c\in\confSet$.
 %We assume \textit{well-formed} executions where each client may invoke one operation ($\act{read}()$, $\act{write}(v)$ or $\act{reconfig}(c)$)  at a time. 
A complete history $\hist{\EX}$ is atomic if there exists some 
total order on the operations of $\hist{\EX}$ such that it respects 
the real-time order $\bef$ of operations, and is consistent with 
the semantics of the two operations. 

More formally, an implementation of a R/W object satisfies the atomicity (linearizability \cite{HW90}) property if the following holds \cite{Lynch1996}. Let the set $\Pi$ contain all complete {read/write} operations in any well-formed execution $\EX$ of $A$. 
Then there exists an irreflexive partial ordering $\prec$ on $\Pi$
satisfying the following:	
	\begin{itemize}
		%\item [\em P1.] No operation has infinitely many other 
		%			operations ordered before it.
		\item [\bf A1.] (Real-Time Ordering) 
		%					The partial order is consistent with the 
		%					external order of invocation and responses, that is, there do 
		%					not exist operations $\op_1$ and $\op_2$, 
		%					such that $\op_1$ completes before $\op_2$ starts, 
		%					yet $\op_2 \prec \op_1$.
		For any operations $\op_1$ and $\op_2$ in $\Pi$,  if $\op_1\bef\op_2$, then it
		cannot be the case that $\op_2\prec \op_1$.
		\item[\bf A2.] (Total-Ordering of Writes)
		If $\pi\in\Pi$ is a write operation and $\pi'\in\Pi$ is any {read/write} operation,  
		then either $\pi\prec \pi'$ or $\pi'\prec \pi$.
		%					All write operations are totally 
		%					ordered and every read operation is ordered with respect 
		%					to all the writes.
		\item[\bf A3.] (Partial-Ordering of Reads with respect to Writes)  
		The value returned by a read operation is the value 
		written by the last preceding write operation according to
		$\prec$ (or the initial value if there is no such write).
		%		Every read operation ordered after any writes returns
		%the value of the last write preceding it in the partial order, and any
		%read operation ordered before all writes returns the initial value
		%of the object.
\end{itemize}

\myparagraph{{\bf Tags.}}
We use logical tags to order operations. A tag $\tg{}$ is defined as a pair $(z, w)$, where $z \in \mathbb{N}$ and $w \in \mathcal{W}$, an ID of a writer.
Let $\mathcal{T}$ be the set of all tags.
Notice that tags could be defined in any totally ordered domain and if this domain is countably infinite, then 
there can be a direct mapping to the domain we assume. 
% and we denote by  $\mathcal{T}$  the set of all possible tags. 
For any  $\tg{1}, \tg{2} \in \mathcal{T}$ we define  $\tg{2} > \tg{1}$ if $(i)$ $\tg{2}.z > \tg{1}.z$ or $(ii)$ $\tg{2}.z = \tg{1}.z$ and $\tg{2}.w > \tg{1}.w$.

%% file: byzantine_rw_algorithm.tex
At the core of the Optimum service lies an algorithm that implements an 
atomic shared memory service over a message-passing environment. 
Our algorithm utilizes the RLNC coding scheme to reduce message complexity and storage cost, tolerate Byzantine failures, and implement multiple objects. 

We present our approach in three stages. First, we  
specify an atomic R/W implementation using RLNC and 
assuming a \emph{static}
environment, where the set of flexnodes is fixed 
and known in advance, and 
%. In this stage, we assume the implementation of
a single R/W object. 
In the second stage, we focus on developing an algorithm for manipulating multiple concurrent R/W atomic objects in a
scalable manner. Finally, in the third stage we examine how to improve the longevity of the service, by considering 
\textit{dynamic} environments, where flexnodes may join and 
%participating flexnodes may 
leave the service. 

% Subsequent sections investigate how 
% the algorithm can be adopted in the dynamic environment, where flexnodes 
% may join or depart the service, and how to manage multiple objects. 

\subsection{Preliminaries}
\label{ssec:rwpreliminaries}
% Before proceeding to the description of the algorithm we present system assumptions, 
% definitions, and notation we will use in this section. 

%We focus on the implementation of a MWMR, atomic R/W memory emulation. 
For this section, we assume a \textit{static} environment and a single flexnode cluster $c$.
Since we assume a single cluster, for the rest of this section we denote by $C= \members{c}$ the set of 
flexnodes in $c$, and by $\qs = \quorums{c}$ the quorum system defined over the cluster $c$. 
%fixed set (cluster) of flexnodes $C$ and each flexnode is aware of $C$. 
We also focus on the implementation of a single 
atomic object, thus $|\objects{c}| = 1$. 
%By the composability property of atomicity, multiple instances of this algorithm may  yield a complete atomic memory. 
Section \ref{sec:recofi} will examine how 
we can extend this algorithm for the dynamic, multi-object, environment. 

The RLNC encoding scheme is used to tolerate flexnode failures. 
Write operations use RLNC to encode each value $v$ to a set of $n=|C|$ coded
elements, out of which any $k$ can be used by read operations to decode $v$. 
Tags are associated with each coded element, to facilitate operation ordering.
We assume the use of \emph{non-forgeable} signatures to sign the tag-coded element pairs: a flexnode $\flexpr{}$ uses the operation $\act{sign}(\flexpr{}, \tup{\tg{},e})$ to generate a signature $s$ for a $\tup{\tg{},e}$ pair, and uses the operation 
$\act{verify}(\flexpr{}, \tup{\tg,e}, s)$ to verify the correct signature. 

% We assume that up to $b$ 
% %$B\subseteq C$, s.t. $b=|B|<\frac{|C|-k}{3}$, of 
% flexnodes in $C$ may 
% fail arbitrarily. We also assume all the write operations are invoked by \emph{correct}
%flexnodes. Therefore, 
Since we assume the use of signatures, any flexnode invoking a write operation
will sign every coded element and its associated tag pair. As a result, Byzantine flexnodes may 
%harm 
violate the safety and liveness of 
executions in the following ways: (i) by failing to reply to value requests, and 
(ii) by suppressing 
%So the only power that 
%the rest of the byzantine flexnodes will have is to suppress 
the receipt of 
a newer tag and replying with an outdated, correctly signed pair.

\subsection{ The Read/Write Protocol}
\label{ssec:rwregister}
% In this section, we present the Optimum's R/W protocol. The protocol
% assumes the following system conditions: (i) a \emph{static} set of flexnodes $C\subset\flexSet$ that it is given to any read/write operation, and (ii) $b < \frac{|C|}{3}$ flexnodes may be Byzantine. Flexnodes behave
% as proxies for external clients that want to perform read/write operations.
% We assume that flexnodes are not Byzantine when invoking write 
% operations. 

% detailed description of the Optimum's R/W algorithm. 
% The algorithm assume that the algorithm runs on top of a given set of flexnodes, $C\subset\flexSet$. As noted in the model, each flexnode in $C$, holds 
% object replicas, and invokes read/write operations. 
% An environmental input to some flexnode $\flexpr{}\in C$, triggers 
% the invocation of a write/read operation at $\flexpr{}$. 

% When assuming that any flexnode that invokes a write is \emph{correct} 
% then we can reuse algorithms for the crash model given that we use signatures. 
% Below we demonstrate this claim by presenting a EC algorithm presented in \cite{ARES19}.
%\subsubsection{Non-Malicious Writers}

To achieve a modular 
algorithm, and make our implementation compatible with \cite{ARES19}, the algorithm is expressed in 
terms of \emph{data access primitives} (DAPs) as defined in \cite{ARES19}. 
The \optimum{} R/W protocol is a generalization of the erasure-coded algorithm 
presented in \cite{ARES19} with the following differences: (i) it uses 
RLNC for encoding and decoding the object value, (ii) it assumes 
Byzantine failures, and (iii) it employs signatures to limit Byzantine behavior, thereby preventing the production of incorrect coded elements and tag pairs. 
The pseudocode of the DAPs 
%presented in Section \ref{sec:byz-rw} 
appears in Algorithms \ref{algo:atomicity:generic1}, \ref{fig:bft-mwmr-optimum-dap}, and \ref{fig:bft-mwmr-optimum}.
Algorithm \ref{algo:atomicity:generic1} shows the generic algorithm, while Algorithm \ref{fig:bft-mwmr-optimum-dap} the DAP implementation and Algorithm \ref{fig:bft-mwmr-optimum} the responses from the data hosts.
% Notice that the algorithms resemble the one presented in \ares{}, enhanced with cryptographic 
% functions to tolerate Byzantine failures. 
%Due to lack of 
% space we present the pseudocode of the DAPs in the optional Appendix.

\myparagraph{State Variables:} Each flexnode $\flexpr{i}$ maintains 
the latest tag-value pair that $\flexpr{i}$ has written or decoded (as a result of a write or read operation resp.), and a set $List$ containing of 
up to $(\delta+1)$ triplets of the form
 \tup{tag, coded element, signature}.  
Essentially, $\delta$ determines the maximum number of concurrent write
operations the algorithm may tolerate without violating safety. 
%The smaller the $\delta$ the fewer concurrent operations
%are allowed in the service. 
When $\delta=1$ the system resembles 
a SWMR model, while when $\delta=|C|$ then all the flexnodes in $C$ 
may invoke a write concurrently. 
% As we see in the proof of correctness
% of the algorithm, the parameter $\delta$ is necessary for the algorithm
% to satisfy atomicity.
%Initially the set at each flexnode $\flexpr{i}$ is empty. 

\myparagraph{Read/Write Operations (DAPs):}
%We assume that the writes are all invoked by nonfaulty flexnodes. 
The write and read protocols 
%(Algorithm \ref{algo:atomicity:generic1}) 
are based on the template 
algorithm presented in \cite{ARES19}; a write discovers the max tag 
in the system using the \dagettag{} primitive, increments that tag, 
and writes the new tag-value pair using the $\daputdata{*}{*}$ primitive; 
the read discovers the latest tag-value pair using the \dagetdata{}
primitive and writes back the pair discovered using the $\daputdata{*}{*}$
before returning that pair. Below we describe the implementation of 
each DAP.

\input{algorithm-aresec-byzantine}

$\dagettag{c}$: During the execution of a  $\dagettag{c}$ primitive, a flexnode queries all the $\members{c}$ for the highest tags in their  $Lists$.
%responses from $\left\lceil \frac{n+k}{2} \right\rceil$ servers.
Upon receiving the {\sc get-tag} request, a flexnode 
responds with the highest $\tup{\tg{},e}$ pair that it stores 
in its local $List$ set, i.e. 
$\tg{max} \equiv \max_{(t,*) \in List} \{t\}$. 
Notice that since each pair in $List$ is signed, and since signatures
are unforgeable, the flexnode $f$  replies with the triple 
$(\tup{\tg{},e}, \act{sign}_f(\tup{\tg{},e}))$ 
where $\tg{}=\tg{max}$. Once the requesting flexnode receives 
replies from a quorum $Q$ it discovers and 
returns the highest, correctly signed tag among those replies.

$\daputdata{c}{\tup{\tg{w}, v}}$: During the  execution of the primitive  $\daputdata{c}{\tup{\tg{w}, v}}$, a flexnode $\flexpr{i}$ encodes the value $v$ using RLNC into a vector $(e_1,\ldots,e_{c})$ of coded elements of length $|\members{c}|$, and sends a {\sc put-data} message with the triple $(\tup{\tg{w},e_j}, \act{sign}_{\flexpr{i}}(\tup{\tg{w},e_j}))$ to the flexnode $\flexpr{j}\in \members{c}$. 
When a flexnode $\flexpr{j}$ receives a {\sc put-data} message with a triple 
$(\tup{\tg{w},e_j}, \act{sign}_{\flexpr{i}}(\tup{\tg{w},e_j}))$, it first verifies the 
signature and then adds the triplet in its local $List$ set,
it removes the elements with the smallest tags exceeding the 
length $(\delta+1)$ of the $List$, and responds with an $\act{ACK}$ to the requesting flexnode $\flexpr{i}$.
Once a quorum replies to $\flexpr{i}$ the operation completes. 

$\dagetdata{c}$: During the execution of a  $\dagetdata{c}$ primitive,
a flexnode $\flexpr{i}$ sends {\sc query-list} messages to every 
flexnode in $\members{c}$, requesting their local variable $List$. 
Once $\flexpr{i}$ receives the $List$ from a quorum $\quo{}$,
it first discovers the correctly signed $\tup{\tg{}, e}$ pairs 
from those sets %(A\ref{fig:bft-mwmr-optimum-dap}:L\ref{fig:bft-mwmr-optimum-dap:decode:getdata:k})
 and then selects the tag $\tg{}$, such that: (i) it appears in more than $k$ 
 %times in the set of 
 verified pairs, 
i.e. its associated value can be decoded, and (ii) it is the highest 
tag among the decodable tags. %((A\ref{fig:bft-mwmr-optimum-dap}: L\ref{fig:bft-mwmr-optimum-dap:max-decodeable-tag}). If it discovers such 
% a tag it decodes the value using the coded elements associated with 
% $k$ pairs containing $\tg{}$ and returns $\tup{\tg{}, v}$ pair. 
If no such a tag exists the operation does not complete. 
 
% \myparagraph{Write Protocol:} A write operation proceeds in two phases. During the first phase it broadcasts a {\sc query} message to all the flexnodes in $C$ and waits for a \emph{b-dissemination} quorum $Q$ to reply. Next, it attempts to discover the maximum, signed, $\tup{\tg{},*}$ pair among the replies, and generates a new tag $\tup{\tg{}.ts+1, \flexpr{}}$, larger 
% than the maximum tag observed. In Phase II the write operation
% encodes the value $v$ using RLNC into $c=|C|$ coded elements 
% $\{e_1,\ldots,e_{c}\}$ and sends the triple $\tup{\tup{\tg{},e_j}, \st_{\flexpr{i}}}$ to the flexnode $\flexpr{j}\in \members{C}$, 
% where $\tg{}$ the generated tag and $\st_{\flexpr{i}}=\act{sign}(\flexpr{i},\tup{\tg{},e_j})$. Once a quorum replies with acknowledgments the write operation completes. 

% \myparagraph{Read Protocol:} A read operation also involves two phases. 
% During Phase I, the read queries a quorum $Q$ of flexnodes in $C$ for their local $List$ variable. When the 

% \myparagraph{Receiver Protocol:} Before proceeding to the description of the read and write protocols let us present the messages that can be processed at a flexnode and the actions the flexnode takes in each message received. 

% \begin{theorem} \label{thm:single-object:safety}
%     Given a set of flexnodes $C$ where up to $b$, $b<\frac{|C|-k}{3}$, flexnodes may be Byzantine, any execution $\EX$ of BFT-MWMR-\optimum{} implements an atomic read/write register. 
% \end{theorem}

\subsection{BFT-MWMR-\optimum{} Correctness}
\input{safety_liveness.tex}

% \subsection{The {\sc ares-ec} with RLNC}
% \label{sec:aresec}
% \input{sec_algorithm_aresec.tex}

% \subsection{The \optimum{} Read/Write Protocol}
% \label{sec:r/w}
% \input{sec_algorithm_optimum.tex}

%% file: algorithm-aresec-byzantine.tex
\begin{algorithm}[!ht]
\begin{algorithmic}[2]
    \begin{multicols}{2}
        {\scriptsize
            %\Part{Generic Algorithm $A_1$}
                
            % \State $op$, a record with fields:
            % \State ~$type\in\{\act{read}, \act{write}\}$, initially $\bot$
            % \State ~$num\in\Nat$, initially 0
            % \State ~$proc\in\flexSet$ initially $\flexpr{i}$
            % \State{at each flexnode $\flexpr{i}\in \members{c}$}   
            % \Statex
            \State{\bf State Variables:}
            \State $List \subseteq  \tgSet \times \ecSet \times \Sigma$, initially  $\emptyset$
            %\State $\beta\in \Nat$ s.t. $1\leq\beta\leq k$, initially $1$
            \State $\tg{}\in \tgSet$, initially $\tup{0,\flexpr{i}}$
            \State $v\in \valSet$, initially $\bot$  
            
            \Statex  
           
            \Operation{read}{$C$} 
            %\State $wCounter\gets wCounter+1$
            \State $\tup{\tg{}, v} \gets \dagetdata{C}$
            \State $\daputdata{C}{ \tup{\tg{},v}}$
            \State return $ \tup{\tg{},v}$
            \EndOperation
            
            \Statex
            
            \Operation{write}{$C, v$} 
            %\State $wCounter\gets wCounter+1$
            \State $\tg{} \gets \dagettag{C}$
            \State $\tg{w} \gets \tup{\tg{}.ts + 1,  w}$
            \label{algo:atomicity:generic1:increment}
            \State $\daputdata{C}{\tup{\tg{w},v}}$
            \EndOperation
            %\EndPart
        }
    \end{multicols}
    \end{algorithmic}
\caption{BFT-MWMR-\optimum{} Read and Write operations at flexnode $\flexpr{i}$ on a given cluster $C$.}
\label{algo:atomicity:generic1}
\vspace{-1em}
\end{algorithm}

    \begin{algorithm*}[!ht]
        \begin{algorithmic}[2]
            {\scriptsize
            \begin{multicols}{2}
                
                \Procedure{get-tag}{C}
                    %\State {\bf send} $(\text{{\sc query-tag}})$ to each  $\flexpr{}\in \members{C}$
                    \State {\bf send} $(\text{{\sc query-tag}})$ to each  $\flexpr{}\in C$
                    %\State {\bf until} $\flexpr{i}$ receives $(\tup{\tg{\flexpr{}}, e_{\flexpr{}}}, \sigma_{\flexpr{}})$ from a set $\quo{r}$ s.t.\WRP $\quo{r}\subseteq C$ and $|\quo{r}| \geq \left\lceil \frac{2|C| + k}{3}\right\rceil$
                    \State {\bf until} $\flexpr{i}$ receives $(\tup{\tg{\flexpr{}}, *}, \act{sig}_f(\tup{\tg{\flexpr{}}, *}))$ from each $f$ \WRP in a quorum $\quo{r}$ 
                    % s.t.\WRP $\quo{r}\in\qs$ and $|\quo{r}| \geq \left\lceil \frac{2|C| + k}{3}\right\rceil$
                    
                    \State $\tg{max} \gets \max( \bigcup_{f\in\quo{r}~ \wedge~\act{verify}(\act{sig}_f(\tup{\tg{\flexpr{}}, *}))=\act{true}\}}
                    \{\tg{\flexpr{}} \})$
%           (\tup{\tg{\flexpr{}}$
                    %\max(\{\tg{\flexpr{}} : (\tup{\tg{\flexpr{}}, -}, \sigma_{\flexpr{}}) \text{ received from }\flexpr{}\in\quo{r}$\WRP 
                %    $\wedge~\act{verify}(\flexpr{}, \tup{\tg{\flexpr{}}, -}, \sigma_{\flexpr{}})=\act{true}\})$
                    %\State \nn{$\tg{max} \gets \max(\{\tg{\flexpr{}} : (\tup{\tg{\flexpr{}}, -}, \sigma_{\flexpr{}}) \text{ received from } \frac{|C|-k}{3}+1 \text{ flexnodes in }\quo{r}$\WRP 
                    % $\wedge~\act{verify}(\flexpr{}, \tup{\tg{\flexpr{}}, e_{\flexpr{}}}, \sigma_{\flexpr{}})=True\})$}
                    \State {\bf return} $\tg{max}$
                \EndProcedure

                \Statex				
                    
                \Procedure{put-data}{$C, \tup{\tg{},v})$}
                    \State $(e_1,e_2,\ldots,e_{|C|})\gets Encode(v)$
                        
                    \State {\bf send} $(\text{{\sc put-data}},\tup{\tg{},e_j}, \act{sig}_{\flexpr{i}}(\tup{\tg{},e_j}))$ \WRP to each $\flexpr{j} \in C$
                    
                    \State {\bf until} $\flexpr{i}$ receives {\sc ack} from a quorum $\quo{r}$ 
                    % s.t. \WRP $\quo{r}\subseteq C\wedge |\quo{r}| \geq \left\lceil \frac{2|C| + k}{3}\right\rceil$
                \EndProcedure
                    
                \Statex
                \Statex
                    
                \Procedure{get-data}{C}
                    \State {\bf send} $(\text{\mtype{query-list}}, \tg{})$ to each  $\flexpr{}\in C$
                    
                    \State {\bf until} $\flexpr{i}$ receives $List_{\flexpr{}}$ from each $f$ in a quorum $\quo{r}$
                    % s.t.\WRP $\quo{r}\subseteq C$ and $|\quo{r}| \geq \left\lceil \frac{2|C| + k}{3}\right\rceil$
                    % \State  $Tags_{*}^{\geq k} = $ set of tags that appears in  $k$ lists	\label{line:getdata:max:begin}

                    %\State\Comment{find verified tag-ec pairs}
                    \State $Pairs_{ver}\gets\{ \tup{\tg{},e}: (\tup{\tg{},e}, \act{sig}_{\flexpr{}}(\tup{\tg{},e})\in List_{\flexpr{}} \wedge \flexpr{}\in\quo{r}$\WRP $ \wedge~ \act{verify}(\act{sig}_{\flexpr{}}(\tup{\tg{}, e}))=~\act{true}\}$
                    
                    \State $L_{\tg{}}=\{List_{\flexpr{}}:  \tup{\tg{}, *}\in Pairs_{ver} \wedge  \tup{\tg{}, *}\in List_{\flexpr{}}\}$
                    \State \label{fig:bft-mwmr-optimum-dap:decode:getdata:k} $Tags_{dec}^{\geq k} \gets\{\tg{}: |L_{\tg{}}| \geq k\}$
                    \State \label{line:getdata:max:end}
                    \If{ $Tags_{dec}^{\geq k}\neq \emptyset$}
                        \State  $\tg{max}^{dec} \gets \max_{\tg{} \in Tags_{dec}^{\geq k}}(\tg{})$       
                       \label{fig:bft-mwmr-optimum-dap:max-decodeable-tag}
                        \State $\tg{} \gets \tg{max}^{dec}$
                        \State $v \gets Decode(\{e: \tup{\tg{max}^{dec},e}\in Pairs_{ver}\})$
                        \label{fig:bft-mwmr-optimum-dap:decode}
                        \State {\bf return $\tup{\tg{max}^{dec},v}$}
                    \Else
                        \State {\bf return $\tup{\tg{},v}$}
                    \EndIf
                \EndProcedure

            \end{multicols}
        }
        \end{algorithmic}	
        \caption{
            %for  template $A_1$ to implement 
            BFT-MWMR-\optimum{} DAPs implementation at flexnode $\flexpr{i}$ on a given cluster $C$.}
        \label{fig:bft-mwmr-optimum-dap}
        \vspace{-1em}
    \end{algorithm*}

\begin{algorithm*}[!ht]
\begin{algorithmic}[2]
    {\scriptsize
    \begin{multicols}{2}
        % \State \nn{at each flexnode $\flexpr{i} \in \members{c}$}
        % %in configuration $c_k$}
        % \Statex
        % \State{\bf State Variables:}	
            
        %     %}\EndPart        
    
        % \Statex
        
        \Receive{{\sc query-tag}}{$\flexpr{i},c$}
            \State $\tg{max} = \max(\tup{\tg{},*} : (\tup{\tg{},*},*) \in List)$
            %\State \nn{$L_{max}\gets \{(\tup{\tg{},e},\st): \tg{}=\tg{max} ~\wedge~(\tup{\tg{},e},\st)\in List\}$}
            %\State \nn{\textbf{send} $(\tup{\tg{},e},\st)\in L_{max}$ to $q$}
            \State \textbf{send} $\tg{max}$ to $q$
        \EndReceive
        
        \Statex

        \Receive{{\sc query-list}, $\tg{}'$}{$\flexpr{i},c$}
            \State \textbf{send} $\{(\tup{\tg{},e}, *) : (\tup{\tg{},e}, *)\in List \wedge \tg{}>\tg{}'\}$ to $q$
              \label{fig:bft-mwmr-optimum:query-list}
        \EndReceive
            
        \Statex
        
        \Receive{{\sc put-data}, $(\tup{\tg{},e},\act{sig}_{\flexpr{i}}(\tup{\tg{},e}))$}{$\flexpr{i},c$}
         \If{$\act{verify}(\flexpr{i}, \tup{\tg{},e}, \st_{\flexpr{i}}))=\act{true}~\wedge~(\tup{\tg,*},*)\notin List$}
           \label{receive:verify:if} 
                \State $List \gets List \cup 
                \{(\tup{\tg{}, e_i},\act{sig}_{\flexpr{i}}(\tup{\tg{}, e_i}))\}$   \label{fig:bft-mwmr-optimum:add-tag}
                \If{$|List| > (\delta+1)$}
                    \State $\tg{min}\gets\min\{\tg{}: \tup{\tg{},*}\in List\}$
                    \State $L_{min}\gets \{\tup{\tg{},e}: \tg{}=\tg{min} ~\wedge~(\tup{\tg{},e},*)\in List\}$
                    
                    \State $List \gets List \backslash~L_{min}$  \label{fig:bft-mwmr-optimum:remove}
                    %\cup \{  (  \tg{min}, \bot)  \}$
                    \label{line:server:removemin}
                    %\State $E \gets E \setminus\{\tup{\tg{},e}: \tg{}=\tg{min} ~\wedge \tup{\tg{},e}\in E\}$
                    %\State $List \gets List  \cup \{  (  \tg{min}, \bot)  \}$\label{line:server:removemin}
                \EndIf
            \EndIf
            \State  \textbf{send} \mtype{ack} to $q$
        \EndReceive
    \end{multicols}
}
\end{algorithmic}	
\caption{BFT-MWMR-\optimum{} response protocols at flexnode $\flexpr{i}$ in a given cluster $C$.}
                \label{fig:bft-mwmr-optimum}
                %\vspace{-1em}
\end{algorithm*}

%% file: safety_liveness.tex
% The pseudocode of the DAPs presented in Section \ref{sec:byz-rw} appears in Algorithms \ref{algo:atomicity:generic1}, \ref{fig:bft-mwmr-optimum-dap}, and \ref{fig:bft-mwmr-optimum}.
% Algorithm \ref{algo:atomicity:generic1} shows the generic algorithm, while Algorithm \ref{fig:bft-mwmr-optimum-dap} the DAP implementation and Algorithm \ref{fig:bft-mwmr-optimum} the responses from the data hosts.
% Notice that the algorithms resemble the one presented in \ares{}, enhanced with cryptographic 
% functions to tolerate Byzantine failures. 

%\input{algorithm-aresec-byzantine}

To prove the correctness of the proposed Algorithm \ref{algo:atomicity:generic1}, we need to show that it is both \textit{safe} and \textit{live}. 
% We first proceed to proof that for any given 
% execution $\EX$ containing operations of the proposed implementation, then examining any pair of operations in $\EX$
% satisfy the DAP consistency properties (i.e. Property~\ref{property:dap}). That is, the tag returned by a $\act{get-tag}()$ operation is larger than the value written 
% by any preceding $\act{put-data}()$ operation, and the 
% value returned by a $\act{get-data}()$ operation is either written by a $\act{put-data}()$ operation or is the initial value of the object. Next, assuming that there cannot be 
% more that $\delta$ $\act{put-data}()$ operations concurrent 
% with a single $\act{get-data}()$ operation, we show that each 
% operation in our implementation terminates. Otherwise 
% a $\act{get-data}()$ operation is at risk of not being 
% able to discover a decodable value and thus fail to terminate
% and return a value.
%
% In this subsection, we discuss the \textit{safety} and \textit{liveness} properties of the atomic read/write object, implemented by Algortihm \ref{algo:atomicity:generic1}.
We prove safety by showing that any execution of the proposed algorithm satisfies the atomicity 
properties A1-A3 presented in Section \ref{sec:model}. Liveness is satisfied 
in two cases: (i) each operation terminates in any execution, given that the 
failure model assumed is satisfied, and (ii) any read operation is able 
to decode the object value given that no more than $\delta$ writes are 
invoked concurrently in the service.

\myparagraph{Liveness:} 
To reason about the liveness of the proposed algorithm, we define a concurrency parameter $\delta$ which captures the maximum number 
of $\act{put-data}$ operations (or max number of writers) in any execution of the algorithm. Termination (and hence liveness) of the DAPs is then guaranteed in an execution $\EX$ over a set of flexnodes $C$, provided that $b$,  $b<\frac{|C|-k}{3}$, flexnodes in $C$ are byzantine, and no more than $\delta$ $\act{put-data}$ may be concurrent at
any point in $\EX$. If the failure model is satisfied, then any operation invoked by a non-faulty flexnode will
collect the necessary replies independently of the progress of any other client process in the system and terminate.
Preserving $\delta$ on the other hand, ensures that any operation will be able to decode a written value. These claims are captured by the following lemmas which lead to the main result of liveness.

\begin{lemma}[Fault Tolerance]
\label{lem:terminate}
In any execution $\EX$ of \optimum{}, every operation terminates
if up to $b$, $b<\frac{|C|-k}{3}$, flexnodes are Byzantine in $C$.
\end{lemma}

\begin{proof}
    This lemma follows directly from the algorithm. A Byzantine flexnode can arbitrarily choose not to reply. Since it is assumed that \( b < \frac{|C| - k}{3} \), we cannot wait for more than \( |C| - b = \frac{2|C| + k}{3} \) flexnodes to respond. Consequently, each DAP waits for \( \frac{2|C| + k}{3} \) replies. Given our failure model, this ensures that the required responses are always received, allowing each operation to terminate successfully in $\EX$. 
\end{proof}

Now that we have demonstrated that any operation terminates based on the messages it receives, we must examine whether an operation terminates when it is required to decode a coded value. According to the algorithm, write operations only encode values and do not decode them. Therefore, in the next lemma, we focus exclusively on whether a read operation can always decode and return the object's value.

\begin{lemma}[Concurrency Factor]
\label{lem:delta}
In any execution $\EX$ of \optimum{}, where $n=|C|$ the number of flexnodes such that at 
 most $b$, $b<\frac{|C|-k}{3}$, flexnodes may be Byzantine and using an $(n,k)$-RLNC code, a read operation $\rd$ 
returns the value $v$ where at most $\delta$ write operations are concurrent with $\rd$ and $k$ encoded elements are sufficient to decode.
\end{lemma}

\begin{proof}
    This lemma follows from the fact that any flexnode has a variable $List$ that maintains 
    the largest tags it received for the object. 
    
    Let us assume by contradiction that there exists an execution, say $\EX$, 
    where $\delta$ write operations are concurrent to $\rd$ and $Tags_{dec}^{\geq k}=\emptyset$. By \optimum{}, $\rd$ 
    collects the coded elements during the $\dagetdata{C}$ DAP, and 
    stores any tag that appears more than $k$ times in the replies in the 
    set $Tags_{dec}^{\geq k}$. Let $\quo{\rd}$ of $\frac{2|C|+k}{3}$ be 
    the set of flexnodes that reply with their local $List$ to the $\dagetdata{C}$ operation of $\rd$. 

    Let now write $\wrt$ be the last non-concurrent operation to $\rd$, i.e. $\wrt\bef\rd$ in $\EX$, and $\Lambda$ be the set of all concurrent writes 
    to $\rd$. Let $\quo{\wrt}$ be the set of flexnodes that replied 
    during the $\daputdata{C}{\tup{\tg{\wrt}, v}}$ operation of $\wrt$. Every 
    flexnode $\flexpr{}\in \quo{\wrt}$ adds a signed $\tup{\tg{\wrt}, e_{\flexpr{}}}$ in its $List$ before replying to $\wrt$. Note that since $|\quo{\wrt}|= \lfloor\frac{2|C|+k}{3}\rfloor$ and since it is not concurrent with $\rd$ 
    then it is completed before $\rd$ is invoked and 
    it follows that $|\quo{\rd}\cap\quo{\wrt}|\geq\frac{|C|+2k}{3}$ reply 
    to $\rd$ as well. Furthermore,  
    since $b <\frac{|C|-k}{3}$ flexnodes can be Byzantine, then at least 
    $\frac{|C|+2k}{3}-\frac{|C|-k}{3}=k$ flexnodes of that intersection 
    are correct.
    
    Let us assume w.l.o.g., that $\tup{\tg{\wrt}, *}$ is the largest tag received by any 
    $\flexpr{}\in\quo{\wrt}$, and hence the largest tag in $List_{\flexpr{}}$.
    Once in the $List_{\flexpr{}}$, the pair $\tup{\tg{\wrt}, e_{\flexpr{}}}$ may be 
    removed from $List_{\flexpr{}}$ only when $\delta+1$ larger tags are 
    added in $List_{\flexpr{}}$ (A\ref{fig:bft-mwmr-optimum}:L\ref{receive:verify:if}). Also, note that $List$ is a variable of type set and the maximum number of concurrent writes is $\delta$ and $List$ always retains the $\delta$ largest tags and coded-element pairs.
    %So, if any flexnode in $\quo{\wrt}$ receives \mtype{put-data} message due to a write $\wrt_i\in\Lambda$ then will attempt to 
    %add the tag written by $\wrt_i$ in the $List_{\flexpr{}}$ set once. 
    However, according to our assumption $|\Lambda|=\delta$, then 
    $\tup{\tg{\wrt}, *}\in List_{\flexpr{}}$ even after $\delta$
    possibly larger tags are added in the set. This holds also for any 
    flexnode $\flexpr{}\in\quo{\wrt}\cap\quo{\rd}$. Thus, for $\rd$ it 
    will hold that $Tags_{dec}^{\geq k}$ will contain at least $\tg{\wrt}$,
    and hence $Tags_{dec}^{\geq k}\neq\emptyset$, contradicting our assumption.
    %\nn{[NN: Note that here we assume that $k$ are decodable using RLNC]}
\end{proof}

\begin{theorem}[Liveness]
    Given a set of flexnodes $C$ and a parameter $\delta$, then every read/write operation $\op$ terminates in any execution $\EX$ of \optimum{}, given that $b$, $b<\frac{|C|-k}{3}$ Byzantine failures exist in $C$, and no more than $\delta$ write operations execute concurrently. 
\end{theorem}

\begin{proof}
    The theorem follows from Lemmas \ref{lem:terminate} and \ref{lem:delta} leading to the liveness of the \optimum{} protocol
    over a static set of flexnodes and a single object $\obj{}$.
\end{proof}

\myparagraph{Safety:}
To proof that Algorithm \ref{algo:atomicity:generic1} guarantees
atomicity we will first proceed to show that the DAPs presented 
satisfy the following two properties (Property 1 in \cite{ARES19}):

 \begin{definition}[DAP Consistency Properties]\label{property:dap}  If $\Pi$ is the set of complete DAP operations in an execution $\EX$, and $C$ the set of flexnodes in $\EX$,  
 then for any $\phi,\pi\in\Pi$ the following must hold: 
 	%be an execution of some algorithm that executes the data-primitives 
 \begin{enumerate}
 \item[]\textbf{DAP1}:  If $\phi$ is  $\daputdata{C}{\tup{\tg{\phi}, v_\phi}}$, 
 %$\tup{\tg{\phi}, v_\phi} \in\tgSet\times\valSet$, % and $v_1 \in \valSet$, and
 $\pi$ is $\dagettag{C}$ (or  $\dagetdata{C}$) 
 %in $\EX$ such that 
 that returns $\tg{\pi} \in \tgSet$ 
 %(or $\tup{\tg{\pi}, v_{\pi}} \in \tsSet \times \valSet$) 
 and $\phi$ completes before $\pi$ is invoked in $\EX$, then $\tg{\pi} \geq \tg{\phi}$.
 \item[] \textbf{DAP2}: \sloppy If $\phi$ is a $\dagetdata{C}$ that returns $\tup{\tg{\pi}, v_\pi } \in \tsSet \times \valSet$, 
 then there exists $\pi$ such that $\pi$ is a $\daputdata{C}{\tup{\tg{\pi}, v_{\pi}}}$ and $\phi$ did not complete before the invocation of $\pi$. 
 If no such $\pi$ exists in $\EX$, then $(\tg{\phi}, v_{\phi})$ is equal to $(t_0, v_0)$.
 \end{enumerate} \label{def:consistency}
 \end{definition}\vspace{-0.5em}

We begin with a lemma that shows that the DAPs of \optimum{} satisfy property \textbf{DAP1} of 
Definition \ref{property:dap}. Note that we need each property to hold,
despite the Byzantine failures in the system. 

\begin{lemma}[Property DAP1]
\label{lem:static:C1}
    In any execution $\EX$ of~\optimum{}, if $\phi$ is a complete
    $\daputdata{C}{\tup{\tg{\phi},v_\phi}}$ DAP in $\EX$ 
    and $\pi$ a $\dagettag{C}$ or $\dagetdata{C}$ DAP 
    that is invoked after $\phi$ is completed in $\EX$, then 
    $\pi$ returns $\tup{\tg{\pi}, v_\pi}$ such that $\tg{\pi}\geq\tg{\phi}$.
\end{lemma}

\begin{proof}
    As mentioned above we assume a static environment and 
    each operation accesses a static set $C$ of flexnodes.
    Let $\EX$ be some execution of \optimum{}, then we consider two cases for $\pi$ for proving property $DAP1$:  (a) $\pi$ is a $\act{get-tag}$, or (b) $\pi$ is a $\act{get-data}$ primitive. 

    Case $(a)$: $\phi$ is $\daputdata{c}{\tup{\tg{\phi}, v_\phi}}$ and  $\pi$ is a $\dagettag{C}$ that returns $\tg{\pi} \in \tsSet$. Let $\flexpr{\phi}$ and $\flexpr{\pi}$ the flexnodes (not necessarily different) that invoked $\phi$ and $\pi$ respectively. Furthermore,
    let $Q_{\phi}, Q_{\pi}\subset C$ be the set of flexnodes that repied 
    to $\phi$ and $\pi$ respectively. By the algorithm the cardinality of the two sets is 
    $|Q_{\phi}|=|Q_{\pi}|=\lfloor\frac{2|C|+k}{3}\rfloor$, where $k$ is derived 
    from the $(n,k)$ LRNC coding scheme used. It follows that the 
    two sets have a non-empty intersection of $|\quo{\phi}\cap\quo{\pi}|\geq\frac{|C|+k}{3}$ flexnodes. 
    Since $\phi$ completes before the invocation of $\pi$, then every
    flexnode $\flexpr{}\in \quo{\phi}\cap\quo{\pi}$ receives a message
    with the pair $\tup{\tg{\phi},e_{\flexpr{}}}$ from $\phi$, 
    before replying to the \mtype{query-tag} request from $\pi$. 

    By our failure model $b$ ($b<\frac{|C|}{3}$) of the flexnodes that reply may be Byzantine. So we need to examine both cases when a replying node 
    $\flexpr{}\in \quo{\phi}\cap\quo{\pi}$ is correct or faulty. 
    
    Let's first assume that $\flexpr{}$ is correct, following the protocol. 
    So, $\flexpr{}$ adds the $\tup{\tg{\phi},e_{\flexpr{}}}$ in its 
    local $List$ before replying to $\phi$ and thus before replying to $\pi$
    as well. There, are two cases to consider: (i) either $\tg{\phi}$ is the 
    largest tag in the state variable $List$ maintained by $\flexpr{}$, or (ii) there is a tag $\tg{}'> \tg{\phi}$ added in the list of $\flexpr{}$ before the invocation of $\pi$.
    In any case, $\pi$ will discover a max tag from each benign flexnode 
    $\tg{max}\geq\tg{}'\geq\tg{\phi}$.

    It remains to examine whether a Byzantine flexnode may force an invalid tag to be returned. An invalid tag is one that it is not generated (and thus written) by any writer. Let $\flexpr{}'\in \quo{\phi}\cap\quo{\pi}$
    be Byzantine. $\phi$ signs the $\tup{\tg{\phi},e_{\flexpr{}'}}$ pair
    before sending it to $\flexpr{}'$.  Since signatures are unforgeable 
    then $\flexpr{}'$ can only send a $\tup{\tg{},e_{\flexpr{}'}}$ pair 
    to $\pi$ that was propagated during a $\daputdata{*}{*}$ operation. 
    The only means to diverge from the protocol is for $\flexpr{}'$ to
    not reply to $\pi$ or send a tag smaller than the one it ever received. 
    As $\flexpr{}'\in \quo{\phi}\cap\quo{\pi}$, then $\pi$ received a reply
    from $\flexpr{}'$. We need to examine the case where $\flexpr{}'$ sends a $\tup{\tg{},e_{\flexpr{}'}}$
    such that $\tg{}<\tg{\phi}$. Since  $|\quo{\phi}\cap\quo{\pi}|\geq\frac{|C|+k}{3}$ and since $b<\frac{|C|-k}{3}$, then there exists 
    %\blue{(delete: at least one non Byzantine)}
    a correct flexnode 
    in $\quo{\phi}\cap\quo{\pi}$ that replies to $\pi$. That flexnode replies
    with a tag at least as large as $\tg{\phi}$, and thus $\pi$ discovers a tag greater or equal to $\tg{\phi}$, returning a tag $\tg{\pi}\geq \tg{\phi}$ as expected.

    Case $(b)$: $\phi$ is   $\daputdata{C}{\tup{\tg{\phi}, v_\phi}}$ and  $\pi$ is a $\dagetdata{C}$ that returns $\tup{\tg{\pi}, v_{\pi}} \in \tsSet \times \valSet$.  Let $\quo{\phi},\quo{\pi}\subseteq C$ be the set of flexnodes that
    replied to $\phi$ and $\pi$ respsectively. Similar to case (a), $|\quo{\phi}\cap\quo{\pi}|\geq\frac{|C|+k}{3}$. Consequently, since $\phi$ 
    completes before $\pi$, then every flexnode $\flexpr{}\in\quo{\phi}\cap\quo{\pi}$
    receives a \mtype{put-data} request from $\phi$ before replying to 
    the \mtype{query-list} request from $\pi$. Let $\tup{\tg{\phi}, e_{\flexpr{}}}$ be the 
    pair that $\phi$ sent to $\flexpr{}$. The receiving node will insert the pair in its 
    $List_{\flexpr{}}$ if the tag is correctly signed and if not another tuple for 
    tag $\tg{\phi}$ already exists in $List_{\flexpr{}}$. 
    So assuming that $\phi$ is executed by a correct flexnode, and since $|\quo{\phi}\cap\quo{\pi}|\geq\frac{|C|+k}{3}$, then we need to 
    %when $\pi$ receives $List_{\flexpr{}}$ from $\flexpr{}$ we need to 
    investigate if $\pi$ discovers $\tup{\tg{\phi},*}\in List_{\flexpr{}}$ for each flexnode $\flexpr{}$ in a set $S\subset \quo{\phi}\cap\quo{\pi}$ such that either: (i) $|S|\geq k$, or (ii) $|S|< k$. 
    
    In Case (i), since $\tup{\tg{\phi},*}$ appears in $|S|\geq k$ lists then $\tg{\phi}\in Tags_{dec}^{\geq k}$ in $\pi$ (A\ref{fig:bft-mwmr-optimum-dap}:L\ref{fig:bft-mwmr-optimum-dap:decode:getdata:k}). Hence, A\ref{fig:bft-mwmr-optimum-dap}:L\ref{fig:bft-mwmr-optimum-dap:decode}, $\pi$ will decode 
    and return the value associated with the maximum tag in $Tags_{dec}^{\geq k}$. Therefore,
    $\pi$ will return a tag $\tg{\pi}\geq\tg{\phi}$ in this case. 
    
    The second case is possible only when $\tup{\tg{\phi},*}$ is removed from the lists 
    of every flexnode $\flexpr{}'$ in the set $(\quo{\phi}\cap\quo{\pi})\setminus S$. 
    By A\ref{fig:bft-mwmr-optimum}:L\ref{fig:bft-mwmr-optimum:remove}, a tag is removed from $List_{\flexpr{}'}$ only when $\delta+1$ tuples
    with tags $\tg{}'>\tg{\phi}$ are added in the $List_{\flexpr{}'}$. Since, by assumption, 
    at most $\delta$ write operations may be concurrent with a read operation (and thus with $\pi$), 
    then by the pigeonhole principle, there must be a tag in $List_{\flexpr{}'}$ propagated 
    by a write operation $\wrt'$ that has completed before the invocation of the read that executes
    $\pi$, and thus before the invocation of $\pi$. Let $\tg{\wrt'}$ be the tag of the last 
    completed write operation before $\pi$. Since $\wrt'$ receives replies from a quorum $\quo{\wrt'}$, then as 
    we explained before it must be the case that $\quo{\wrt'}\cap\quo{\pi} > \frac{|C|+1}{3}$. 
    Since $b < \frac{|C|-k}{3}$ flexnodes may fail, then $\pi$ will discover $\tg{\wrt'}$ in at
    least $k$ lists. With similar arguments as in Case (i) $\pi$ will return a tag $\tg{\pi}\geq \tg{\wrt'}\geq\tg{\phi}$ and that completes the proof.
\end{proof}

\begin{lemma}[Property DAP2]
\label{lem:static:C2}
    In any execution $\EX$ of \optimum{}, if $\pi$ a complete
    $\dagetdata{C}$ DAP in $\EX$ that returns $\tup{\tg{\pi},v_\pi}$, then there exists $\phi$ such that $\phi$ is a $\daputdata{C}{\tup{\tg{\pi} ,v_{\pi}}}$ in $\EX$ and $\pi$ did not complete before the invocation of $\phi$ in $\EX$. If no such $\pi$ exists in $\EX$, then $(\tg{\phi}, v_{\phi})$ is equal to $(t_0, v_0)$.
\end{lemma}

\begin{proof}
    By \optimum{} the $List$ variable at each flexnode is initialized with a single tuple 
    $\tup{\tg{0},\bot}$.  
    Moreover, new tuples are inserted in the $List_{\flexpr{}}$ of some flexnode $\flexpr{}$, when 
    $\flexpr{}$ receives a \mtype{put-data} message, and the tuple  
    is signed and the signature is verified.

    When $\flexpr{}$ receives 
    %either 
    % a \mtype{query-tag} message, sent by a $\dagettag{C}$ operation
    % (Line \ref{}), or 
    a \mtype{query-list} message, sent by a $\dagetdata{C}$ operation (A\ref{fig:bft-mwmr-optimum}:L\ref{fig:bft-mwmr-optimum:query-list}), 
    $\flexpr{}$ responds with 
    %either the max tag in its $List$ variable or the whole 
    its $List$ variable.    
    Since each $List$ is initialized with an initial value $\tup{\tg{0},\bot}$ (known to every flexnode), then $\pi$ should return either $\tg{\pi}=\tg{0}$ or a signed tag $\tg{\pi}>0$. In the 
    case $\tg{\pi}=0$ there is nothing to prove. If $\tg{\pi}>0$, then it must be the case that
    $\pi$ discovered $\tg{\pi}$ in the $List_{\flexpr{}}$ of some flexnode $\flexpr{}$.  
    Every tuple must be signed before being added in each flexnode's list by an honest writer. 
    Therefore, Byzantine flexnodes cannot forge and arbitrarily
    introduce new tuples for other flexnodes in their $List$ variables.
    Hence, $\pi$ may return a tag $\tg{\pi}>0$ only if it discovers $\tg{\pi}$ in $k$ $List$s. This means that each flexnode $\flexpr{}$ adding $\tg{\pi}$ in their 
    list received a \mtype{pu-data} request with a signed
    $\tup{\tg{\pi}, *}$ pair before replying to $\pi$. Therefore, it must be 
    the case that the $\phi$ operation was invoked before $\pi$ communicated with $\flexpr{}$ and thus, before $\pi$ completed.
\end{proof}

The next lemmas show the consistency between read and write operations. These 
lemmas will eventually lead to the main result of this section. The first 
lemma examines the consistency of two write operations. 

\begin{lemma}
\label{lem:static:write:write}
    Given a set of flexnodes $C$ where $b$, $b<\frac{|C|-k}{3}$, flexnodes may be Byzantine, any execution $\EX$ of Algorithm \ref{algo:atomicity:generic1}, if 
    $\pi, \wrt$ two complete operations in $\EX$ such that $\pi$ is a read or write and $\wrt$ a write operation, $\pi\bef\wrt$
    and $\pi$ executes a $\daputdata{C}{\tup{\tg{\pi}, v_{\pi}}}$, then 
    $\wrt_2$ executes $\daputdata{C}{\tup{\tg{\wrt}, v_{\wrt}}}$ such that
    $\tg{\wrt}>\tg{\pi}$.
\end{lemma}

\begin{proof}
    By Algorithm \ref{algo:atomicity:generic1}, a write operation has two
    phases for writing a value $v\in\valSet$: in the first phase it queries a quorum of flexnodes in $C$ to discover 
    the maximum tag in the system using a $\dagettag{C}$, and in the second
    phase it generates a new tag $\tg{}$ and propagates the value in $C$
    by invoking a $\daputdata{C}{\tup{\tg{},v}}$ operation. Similarly a read operation invokes $\daputdata{C}{\tup{\tg{},v}}$, to propagate the pair 
    that is about to return. 
    Since, $\pi\bef\wrt$ it follows that $\pi$ completed before $\wrt$ 
    was invoked. Consequently, the $\daputdata{C}{\tup{\tg{\pi}, v_{\pi}}}$
    (either if $\pi$ is a read or write)
    invoked by $\pi$ was also completed before 
    the invocation of the $\dagettag{C}$ operation at $\wrt$. 
    So, by property $DAP1$ and Lemma \ref{lem:static:C1}, the $\dagettag{C}$
    operation at $\wrt$ will return a tag $\tg{}\geq\tg{\pi}$. In A\ref{algo:atomicity:generic1}:L\ref{algo:atomicity:generic1:increment} however, $\wrt$ increments the integer part of $\tg{}$
    returned by the \act{get-tag} operation,
    essentially 
    generating $\tg{\wrt}$. By the definition of tag comparison, since 
    $\tg{\wrt}.ts > \tg{}.ts$, then $\tg{\wrt}>\tg{}$. By the previous 
    claim we know that $\tg{}\geq\tg{\pi}$, hence it follows that 
    $\tg{\wrt}>\tg{\pi}$ completing the proof.
\end{proof}

The second lemma examines the consistency between read and write operations. 

\begin{lemma}
\label{lem:static:write:read}
    Given a set of flexnodes $C$ where up to $b$, $b<\frac{|C|-k}{3}$, flexnodes may be Byzantine, any execution $\EX$ of Algorithm \ref{algo:atomicity:generic1}, if 
    $\wrt$ and $\rd$ complete write and read operations respectively in $\EX$ such that $\wrt\bef\rd$
    and $\wrt$ executes a $\daputdata{C}{\tup{\tg{\wrt}, v_{\wrt}}}$, then 
    $\rd$ returns a value associated with a tag $\tg{\rd}$ such that
    $\tg{\rd}\geq\tg{\wrt}$.
\end{lemma}

\begin{proof}
    With similar claims as in Lemma \ref{lem:static:write:write}, since 
    $\wrt\bef\rd$ in $\EX$, then it must be the case that the $\daputdata{C}{\tup{\tg{\wrt}, v_{\wrt}}}$ operation completed before the invocation 
    of $\dagetdata{C}$ operation in $\rd$. By Lemma \ref{lem:static:C1}, 
    we showed that the $\dagetdata{C}$ operation at $\rd$ will return 
    a tag $\tg{}\geq\tg{\wrt}$. Since $\rd$ propagates in the second 
    phase and returns the tag-value pair returned by the \act{get-data} operation
    then $\tg{\rd}=\tg{}$ and hence $\tg{\rd}\geq\tg{\wrt}$ as required. 
\end{proof}

The final lemma examines the consistency between two read operations. 

\begin{lemma}
\label{lem:static:read:read}
    Given a set of flexnodes $C$ up to  $b$, $b<\frac{|C|-k}{3}$, flexnodes may be Byzantine, any execution $\EX$ of Algorithm \ref{algo:atomicity:generic1}, if 
    $\rd_1, \rd_2$ two complete read operations in $\EX$ such that $\rd_1\bef\rd_2$, If $\rd_1$ returns a value associated with a tag 
    $\tg{\rd_1}$ and $\rd_2$ returns a value associated with a tag $\tg{\rd_2}$,
    then it must hold that $\tg{\rd_2}\geq\tg{\rd_1}$.
\end{lemma}

\begin{proof}
    In the case of two read operations, notice that according to the algorithm,
    $\rd_1$ will perform a $\daputdata{C}{\tup{\tg{\rd_1},v_{\rd_1}}}$ before 
    completing. With similar reasoning as in the previous lemmas, $\rd_2$ 
    will perform a $\dagetdata{C}$ following the \act{put-data} operation 
    from $\rd_1$ and thus by Lemma \ref{lem:static:C1} will obtain, propagate and return the value associated with a tag $\tg{\rd_2}\geq\tg{\rd_1}$. 
    Notice here that by Lemma \ref{lem:static:C2} both $\rd_1$ and $\rd_2$
    will return a tag that was previously written by a write operation. 
\end{proof}

The main safety result is as follows. 

\begin{theorem} \label{thm:single-object:safety}
    Given a set of flexnodes $C$ where up to $b$, $b<\frac{|C|-k}{3}$, flexnodes may be Byzantine, any execution $\EX$ of Algorithm \ref{algo:atomicity:generic1} implements an atomic read/write register. 
\end{theorem}

\begin{proof}
    We need to show that in any execution $\EX$ of Algorithm \ref{algo:atomicity:generic1} all three properties (A1-A3) of atomicity, as 
    those are presented in Section \ref{sec:model}, hold.
    Algorithm \ref{algo:atomicity:generic1} uses tags to order operations. So we can express the 
    partial order required by atomicity properties in terms 
    of tags written and returned. More precisely, for each execution $\EX$ of the algorithm there must exist a partial order $\prec$ on  the set of completed operations $\Pi$ that satisfy conditions A1, A2, and A3. Let $\tg{\op}$ be the value of the tag at the
    completion of $\op$. With this, we denote the partial order on operations as follows. For two operations $\op_1$ and $\op_2$, when $\op_1$ is any operation and $\op_2$ is a write, we let $\op_1\prec\op_2$ if $\tg{\op_1}<\tg{\op_2}$. For two operations $\op_1$ and $\op_2$, when $\op_1$ is a write and $\op_2$ is a read we let $\op_1\prec\op_2$ if $\tg{\op_1}\leq\tg{\op_2}$. The rest of the order is established by transitivity and reads with the same timestamps are not ordered. We are now ready to reason about each of the 
    atomicity properties given the lemmas stated above. 
    
   \begin{itemize}
       \item[A1] For any $\op_1,\op_2\in\Pi$ such that $\op_1\bef\op_2$ it cannot be that $\op_2\prec\op_1$.
        This property follows directly from Lemmas \ref{lem:static:write:write}-\ref{lem:static:read:read}. In the lemmas we showed that for any combination of operations (reads or writes) it holds that either $\tg{\op_1}<\tg{\op_2}$ or $\tg{\op_1}=\tg{\op_2}$. Therefore, by the definition of $\prec$, either $\op_1\prec\op_2$ or the relation cannot be defined.

        \item[A2] For any write $\wrt\in\Pi$ and any operation $\op\in\Pi$, either $\wrt\prec\op$ or $\op\prec\wrt$. If $\tg{\wrt} < \tg{\op}$ or $\tg{\wrt} > \tg{\op}$, then $\wrt\prec\op$ or $\op\prec\wrt$ follows directly. If
        $\tg{\wrt}=\tg{\op}$, then it must be that $\op$ is a read. 
        By Algorithm \ref{algo:atomicity:generic1} a write operation $\wrt'$ generates a tag that includes
        a timestamp larger than the one stored in the writer $w$ and includes $w$'s unique identifier. Then it holds: (i) the new tag is larger than any tag previously generated by $w$, and (ii) is either lexicographically smaller or bigger than any tag generated by any other writer $w'$. Therefore,  any tag generated by $\wrt$ cannot be equal to any tag generated during any other write operation. Thus, $\tg{\op}=\tg{\wrt}$ only holds if $\op$ is a read, and $\wrt\prec\op$ follows.

        \item[A3] Every read operation returns the value of the last write preceding it according to $\prec$ (or the initial value if there is no such write).
        Let $\wrt$ be the last write preceding read $\rd$ by $\prec$. From our definition it follows that $\tg{\rd}\geq\tg{\wrt}$. If $\tg{\rd} = \tg{\wrt}$, then $\rd$ returned the value written by $\wrt$. If $\tg{\rd}>\tg{\wrt}$, then it means that $\rd$ obtained a larger timestamp. However, by the DAP Property \ref{property:dap} and Lemma \ref{lem:static:write:write} the larger timestamp can only be originating from a write $\wrt'$, s.t. $\tg{\wrt'} > \tg{\wrt}$. From the definition it follows that $\wrt\prec\wrt'$ and $\wrt'\prec\rd$. Thus, $\wrt$ is not the last preceding write contradicting our assumption. So it cannot be that $\tg{\rd}>\tg{\wrt}$ and hence each read returns a tag (and hence the value) equal to the last preceding write. Lastly, if $\tg{\rd} = \tg{0}$, no preceding write exist, and $\rd$ returns the initial value.
   \end{itemize}
\end{proof}

%% file: multi_object_manipulation.tex
In the previous section, we presented a protocol for implementing a single  
wait-free 
atomic read/write object over a fixed set of storage locations. 
% This protocol assumes a fixed set of 
% %logical 
% storage locations, and the implementation of a single atomic object. 
%with each logical location represented by servers in the system. However, since the actual data resides on physical nodes, each logical location’s data is stored across specific physical nodes.
%
However, in practical systems, a shared memory is expected to maintain multiple objects. Due to the composable nature of atomicity each object can be independently implemented. Initiating multiple instances of a DSM protocol however (one per atomic object) is impractical and not 
yet well defined.

In single-object shared memory algorithms \cite{ABD96, LS97, ARES19}, a
single set of hosts maintains the object replicas. Such a single cluster would quickly become a scalability 
bottleneck if it maintains multiple objects. Hence, 
for ensuring scalability, multi-object solutions
must devise a strategy for placing each object in a subset of nodes 
%and identify which set of nodes 
that will be responsible 
for each object. To tackle this problem we use a consistent hashing approach to match objects with nodes (\cite{StoicaMKKB01}).
More formally, given a set of objects $\objSet$, and a set of flexnodes $\flexSet$ we need to devise
a function $\mathfrak{D}$ such that for each $\obj{}\in\objSet$:
\[
    \mathfrak{D}(\obj{},\flexSet)\mapsto C
\]
where $C\subseteq \flexSet$. We assume that  
%$\mathfrak{D}(\obj{1}, \flexSet) = n$ 
$|C| = n$ for a fixed parameter, which we refer to as the \textit{coded replication factor} (CRF). 
% We use a gossiping layer to update an approximately active and participating set of nodes in the network (Fig. ~\ref{fig:coded-storage-decoding}).
In our RLNC encoding scheme the CRF is the encoding parameter $n$, from the $[n,k]$ encoding 
parameters used. 
%

% Furthermore, 
% %let each object has a unique identifier derived by
% %from its byte sequence, such as through 
% by applying a collision-resistant hash function to the object's byte sequence. 
% For an object $o$, we define its identifier as $id_o = \hash{}(o)$, where $\hash{}$ 
% is a collision-resistant hash function that outputs a 256-bit string, and the binary representation of $o$.
% %, ensuring uniqueness of the locations.

% There are two primary challenges in managing multiple atomic memory objects in dynamic environments: (i) distributing objects among participating flexnodes, and (ii) maintaining the atomicity of read and write operations as new flexnodes join and existing ones leave the network. To tackle (i) we use a consistent hashing approach to identify the nodes that are responsible for objects (\cite{StoicaMKKB01}). To handle (ii) we introduce 
% a new join/depart protocol, which relies on a replicated state machine (e.g. a blockchain implemented using Cosmos
% SDK \cite{CosmosSDK}) to handle the order of flexnode joins and departures. 
% allow new flexnodes to participate in the system and existing flexnodes to depart gracefully.
%
% Each object is mapped to a unique sequence of virtual locations within a large universe of possible locations. In this setup, a virtual location can occupy any of $2^{160}$ possible addresses, corresponding to the same key space defined in the Kademlia protocol. 
\begin{figure}[h]
    \centering
    % \begin{minipage}{0.55\textwidth}
    %     \centering
    %     \includegraphics[width=\textwidth]{2024-OPTIMUM/ICDCS26/images/flexnode-architecture.png}
    %     \caption{A flexnodes consist of three key components: gossiping, read/write operations, and the replicated state machine.}
    %     \label{fig:coded-storage-decoding}
    % \end{minipage}
    % \hfill
     \begin{minipage}{0.45\textwidth}
         \centering
        \includegraphics[width=\textwidth]{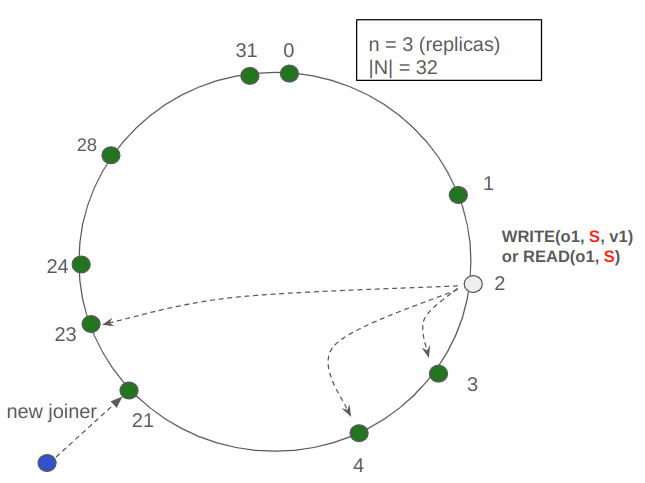}
        \caption{The flexnodes and objects are logically placed in a ring. Here for the example we are assuming $N=2^5$.}
        \label{fig:consistent-hashing}
    \end{minipage}
\end{figure}
% In contrast to single-object shared memory algorithms \cite{ABD96, LS97, ARES19} where a
% single set of hosts maintains the object replicas, it is inevitable for multi-object solutions
% to devise a strategy for identifying the set of nodes responsible for each object in the system.
% More formally, given a set of objects $\objSet$, and a set of flexnodes $\flexSet$ we need to devise
% a function $\mathfrak{D}$ such that for each $\obj{}\in\objSet$:
% \[
%     \mathfrak{D}(\obj{},\flexSet)\mapsto C
% \]
% where $C\subseteq \flexSet$. We assume that  
% %$\mathfrak{D}(\obj{1}, \flexSet) = n$ 
% $|C| = n$ for a fixed parameter, which we refer to as the \textit{coded replication factor} (CRF). 
% We use a gossiping layer to update an approximately active and participating set of nodes in the network (Fig. ~\ref{fig:coded-storage-decoding}).
% In our RLNC encoding scheme the CRF is the encoding parameter $n$, from the $[n,k]$ encoding 
% parameters used. 

\optimum{} adopts a \textit{distance} based approach between nodes and objects, and we store the 
data of an object $\obj{}$ to its $n$ "closest" nodes. Using a collision-resistant hash function $\hash{}$, we map both 
node and object identifiers to a 256-bit space supporting $N=2^{256}$ elements. In turn, given two identifiers $a,b\in\objSet\cup\flexSet$ with hashes $id_a=\hash{}(a)$ and $id_b = \hash{}(b)$ their distance is measured \textit{clockwise} as if those identifiers are 
placed on a ring (similar to Chord \cite{StoicaMKKB01}) as:
\remove{
\[
    \distance{id_a, id_b} = \begin{cases} 
id_b - id_a & \text{if } id_b \geq id_a \\
id_b - id_a + N & \text{if } id_b < id_a 
\end{cases}
\]  
% The distance definition resembles the one used by the Chord protocol \cute{}, where the 
% distance between two points is measured \textit{clockwise} on a ring. 
%
Alternatively, this can be expressed compactly using modulo arithmetic:}
\[
\distance{id_a, id_b} = (id_b - id_a)\mod N
\]
This formula ensures that the distance is always \textit{non-negative} and measured clockwise;
if \(id_b \geq id_a\), the distance is simply \(id_b - id_a\), while 
if \(id_b < id_a\), the calculation wraps around the ring by adding the total size of the identifier space, \(2^{256}\), ensuring a clockwise measurement. Given the distance metric, we can now define 
our mapping function as follows: 
\[
\mathfrak{D}_{suc}(o,\flexSet)=\{s^0,\ldots, s^n\}
\]
where $s^i\in\flexSet $ and $\distance{\hash{}(o),\hash{}(s^i)} < \distance{\hash{}(o),\hash{}(s')}$, for $0\leq i\leq n$ and  $\forall s'\in \flexSet\setminus \mathfrak{D}(o,\flexSet)$.

So the set of nodes that will host the data for object $o$ are the $n$ nodes in 
$\flexSet$ with the minimum distance from the hashed identifier of $o$. 
%Since the distance is non-negative, and since the hashes of the identifiers
%are collision-resistant, then with a very high propability $|\mathfrak{D}(o,\flexSet)|=k$ as expected. 
The same mapping can be used as $\mathfrak{D}(\flexpr{},\flexSet)$ to find the $n$ 
closest nodes in the neighborhood of a flexnode $\flexpr{}\in\flexSet$. Notice that all of 
the $n$ closest nodes are successors of $\flexpr{}$ (i.e. appear after $\flexpr{}$) on the ring. 
In a similar fashion, we can define the set of $n$ nearest \textit{predecessors} of $\flexpr{}$:
\[
\mathfrak{D}_{pre}(\flexSet, \flexpr{})=\{s^0,\ldots, s^n\}
\]
where $s^i\in\flexSet$ and $ \distance{\hash{}(s^i),\hash{}(\flexpr{})} < \distance{\hash{}(s'), \hash{}(\flexpr{})}$, for $0\leq i\leq n$ and  $\forall s'\in \flexSet\setminus \mathfrak{D}(\flexSet, \flexpr{})$.

%% file: multi_object_v2.tex
Given the distance functions presented in the previous section, we can now derive 
a dynamic atomic read/write shared memory protocol supporting multiple objects.

The main challenges arising from the dynamic environment are: (i) how to redistribute existing objects, 
and (ii) ensure atomicity of ongoing read/write operations as nodes join and leave the service.  

In this section we describe the \optimum{} algorithm that implements an atomic DSM while supporting
multiple objects and dynamic participation. 
Algorithm \ref{algo:signature}, presents the signature of \optimum{},
including the state variables and the operations supported. 

% To tackle (i) we use the consistent hashing approach of Section to identify the nodes that are responsible for objects (\cite{StoicaMKKB01}). To handle those challenges we introduce 
% a new join/depart protocol, which relies on a replicated state machine (e.g. a blockchain implemented using Cosmos
% SDK \cite{CosmosSDK}) to handle the order of flexnode joins and departures. 
% allow new flexnodes to participate in the system and existing flexnodes to depart gracefully.

\input{agorithm-optimum-signature}

In high level, a $\act{read}(o)$ or $\act{write}(o,v)$ operation use the $\mathfrak{D}_{suc}(o,\flexSet)$
distance function to identify the nodes that host object $o$ before communicating with them. 
Similarly the $\act{join}$ and $\act{depart}$ operations, use the distance function to 
specify object responsibility and migrate the data appropriately before adding or removing 
a node respectively.

\myparagraph{External Services:}
We assume the presence of two external services: (i) an \textit{RLNC} service, and 
(ii) a replicated state machine oracle $SMR$ (e.g. a blockchain). 

The RLNC service is used to \textit{encode} a value to coded elements, 
\textit{decode} a set of coded elements to the original value,
and \textit{recode} a set of signed coded elements to generate a new, 
%\textit{linear combination} 
coded element along with its signature.
We assume that the service uses a $[n,k]$ coding scheme, i.e., the 
encode procedure outputs $n$ coded elements, and the decode procedure
expects $k$ coded elements to output the decoded value. 

% We assume the existence of an external replicated state machine oracle $SMR$ (e.g. a blockchain) that 
The SMR oracle is used as a \textit{node registry} to keep track of the 
joins and removals in the system. 
We assume that a flexnode $s\in\flexSet$ joins the service once.
In practice if a node wants to re-join the service will generate a new identifier $s'\in\flexSet$.
In particular, the oracle maintains 
a set of flexnode \textit{additions} and {\em removals} in $S^{\pm}$, $S^{\pm}\subseteq\{+,-\}\times\idSet$,
%, we call service configuration, with the identifiers of the 
%active participants in the service. The oracle 
and supports two operations:
(i) $\act{add}(change)$, such that $change\in \{+,-\}\times\idSet$, and results in 
$S^{\pm}=S^{\pm}\cup \{change\}$, and 
(ii) $\act{get}()$ which returns the set $S^{\pm}$.
%
% (i) $\act{join}(N)$ which results in $S=S\cup \{\tup{+,i}:i\in N\}$, 
% (ii) $\act{remove}(N)$ which results in $S=S\cup \{\tup{-,i}:i\in N\}$, and 
% (iii) $\act{register}(i)$ which returns returns $S$ if $\tup{+,i}\notin S$, and $\bot$ otherwise, and
%$\members{S}=\{i: \tup{+,i}\in S\wedge \tup{-,i}\notin S\}$. 
The oracle provides these semantics in any execution $\EX$:
\begin{itemize}
    \item {\bf Total Order:} If $\op_1$ and $\op_2$ any two operations 
    then either $\op_1\bef\op_2$ or $\op_2\bef\op_1$ in $\EX$,
    
    \item {\bf Validity:} If $\op_{1}$ is a $\act{get}()$ operation that returns $S^{\pm}$ in $\EX$, then $\forall ch \in S^{\pm}$, there exists an $\act{add}(ch)$ operation  $\op_{2}$ in $\EX$, such that $\op_2\bef\op_1$ 
    %the invocation of $\op_2$ appears before the invocation of $\op_1$ 
    in $\EX$.
    
    \item {\bf Inclusion:} if $\op_1$ and $\op_2$ two $\act{get}$ operations that return $S$ and $S'$ respectively, such that $\op_1\bef\op_2$ in $\EX$, then $S\subseteq S'$.
\end{itemize}

Given the set of additions and removals $S^{\pm}$, we may obtain the identifiers of the flexnodes $S^+\subseteq\flexSet$ participating in the service as follows:
\[
    S^+ = \{s : \tup{+,s}\in S^{\pm}~\wedge~\tup{-,s}\notin S^{\pm}\}
\]
To ensure the liveness of the service it must hold that at any point in any execution, 
%of the service 
$|S^+|\geq n$, where $n$ the RLNC coding parameter. We assume 
%that the oracle can satisfy the constraint on the minimum number of participants. Moreover, we assume 
that the changes recorded by the oracle are signed so no process may generate an invalid change.

As we examine later, the oracle does not affect the safety of our protocol
but rather ensures the liveness and longevity of the service. Furthermore, the oracle does not affect the latency of read and write operations but rather, will only delay the addition of new flexnodes. 

\subsection{The Join/Depart Protocols}
\label{ssec:join}
\input{sec_algorithm_reconfig_v2.tex}

\subsection{The Read/Write Protocols}
\label{ssec:dyn:rw}
\input{sec_algorithm_dynamic_rw.tex}

%% file: agorithm-optimum-signature.tex
\begin{algorithm}[!h]
    \begin{algorithmic}[2]
      {\scriptsize  
        \begin{multicols}{2}
                
            \Variables{}
                \State $Changes\subseteq \{+,-\}\times\idSet$ init. $\emptyset$
                \State $S\subseteq \idSet$ init. $\emptyset$ 
                \State \Comment{set of active flexnodes}
                \State $D\subseteq \objSet$ init. $\emptyset$ 
                \State \Comment{set of local object ids}
                %\State ~ ~ $J\subseteq \idSet$ initially $\emptyset$ \Comment{set of joining flexnodes}
                \State for each object $o\in D$:
                \State ~ ~ $\obj{}.List \subseteq  \tgSet \times \ecSet \times \Sigma$, init.  $\emptyset$
                \State ~ ~ $\obj{}.E\subseteq\tgSet\times2^\ecSet$, init. $\emptyset$
                \State ~ ~ $\obj{}.\tg{}\in \tgSet$, init. $\tup{0,\flexpr{i}}$
                \State ~ ~ $\obj{}.v\in \valSet$
             \EndVariables
             
            \Statex

            \Operations{}
                \State $\act{read}(\obj{}), \obj{}\in\objSet$
                \State $\act{write}(\obj{},v), \obj{}\in\objSet, v\in\valSet$
                \State $\act{join}()$
                \State $\act{depart}()$
           \EndOperations

        \end{multicols}
        }
    \end{algorithmic}
\caption{\optimum{} Algorithm Signature}
\label{algo:signature}
\end{algorithm}

%% file: sec_algorithm_reconfig_v2.tex
In \optimum{} we handle joins and departures of individual nodes and not the modification of entire 
configurations as done by previous algorithms \cite{ARES19,GLS03}.
Here we present the $\act{join}$ and $\act{depart}$ operations that are responsible to 
register and de-register nodes on the oracle, migrate the data to the nodes based on the joins and departures, 
and update the service state after a join or a departure. The protocols build upon the algorithm presented
in Section \ref{sec:byz-rw} where coded elements are associated with tags and every node maintains a list 
of the latest tags and codes it receives. 

% In this section, we describe the reconfiguration protocol that involves
% the modification, either by adding or removing nodes from the set of flexnodes in the service. We will consider the case of joins and departures separately. 

\myparagraph{The Join Protocol:}
Flexnodes join the service via the $\act{join}()$ operation. There are three main steps
in the join operation that a node may follow before it is fully included in the \optimum{}
service: (i) {\em data collection and code generation}, (ii) {\em addition} in the set of joined flexnodes, 
and (iii) {\em join finalization}.
% Registration ensures that the flexnode is secured by the replicated state machine and its 
% request is propagated to active flexnodes; 
%During the 
Data collection ensures that the newly join flexnode 
receives data for the objects it is expected to maintain in the service. This data allows the 
new flexnode to generate a new RLNC coded element without decoding the object value, speeding 
up the object collection process. 
%generates appropriate RLNC coded elements for them; 
Once the flexnode is up to date with 
the data, it initiates an \act{add} request to the oracle so to be %it's identifier is 
added in the list of joined flexnodes. Finally, %once added in the set of joined nodes,
it informs its neighbors to remove them from their join sets and add its identifier in their local estimates. 

% \begin{algorithm*}[!h]
% 	%\hrule \F
% 	\begin{algorithmic}[2]
% 		\begin{multicols}{2}
%             {\scriptsize
% 			\State at each flexnode $\flexpr{i}$
		      
% %  %                \Statex
                
%                 \State{\bf State Variables:}
% 			\State $S\subseteq \idSet$ initially $\emptyset$ \Comment{estimated flexnode participants}
%                 \State $D\subseteq \objSet$ initially $\emptyset$ \Comment{local object ids}
%                 \State {$J\subseteq \idSet$ initially $\emptyset$} \Comment{joining nodes}
%                 \State $\tg{}\in \tgSet$, initially $\tup{0,\flexpr{i}}$
%                 \State $v\in \valSet$, initially $\bot$ \Comment{Weird}
% %  %                    %\State $c, c'\in C$, initially $\bot$
%             \Operation{join}{}\EndOperation
% 	}
%         \end{multicols}	
% 	\end{algorithmic}
% 	% %\hrule \B
% 	\caption{The Flexnode State Variables and External Signature.}
% 	\label{algo:state}
% \end{algorithm*}
\input{algorithm-optimum-join-v2}

Algorithm \ref{algo:optimum:join} presents the detailed pseudocode of the join protocol. The three phases of the join protocol are as follows. First,
% joining node $j$ issues a \act{register} operation to the $Con$, requesting 
% to be accepted and retrieving the latest flexnode participation estimate $S$ (Line \ref{}). Once, it 
the joining node, say $\flexpr{j}$, 
%discovers $S$ it 
proceeds to collect data for the objects $\flexpr{j}$ will be responsible for 
Algorithm \ref{algo:optimum:join}, line \ref{line:join:collect}
(A\ref{algo:optimum:join}:L\ref{line:join:collect}). 
To do so, it first computes its closest \textit{successor} amd \textit{predecessor} neighbors
(A\ref{algo:optimum:join}:L\ref{line:join:neighbors}) based on their consistent hashing distance, 
and then it sends a \mtype{fetch-obj-list} message to all of them 
requesting the objects 
%(along with their encoded elements) 
that $\flexpr{j}$ will 
store. When a flexnode $\flexpr{i}$ receives such a request, 
%it waits to receive a \mtype{sub-reg} message from the blockchain, verifying that the joining node $j$ has indeed successfully registered. This helps $q$ replying to a malicious node.   When the \mtype{sub-reg} message is received, 
$\flexpr{i}$ adds $\flexpr{j}$ in a set of joining flexnodes $J$, and computes the set of objects $Objs$, from its local object set,  
%which of the objects it stores locally 
that should be sent to $\flexpr{j}$ (based on the object's consistent hashed 
distance to $\flexpr{j}$). The identifier of each object in $Objs$ 
along with its local $o.List$ at $\flexpr{i}$ is sent to $\flexpr{j}$.
% It prepares  all those objects 
% along with its largest known tag and coded element and sends this info to $j$. 
Once flexnode $\flexpr{j}$ collects at least $\frac{2|C|+\beta}{3}$ replies, for some constant $\beta\leq k$ enough to recode,
it computes its local set of objects $D$. For each object $o$, it discovers the 
$\delta$ maximum tags associated with encoded elements of $o$, which were found
in at least $\beta$ replies. This ensures that such tags were sent by at least a single
correct flexnode. Next we discover the coded elements associated with those tags
and we use them to generate a new coded element (recode) for $\flexpr{j}$.\footnote{Note that since we make use of signatures, and thus flexnodes cannot
generate arbitrary coded elements, thus we can use any number of coded elements for 
the recoding and not necessarily $b+1$.} The new element is associated with the running tag 
and the pair is inserted in the local $o.List$ variable. By the end of this process
the $o.List$ variable will contain at most $\delta$ new coded elements (A\ref{algo:optimum:join}:L\ref{line:join:loop:start}-A\ref{algo:optimum:join}:L\ref{line:join:loop:end}). 

Once the joining flexnode collects and stores the data for every object it needs 
to store, it invokes an $\act{add}(\tup{+,\flexpr{j}})$ operation to the SMR oracle, inserting $\tup{+,\flexpr{j}}$
to the oracle's changeset. By the properties of the oracle, 
this action sets $\flexpr{j}$ discoverable by any process 
that invokes a $\act{get}$ action at the oracle. Finally, 
$\flexpr{j}$ sends a \mtype{fin-join} message
to inform its neighbors about its successful join. Every flexnode 
that receives a \mtype{fin-join} message verifies that the node is indeed recorded
in the SMR oracle by invoking a $\act{get}$ request, it removes $\flexpr{j}$ from the 
joining set (if found), and updates its local estimate $S$ before replying to $\flexpr{j}$.
Once $\flexpr{j}$ receives enough replies it updates its local $Changes$ variable
that is going to be used for computing its local participation estimate $S$ in subsequent actions.

\input{algorithm-optimum-depart-v2}

\myparagraph{The Depart Protocol:} Departures cannot be distinguished from 
crash failures in the system as in both cases departing/crashed flexonodes
stop responding to any operation requests. Persistent failures, however, 
may affect the liveness of the system when each operation may not be able 
to receive enough replies to complete. The 
depart protocol allows the \textit{graceful removal} of flexnodes from the network 
allowing the remaining flexnodes to utilize existing resources for their 
read/write operations. The depart protocol disseminates information 
about the departure of the flexnode from the system, and it mobilizes any flexnodes that may assume 
control of the objects previously occupied by the departing flexnode to retrieve their data.
% in \optimum{} is similar with 
% the join protocol, as it attempts to propagate the latest known value of 
% each object the departed node maintains to the nodes that will be responsible
% for that object once the flexnode departs. 
So according to the pseudocode 
of Algorithm \ref{algo:optimum:depart} the depart operation takes two phases
for each object of the departing flexnode $\obj{}\in D$: 
(i) the departing node $\flexpr{d}$ sends a depart message to all the \textit{remaining} 
flexnodes that will maintain $\obj{}$, and 
(ii) any flexnode that receives such a message removes 
%the departing node 
$\flexpr{d}$ from its local estimates, and if it does not have the object $\obj{}$, 
it attempts to collect its recent data and generate a new coded element 
by invoking the \act{collect-data} action on the flexnodes that maintain $\obj{}$ before $\flexpr{d}$'s  departure.
% (i) the departing node reads the value of $\obj{}$
% from his latest known estimate $S$ in phase 1, and (ii) propagates the 
% value to all the nodes that will be responsible for the object once it 
% departs, i.e. $\mathfrak{D}_{suc}(\obj{},S\cup\{\tup{-,\flexpr{i}}\})$. 

%% file: algorithm-optimum-join-v2.tex
\begin{algorithm*}[!h]
	\begin{algorithmic}[2]
		\begin{multicols}{2}{\scriptsize
				\Operation{join}{} 
                % \State $Changes \gets SMR.\act{get}()$ \label{line:join:changes1}
                % \State $S \gets \{s:\tup{+,s}\in Changes~\wedge~\tup{-,s}\notin Changes\}$
                %%%%%%%%%%%%%%%%%%%%%%%%%%
                %% SOS: Needs to be inserted for removing the bound on the depart protocol -- Pre-Registration
                %% Did not apply that in the TLDR but needs to be inserted for future submissions
                %%\State $SMR.\act{add}(\tup{\bot,\flexpr{i}})$ \label{line:join:register}
                \State $\text{\act{collect-data}}(\{\})$ \label{line:join:collect}
				%\State $S\gets SMR.\act{join}(\{\flexpr{i}\})$
				\State $SMR.\act{add}(\tup{+,\flexpr{i}})$ \label{line:join:add}
                \State $\act{finalize-join}()$ \label{line:join:finalize}
                %\EndIf
			\EndOperation
				\Statex
				\Procedure{collect-data}{$Obj$}
                    \State $Changes \gets SMR.\act{get}()$ \label{line:join:changes1}
                    \State $S \gets \{s:\tup{+,s}\in Changes~\wedge~\tup{-,s}\notin Changes\}$
                    \If{$Obj = \emptyset$}
                    \State $C \gets \mathfrak{D}_{suc}(\flexpr{i}, S)\cup\mathfrak{D}_{pre}(S, \flexpr{i})$\label{line:join:neighbors}
                    \Else
                    \State $C \gets \bigcup_{o\in Obj} \mathfrak{D}_{suc}(o, S)$                    
                    \EndIf
                    % send fetch and wait for replies
                    \State \textbf{send} $(\mtype{fetch-obj-list}, Obj)$ to each $p\in C$
                    \State \textbf{wait until} $\exists \quo{r}\subseteq C$, s.t.  $|\quo{r}|\geq \ceil{\frac{2|C|+\beta}{3}}$ \WRP $\wedge \flexpr{i}$ received  $O_{\flexpr{}}$ from all $\flexpr{}\in\quo{r}$
                    \label{line:join:quorum}
                    % discover max tag for each object received
                    %\State $Data \gets \bigcup_{p\in\idSet_r}O_p$
                    %\State $Obj \gets \{ o : \tup{o,*}\in Data\}$
                    \State $RObj \gets \{ o : \tup{o,*}\in \bigcup_{\flexpr{}\in\quo{r}}O_{\flexpr{}}\}$
                    \For{$\obj{} \in RObj$}    \label{line:join:loop:start}
                        % \State $C'\gets\emptyset$
                        % \State $C \gets \mathfrak{D}_{suc}(\obj{}, S)$
                        \State $Lists_{\obj{}}\gets \{L: \tup{o, L} \in  \bigcup_{\flexpr{}\in\quo{r}}O_{\flexpr{}}\}$
                        \State $\obj{}.E[\tg{}]\gets\{\tup{\tg{},*}: \tup{\tg,*}\in L~\wedge~ L\in Lists_{\obj{}}\}$ \label{line:join:collecttags}
                        \State $Tag_{cor} = \{\tg{}: |\obj{}.E[\tg{}]| \geq \beta\}$
                        \State $maxTag_{\delta} = maxdeltatags(\{\tg{} : \tg{} \in Tag_{cor}\})$\label{line:join:deltamax}          
                        \For{$\tg{} \in maxTag_{\delta}ß$}
                            \State $\tup{e_{new}, \st_{new}} = \act{RLNC.recode}(\obj{}.E[\tg{}])$
                            \State $\obj{}.List \gets \obj{}.List \cup \{\tup{\tg{}, e_{new}, \st_{new}}\}$ \label{line:join:loop:end}
                        \EndFor
                        \State $D\gets D\cup\{\obj{}\}$
                \EndFor 
		\EndProcedure
		\Statex	
            \Procedure{finalize-join}{}
                \State $C \gets \mathfrak{D}_{suc}(\flexpr{i}, S)\cup\mathfrak{D}_{pre}(S, \flexpr{i})$
                \State \textbf{send} $(\mtype{fin-join})$ to each $\flexpr{}\in C$
                \State \textbf{wait until} $\exists \quo{r}\subseteq C$, s.t.  $|\quo{r}|\geq \ceil{\frac{2|C|+1}{3}}$ \WRP $\wedge~\flexpr{i}$ received  $\mtype{fin-ack}$ from all $\flexpr{}\in\quo{r}$
                \State $Changes\gets Changes\cup\{+,\flexpr{i}\}$
            \EndProcedure
                \Statex
			\Receive{{\sc fetch-obj-list}, $O'$}{}
                %\State \textbf{wait until} $\flexpr{i}\in J$
                \State $Changes \gets SMR.\act{get}()$\label{line:join:chupdate}
                \State $S \gets \{s:\tup{+,s}\in Changes~\wedge~\tup{-,s}\notin Changes\}$
                \If{$O'\neq \emptyset$}
                    \State $Objs\gets \{ \tup{o, o.List} : o\in O' \}$
                \Else 
                    \State $Objs\gets \{ \tup{o, o.List} : o\in D \wedge \flexpr{j}\in \mathfrak{D}_{suc}(o,S\cup\{\flexpr{j}\}) \}$
                \EndIf
                %\State $J\gets J\cup \{\flexpr{j}\}$
                %\State $Objs \gets \{ \tup{o, \tup{\tg{}, e, \st}} : o\in Candidates $\WRP $\wedge 
                %\tup{\tg{}, e, \st} = \max_{\tg{}}(o.List) \}$
                \State \textbf{send} $\tup{\mtype{fetch-ack}, Objs}$ to $\flexpr{j}$
			\EndReceive
                \Statex
            \Receive{\mtype{fin-join}}{}
                %\State \textbf{wait until} $\flexpr{i}\in J$
                \State $Changes \gets SMR.\act{get}()$\label{line:join:chupdate}
                % \If{$\{+,\flexpr{i}\}\in Changes$}
                %     \State $J\gets J\setminus \{\flexpr{j}\}$
                % \EndIf
                \State $S \gets \{s:\tup{+,s}\in Changes~\wedge~\tup{-,s}\notin Changes\}$
                \State \textbf{send} $\tup{\mtype{fin-ack}}$ to $\flexpr{j}$
			\EndReceive	
		}\end{multicols}	
	\end{algorithmic}
	%\hrule \B
	\caption{\optimum{} Join Protocol at flexnode $\flexpr{i}$.}
	\label{algo:optimum:join}
	\vspace{-1em}
\end{algorithm*}

%% file: algorithm-optimum-depart-v2.tex
\begin{algorithm*}[!h]
	%\hrule \F
	\begin{algorithmic}[2]
		\begin{multicols}{2}{\scriptsize

            \Operation{depart}{} 
                %\State $Changes\gets SMR.\act{get}()$
                %\State $Changes\gets Changes \cup\{\tup{-,\flexpr{i}}\}$
                %\State $S \gets \{s:\tup{+,s}\in Changes~\wedge~\tup{-,s}\notin Changes\}$
                %\If{$|S|\geq n$}
                \State $\text{\act{push-data}}()$
                \State $SMR.\act{add}(\tup{-,\flexpr{i}})$
                \State $\act{finalize-depart}()$
                %\EndIf
            \EndOperation

            \Statex
				
            \Procedure{push-data}{}
                \For{$\obj{} \in D$}
                    \State $C\gets \mathfrak{D}_{suc}(\obj{}, S\setminus\{\flexpr{i}\})$
                    \State \textbf{send} $(\mtype{push}, \obj{})$ to each $\flexpr{}\in C$
                    \State \textbf{wait until} $\exists \quo{r}\subseteq C$, s.t.  $|\quo{r}|\geq \ceil{\frac{2|C|+k}{3}}$ \WRP $\wedge \flexpr{i}$ received  $O_{\flexpr{}}$ from all $\flexpr{}\in\quo{r}$
                    % \State $C'\gets\emptyset$
                    % \State $C \gets \mathfrak{D}_{suc}(\obj{}, S)$
                    % \State $(\tg{},v)\gets\act{get-data}(C, \obj{})$
                    % \State $C\gets \mathfrak{D}_{suc}(\obj{}, S\cup\{\tup{-,\flexpr{i}}\})$
                    % \While{$C\neq C'$}\Comment{until no new flexnodes seen}
                    %     \State $\act{put-data}(C, \obj{},\tup{\tg{},v})$
                    %     \State $C'\gets C$
                    %     \State $C\gets \mathfrak{D}_{suc}(\obj{}, S\cup\{\tup{-,\flexpr{i}}\})$
                    % \EndWhile
                \EndFor
            \EndProcedure

            \Statex

            \Procedure{finalize-depart}{}
                \State $C \gets \mathfrak{D}_{suc}(\flexpr{i}, S)\cup\mathfrak{D}_{pre}(S, \flexpr{i})$
                \State \textbf{send} $(\mtype{fin-depart})$ to each $\flexpr{}\in C$
                \State \textbf{wait until} $\exists \quo{r}\subseteq C$, s.t.  $|\quo{r}|\geq \ceil{\frac{2|C|+1}{3}}$ \WRP $\wedge \flexpr{i}$ received  $\mtype{fin-ack}$ from all $\flexpr{}\in\quo{r}$
                %\State $Changes\gets Changes\cup\{-,\flexpr{i}\}$
            \EndProcedure

            \Statex
            
            %\State\Comment{receive a depart msg from $\flexpr{j}$}
            \Receive{\mtype{push}, $\obj{}$}{}
                %\State $J\gets J\setminus \{\flexpr{j}\}$
                %\State $Changes\gets Changes\cup \{\tup{-,\flexpr{j}}\}$
                %\State $S \gets \{s:\tup{+,s}\in Changes~\wedge~\tup{-,s}\notin Changes\}$ 
                \If{$\obj{}\notin D$}
                    \State $\text{\act{collect-data}}(\{\obj{}\})$ \label{line:depart:collect}
                \EndIf
                \State \textbf{send} $\tup{\mtype{push-ack}, \obj{}}$ to $\flexpr{j}$
            \EndReceive

            \Statex

            \Receive{\mtype{fin-depart}}{}
                
                %\State \textbf{wait until} $\flexpr{i}\in J$
                \State $Changes \gets SMR.\act{get}()$\label{line:join:chupdate}
                % \If{$\{+,\flexpr{i}\}\in Changes$}
                %     \State $J\gets J\setminus \{\flexpr{j}\}$
                % \EndIf
                \State $S \gets \{s:\tup{+,s}\in Changes~\wedge~\tup{-,s}\notin Changes\}$
                \State \textbf{send} $\tup{\mtype{fin-ack}}$ to $\flexpr{j}$
			\EndReceive

		}\end{multicols}	
	\end{algorithmic}
	%\hrule \B
	\caption{\optimum{} Depart Protocol at flexnode $\flexpr{i}$.}
	\label{algo:optimum:depart}
	\vspace{-1em}
\end{algorithm*}

%% file: sec_algorithm_dynamic_rw.tex
In Section \ref{sec:byz-rw} we presented an atomic read/write protocol
while assuming a static set of flexnodes and a single object.
%, on top of a message passing, asynchronous environment, where nodes
%may act maliciously (i.e., exhibit Byzantine behavior). Given the approaches 
Utilizing the join and depart protocols in
%presented in 
Sections $\ref{ssec:join}$,
we now extend the algorithm presented in Section \ref{sec:byz-rw} to provide atomic read/write operations
on multiple objects in a dynamic environment where flexnodes may join and depart 
the service. 
% Algorithms \ref{algo:dynamic:rw}-\ref{algo:dynamic:responses}, 
% present the detailed pseudocode for the extended algorithm. 

Algorithm \ref{algo:dynamic:rw} illustrates the pseudocode for the read
and write operations. Both operations  are structured similarly to the generic protocol.
%Algorithm \ref{algo:atomicity:generic1}. Unlike Algorithm \ref{algo:atomicity:generic1}
However, operations are not invoked on a single object nor on a fixed set of flexnodes. To cope with the changing set of flexnodes, each flexnode maintains the 
following state variables that are used to construct a local
estimate on the set of participating flexnodes:
% Due to the dynamic nature of the environment, the revised algorithm
% implements a mechanism to keep track of the flexnode participation
% in the system. By the algorithm signature, Algorithm \ref{algo:signature},
% we maintain four state variables at each flexnode for this purpose:
% % To this end, we introduce three sets in the state variable of each flexnode $\flexpr{}$ (Lines \ref{}): 
(i) the set $Changes$, for recording the flexnode additions and removals from the system,
(ii) the set $S$ to 
maintain the identifiers of participants based on $Changes$.
%local estimate at $\flexpr{}$ on the current participants in the service,
% (iii) the set $J$ to maintain the identifiers that $\flexpr{}$ knows are joining the service. 
To cope with the multiple objects, each flexnode maintains the set $D$ with the identifiers of the objects that $\flexpr{}$ is storing. Using these state 
variables the two operations are formed as follows.

% Using these sets, the main idea of the enhanced protocol, is to allow flexnodes to discover and invoke read/write operations on 
% the latest participants in the service. 
% %and invoke read/write operations on the latest participants in the service. 
% % This will 
% % ensure that no operation will be invoked on an outdated 
% % set of participants, preserving this way the safety (i.e., atomicity)  of the service. 

% Rather,
% each operation \textit{recovers} $C$ from its local estimate of the participation of the flexnode in the system, which in turn
% is updated either during join requests (see A\ref{algo:optimum:join}:L\ref{line:join:chupdate}), or through information
% attached to read/write messages. More precisely, the two operations are constructed
% as follows.

% More precisely, read and write operations are structured similar to the generic
% Algorithm \ref{algo:atomicity:generic1}. Read operations discover the latest tag-value  ($\tg{},v$) of an object $\obj{}$ by calling $\dagetdata{\obj{}, C}$ and then propagate 
% that value in a quorum using $\daputdata{\obj{},C}{\tup{\tg,v}}$. Pn the other hand, a write operation finds the tag associated with the latest value of an object $\obj{}$,
% by calling 

% Below we examine the algorithmic techniques 
% we use in more detail, through the description of the read/write operations.

\myparagraph{Read Operation:} 
%As in Algorithm \ref{algo:atomicity:generic1},
A read operation has two phases: a phase to discover the latest value of the object and 
a phase to propagate that value to a quorum of other flexnodes. To
handle multiple objects the read accepts an object identifier $\obj{}\in\objSet$,
and stores each object to a different subset of flexnodes. 
Before performing a read, the flexnodes specify the cluster that 
stores $\obj{}$ by calling
%as a parameter 
%and proceeds in two phases. 
% During Phase I, 
% it first computes the set of flexnodes $C$ that store $\obj{}$ using
% the function 
$\suc{\obj{}, S}$ over its local estimate $S$. This essentially, allows
to perform any dap operations on the determined set of flexnodes $C$.
To cope with dynamism, each phase is repeated until no new flexnodes 
are discovered in the system. In particular, the read 
invokes $\dagetdata{\obj{}, C}$ to obtain the latest
value of $\obj{}$ from the flexnodes in $C$. This is repeated while 
new flexnodes are added in the local estimate $S$, and in turn the 
cluster $C$ computed by $\suc{\obj{}, S}$ changes (A\ref{algo:dynamic:rw}:L\ref{line:dyn:read:phase1:start}- A\ref{algo:dynamic:rw}:L\ref{line:dyn:read:phase1:end}). 
Once the system stabilizes, 
the read operation terminates phase I. The read skips the second phase 
and returns the maximum tag-value pair discovered in phase I,
if a quorum of flexnodes, replied with the same maximum tag.
Otherwise, the read performs the propagate phase, where it repeatetly 
invokes $\daputdata{\obj{},C}{\tup{\tg,v}}$ until the local estimate
$S$ and thus $C=\suc{\obj{}, S}$ do not change (A\ref{algo:dynamic:rw}:L\ref{line:dyn:read:phase2:start}- A\ref{algo:dynamic:rw}:L\ref{line:dyn:read:phase2:end}). 

% As we explain
% later, the get-data operation collects information about the
% flexnodes' participation along with the object's value,
% updating the local estimate $S$. So, before accepting the
% tag-value pair returned by the get-data operation, we recompute
% the participation of the flexnodes based on a possibly updated estimate $S$. If no change is determined we proceed to the 
% second phase. During Phase II, we first compute the set of 
% flexnodes $C$ over the union of the local estimate and the joining 
% nodes $S\cup J$ (Line \ref{}). As we show in the proof of the 
% protocol, this is necessary for safety as we transfer the latest 
% value of the object to the flexnodes that are joining and will be responsible to store $\obj{}$. We then attempt to write the value 
% we discovered during the first phase if the following cases 
% are true: (i) either the set $C$ is different from the latest 
% set we obtained during Phase I (i.e., some nodes joining will host $\obj{})$, or (ii) not "enough" flexnodes replied with the highest 
% tag $\tg{}$ discovered in Phase I. In either case we proceed with 
% a $\daputdata{\obj{},C}{\tup{\tg,v}}$ operation and we repeat it 
% as long as we observe changes in the sets $S$ and $J$. Otherwise,
% the read is \textit{fast} and completes in a single phase. 

\myparagraph{Write Operation:}
The write protocol is similar to the read protocol, with the main 
differences being the request of the max tags during phase I, and 
the generation and propagation of a new tag-value pair, during 
phase 2 (A\ref{algo:dynamic:rw}:L\ref{line:dyn:write:inc}). 
As in the read operation, the write establishes $C = \suc{\obj{}, S}$ over 
the local participation estimate $S$ and repeats both phases while 
$S$ and thus $C$ are changing (due to joins). 
% for Phase I, while it 
% writes the new value on $C$ computed over the union $S\cup J$ in 
% Phase II. 
Unlike the read operation, the write performs both phases in all executions. 

\begin{algorithm}[!ht]
\begin{algorithmic}[2]
    \begin{multicols}{2}
        {\scriptsize   
            % \State{\bf State Variables:}
            % \State $Changes\subseteq \{+,-\}\times\idSet$ initially $\emptyset$
            % \State $S\subseteq \idSet$ initially $\emptyset$ \Comment{set of active flexnodes}
            % \State $D\subseteq \objSet$ initially $\emptyset$ \Comment{set of local object ids}
            % \State {$J\subseteq \idSet$ initially $\emptyset$} \Comment{set of joining flexnodes}
            % \State for each object $o\in D$:
            % \State ~ ~ $\obj{}.List \subseteq  \tgSet \times \ecSet \times \Sigma$, initially  $\emptyset$
            % \State ~ ~ $\obj{}.E\subseteq\tgSet\times2^\ecSet$, initially $\emptyset$
            % \State ~ ~ $\obj{}.\tg{}\in \tgSet$, initially $\tup{0,\flexpr{i}}$
            % \State ~ ~ $\obj{}.v\in \valSet$

            % \Statex
            
            \Operation{read}{$\obj{}$} 
            %\State $wCounter\gets wCounter+1$
            \State $C\gets \suc{\obj{},S}$
            \State $C'\gets \emptyset$
            \State \Comment{catch up with ongoing joins}
            \While {$C\neq C'$} \label{line:dyn:read:phase1:start}
                \State $\tup{\tg{}, v} \gets \dagetdata{\obj{},C}$
                \State $C'\gets C$
                \State $C\gets \suc{\obj{},S}$
            \EndWhile\label{line:dyn:read:phase1:end}

            \State $C\gets \suc{\obj{},S}$
            \State \Comment{if all have $\tg{}$ return it}
            \If{$C'\neq C~\vee~|o.E[\tg{}]|<\frac{2|C|+k~}{3}$} \label{line:dyn:read:phase2:start}
                \State $C'\gets \emptyset$
                \While {$C\neq C'$}
                    \State $\daputdata{\obj{},C}{ \tup{\tg{},v}}$
                    \State $C'\gets C$
                    \State $C\gets \suc{\obj{},S}$
                \EndWhile\label{line:dyn:read:phase2:end}
            \EndIf
            \State return $ \tup{\tg{},v}$
            \EndOperation
            
            \Statex
            
            \Operation{write}{$\obj{},v$} 
            %\State $wCounter\gets wCounter+1$
            \State $C\gets \suc{\obj{},S}$
            \While {$C\neq C'$} \label{line:dyn:write:phase1:start}
                \State $\tg{} \gets \dagettag{\obj{},C}$
                \State $C'\gets C$
                \State $C\gets \suc{\obj{},S}$
            \EndWhile \label{line:dyn:write:phase1:end}
            \State \Comment{new tag generation}
            \State $\tg{w} \gets \tup{\tg{}.ts + 1,  w}$
            \label{line:dyn:write:inc}
            \State $C\gets \suc{\obj{},S}$
            \State $C'\gets \emptyset$
            \While {$C\neq C'$} \label{line:dyn:write:phase2:start}
                \State $\daputdata{\obj{},C}{\tup{\tg{w},v}}$
                \State $C'\gets C$
                \State $C\gets \suc{\obj{},S}$
            \EndWhile \label{line:dyn:write:phase2:end}
            \EndOperation
            %\EndPart
        }
    \end{multicols}
    \end{algorithmic}
\caption{\optimum{} Multi-Object Read and Write operations}
\label{algo:dynamic:rw}
\vspace{-1em}
\end{algorithm}

The primitives $\dagettag{},~\dagetdata{},~\daputdata{*}{*}$ are almost
identical to the ones presented in Section \ref{sec:byz-rw} since
they are given a single object and a fixed set of flexnodes. What sets the procedures in Algorithm \ref{algo:dynamic:dap} apart, is their responsibility to 
propagate the object $\obj{}$ on which they operate, as well as the set 
of flexnodes responsible for $\obj{}$, i.e., $C$. Each reply received 
contains a $Ch$ set, which includes the changes the responding flexnode 
determined to $C$. Those changes are incorporated to the local $Changes$ 
set, and in turn to the computation of the new participation estimate $S$.

% They are also 
% responsible to collect new active and joining flexnodes that host
% $\obj{}$ though the received replies, and update the local information at $\flexpr{i}$. This is done through the $\act{update-sets}$ procedure.

    \begin{algorithm*}[!ht]
        \begin{algorithmic}[2]
            {\scriptsize
            \begin{multicols}{2}
                
                \Procedure{get-tag}{$\obj{}, C$}
                    %\State {\bf send} $(\text{{\sc query-tag}})$ to each  $\flexpr{}\in \members{C}$
                    \State {\bf send} $(\text{{\sc query-tag}}, \obj{},C)$ to each  $\flexpr{}\in C$
                    \State {\bf until} $\flexpr{i}$ receives $(\tup{\tg{\flexpr{}}, e_{\flexpr{}}}, \act{sign}_{\flexpr{}}(\tup{\tg{\flexpr{}}, e_{\flexpr{}}}), 
                    Ch_{\flexpr{}})$ from \WRP a quorum $\quo{r}$ 
                    %s.t.\WRP $\quo{r}\subseteq C$ and $|\quo{r}| \geq \left\lceil \frac{2|C| + k}{3}\right\rceil$

                    % \State \nn{$S' \gets \bigcup_{\flexpr{}\in\quo{r}} S_{\flexpr{}} $}
                    % \State \nn{\act{update-sets}($S'$)}
                    \State $Changes\gets Changes \cup \left(\bigcup_{\flexpr{}\in\quo{r}} Ch_{\flexpr{}}\right)$ 
                    \State $S \gets \{s:\tup{+,s}\in Changes~\wedge~\tup{-,s}\notin Changes\}$
                    \State $\tg{max} \gets \max(\{\tg{\flexpr{}} : (\tup{\tg{\flexpr{}}, e_{\flexpr{}}}, \act{sign}_{\flexpr{}}(\tup{\tg{\flexpr{}}, e_{\flexpr{}}})) $\WRP$
                    \text{ received from }\flexpr{}\in\flexSet_r$\WRP 
                    $\wedge~\act{verify}(\act{sign}_{\flexpr{}}( \tup{\tg{\flexpr{}}, e_{\flexpr{}}}))=\act{true}\})$
                    \State {\bf return} $\tg{max}$
                \EndProcedure

                \Statex				
                    
                \Procedure{put-data}{$\obj{}, C, \tup{\tg{},v}$}
                    \State $\{e_1,e_2,\ldots,e_{|C|}\}\gets RLNC.encode(v)$
                        
                    \State {\bf send} $(\text{{\sc put-data}},\tup{\tg{},e_j}, \act{sig}_{\flexpr{i}}(\tup{\tg{},e_j}),\obj{}, C)$ \WRP to each $\flexpr{j} \in C$
                    
                    \State {\bf until} $\flexpr{i}$ receives ({\sc ack}, $ Ch_{\flexpr{}}$) from a quorum $\quo{r}\subseteq C$ 
                    %s.t.\WRP $\quo{r}\subseteq C$ and $|\quo{r}| \geq \left\lceil \frac{2|C| + k}{3}\right\rceil$
                    
                    %\State \nn{\act{update-sets}($\quo{r}$)}
                    \State $Changes\gets Changes \cup \left(\bigcup_{\flexpr{}\in\quo{r}} Ch_{\flexpr{}}\right)$
                    \State $S \gets \{s:\tup{+,s}\in Changes~\wedge~\tup{-,s}\notin Changes\}$
                \EndProcedure
                    
                \Statex
                    
                \Procedure{get-data}{$\obj{}, C$}
                    \State {\bf send} $(\text{{\sc query-list}}, \obj{}, C)$ to each  $\flexpr{}\in C$
                    
                    \State {\bf until} $\flexpr{i}$ receives $\tup{List_{\flexpr{}}, Ch_{\flexpr{}}}$ from a quorum $\quo{r}\subseteq C$ 
                    %s.t.\WRP $\quo{r}\subseteq C$ and $|\quo{r}| \geq \left\lceil \frac{2|C| + k}{3}\right\rceil$
                    % \State  $Tags_{*}^{\geq k} = $ set of tags that appears in  $k$ lists	\label{line:getdata:max:begin}

                    %\State \nn{\act{update-sets}($\quo{r}$)}
                    \State $Changes\gets Changes \cup \left(\bigcup_{\flexpr{}\in\quo{r}} Ch_{\flexpr{}}\right)$
                    \State $S \gets \{s:\tup{+,s}\in Changes~\wedge~\tup{-,s}\notin Changes\}$
                    
                    %\State\Comment{find verified tag-ec pairs}
                    \State $Pairs_{ver}\gets\{ \tup{\tg{},e}: 
                    (\tup{\tg{},e}, \act{sign}_{\flexpr{}}(\tup{\tg{},e}))\in List_{\flexpr{}}$ 
                    \WRP $ \wedge~\flexpr{}\in\flexSet_r~\wedge~\act{verify}(\act{sign}_{\flexpr{}}(\tup{\tg{}, e}))=\act{true}\}$
                    
                    \State $\obj{}.E[{\tg{}}]\gets\{e:  \tup{\tg{}, e}\in Pairs_{ver}\}$
                    \State  $Tags_{dec}^{\geq k} \gets\{\tg{}: |\obj{}.E[{\tg{}}]| \geq k\}$
                    
                    \State  $\tg{max}^{dec} \gets \max_{\tg{}}\{\tg{}: \tg{} \in Tags_{dec}^{\geq k}\}$ 
                    \label{line:getdata:max:end}
                    \If{ $Tags_{dec}^{\geq k}\neq \emptyset$}
                        \State $\obj{}.\tg{} \gets \tg{max}^{dec}$
                        \State $\obj{}.v \gets RLNC.decode(\obj{}.E[\tg{max}^{dec}])$
                        \State {\bf return $\tup{\obj{}.\tg{},\obj{}.v}$}
                    \EndIf
                \EndProcedure

                % \Statex

                % \nn{
                % \Procedure{update-sets}{$S'$}
                %     \For{$s\in S'$}
                %         \State \Comment{if the flexnode not in the local changes}
                %         \If{$\tup{*,s}\notin Changes$}
                %             \State $Changes\gets Changes \cup\{+,s\}$
                %         \EndIf
                %     \EndFor
                %     \State // add new flexnodes in $S$
                    
                %     \State $S\gets S\cup\bigcup_{\flexpr{}\in\quo{}}S_{\flexpr{}}$
                %     \State // add new joiners in $J$ and remove nodes moved to $S$
                %     \State $J\gets (J\cup\bigcup_{\flexpr{}\in\quo{}}J_{\flexpr{}})\setminus S$
                % \EndProcedure
                % }

            \end{multicols}
        }
        \end{algorithmic}	
        \caption{
            %for  template $A_1$ to implement 
            Dynamic BFT-\optimum{} DAPs implementation at flexnode $\flexpr{i}$}
        \label{algo:dynamic:dap}
        \vspace{-1em}
    \end{algorithm*}
		
Key to the algorithm is the computation of $Ch$ before replying to any request. 
As seen in Algorithm \ref{algo:dynamic:responses}, given the object $\obj{}$ 
and the set $S'$ received in the request (i.e., the cluster $C$ computerd at the 
sender), the procedure \act{calculate-changes} first calculates the hosts 
of $\obj{}$ using the $S_{\obj{}}\gets \suc{\obj{}, S}$, based on the local 
estimate $S$ and then compute the changes to be transmitted in the replies (A\ref{algo:dynamic:responses}:L\ref{line:dyn:res:ch}). 
In particular, it adds a $\tup{+,s}$ entry in $Ch$ for any identifier that is included 
in $S_{\obj{}}$ but not in $S'$. As $S'$ includes only flexnodes that are not 
removed according to $\flexpr{j}$'s $Changes_{\flexpr{j}}$ set, note that for each $s\in S_{\obj{}}\setminus S'$, either: (i) $\tup{+,s}\notin Changes_{\flexpr{j}}$, or
(ii) $\tup{+,s}\in Changes_{\flexpr{j}}$ and it was removed or was replaced by a newly joined flexnode. Either way merging $\{\tup{s,+}\}$ to $Changes_{\flexpr{j}}$ would either 
introduce a new flexnode or the set will remain induct. A $\tup{-, s}$ is added in
$Ch$ for any identifier $s$ found in $S'$ and for which a $\tup{-, s}$ entry is
recorded in the local $Changes$ set. This change will cause the insertion of 
$\tup{-,s}$ in $Changes_{\flexpr{j}}$, and thus the removal of $s$ from the next
cluster computation. This procedure is critical for propagating
the changes in the network to the flexnodes in the system.

\begin{theorem}
    \optimum{} implements a multi-object, dynamic, atomic R/W distributed shared memory, 
    using $[n,k]$ RLNC, and allowing $b$, $b<\frac{n-k}{3}$, flexnodes to be Byzantine for each object.
\end{theorem}
% Responses from flexnodes are also enhanced. Before replying to any request, a flexnode $\flexpr{q}$ computes the flexnodes within its estimate $S_{q}$
% as well as within the joining set $J_{q}$ that should host $\obj{}$,
% getting the sets $S_{\obj{}}, J_{\obj{}}$ respectively.
% Flexnode $\flexpr{q}$ attaches the difference of $S_{\obj{}}$ with $C$ 
% and $J_{\obj{}}$ to any reply. This way $\flexpr{q}$ transmits any flexnodes not known to $\flexpr{i}$, enhancing the knowledge of 
% $\flexpr{i}$ about the participation in the system and the hosts 
% of object $\obj{}$.

\begin{algorithm*}[!ht]
\begin{algorithmic}[2]
    {\scriptsize
    \begin{multicols}{2}
        % \State \nn{at each flexnode $\flexpr{i} \in \members{c}$}
        % %in configuration $c_k$}
        % \Statex
        % \State{\bf State Variables:}	
            
        %     %}\EndPart        
    
        % \Statex
        
        \Receive{{\sc query-tag}, $\obj{}, S'$}{$\flexpr{i}$}
            % \State $\tg{max} = \max(\tg{} : (\tup{\tg{},*},*) \in \obj{}.List)$
            % \State $L_{max}\gets \{(\tup{\tg{},e},\st): \tg{}=\tg{max} ~\wedge~(\tup{\tg{},e},\st)\in \obj{}.List\}$
            \State $P_{max}\gets \max_{\tg{}}(\{(\tup{\tg{},e}, *): (\tup{\tg{},e}, *)\in \obj{}.List\})$
            \State $Ch\gets \act{calculate-changes}(\obj{}, S')$
            \State \textbf{send} $(\mtype{qt-ack},P_{max},Ch)$ to $\flexpr{j}$
        \EndReceive
        
        \Statex

        \Receive{{\sc query-list},$\obj{}, C$}{$\flexpr{i}$}
            % \State \nn{$S_{\obj{}}\gets \suc{\obj{}, S}$}
            % \State \nn{$J_{\obj{}}\gets \suc{\obj{}, J}$}
            \State $Ch\gets \act{calculate-changes}(\obj{}, S')$
            \State \textbf{send} ($\obj{}.List, Ch$) to $\flexpr{j}$
            %S_{\obj{}}\setminus C, J_{\obj{}})$ to $q$}
        \EndReceive
            
        \Statex
        \Statex
        
        \Receive{{\sc put-data}, $(\tup{\tg{},e},\act{sign}_{\flexpr{i}}(\tup{\tg{},e})),\obj{}, C$}{$\flexpr{i}$}
            \If{$\act{verify}(\tup{\tg{},e}, \act{sign}_{\flexpr{i}}(\tup{\tg{},e}))~\wedge~(\tup{\tg,*},*)\notin\obj{}.List$}
                \State $\obj{}.List \gets \obj{}.List \cup \{(\tup{\tg{}, e},\act{sign}_{\flexpr{i}}(\tup{\tg{}, e})) \}$ 
                \If{$|\obj{}.List| > (\delta+1)$}
                    \State $\tg{min}\gets\min\{\tg{}: \tup{\tg{},*}\in \obj{}.List\}$
                    \State $L_{min}\gets \{\tup{\tg{},e}: \tg{}=\tg{min} ~\wedge~(\tup{\tg{},e},*)\in \obj{}.List\}$
                    
                    \State $\obj{}.List \gets \obj{}.List \backslash~L_{min}$ 
                    %\cup \{  (  \tg{min}, \bot)  \}$
                    \label{line:server:removemin}
                    %\State $E \gets E \setminus\{\tup{\tg{},e}: \tg{}=\tg{min} ~\wedge \tup{\tg{},e}\in E\}$
                    %\State $List \gets List  \cup \{  (  \tg{min}, \bot)  \}$\label{line:server:removemin}
                \EndIf
            \EndIf
            % \State \nn{$S_{\obj{}}\gets \suc{\obj{}, S}$}
            % \State \nn{$J_{\obj{}}\gets \suc{\obj{}, J}$}
            \State $Ch\gets \act{calculate-changes}(\obj{}, S')$
            \State \textbf{send} (\mtype{ack},$Ch$) to $\flexpr{j}$
            %$S_{\obj{}}\setminus C, J_{\obj{}})$ to $q$}
        \EndReceive

        \Statex

        \Procedure{calculate-changes}{$\obj{}, S'$}
            \State $S_{\obj{}}\gets \suc{\obj{}, S}$
            %\State \nn{$J_{\obj{}}\gets \suc{\obj{}, J}$}
            % \State \Comment{suggest add: flexnodes that are not included in the sender but exist in the receiver}
            % \State \Comment{suggest remove: flexnodes contained in the sender but a removal is recorded in the changes of the receiver}
            \State {\bf return} $\{\tup{+,s}: s\in S_{\obj{}}\setminus S'\} ~ \cup$ \WRP$~ \{\tup{-, s} : s\in S'\setminus S_{\obj{}}~ \wedge ~ \{-, s\}\in Changes\}$ \label{line:dyn:res:ch}
        \EndProcedure
    \end{multicols}
}
\end{algorithmic}	
\caption{Dynamic BFT-\optimum{} response protocols at flexnode $\flexpr{i}$}
                \label{algo:dynamic:responses}
                %\vspace{-1em}
\end{algorithm*}

%% file: sec_optimum_correct.tex
% \subsection{Correctness of the Reconfig Protocol}
% \label{ssec:correct:recon}

%\nn{[TODO: needs revision]}

In this section, we examine the correctness of 
\optimum{}. Actions in \optimum{} are 
performed in a per object basis so, due to composability of atomic objects, it is sufficient to show safety when considering 
a single object. Thus, for the sequel, we refer to the object only whenever necessary. Before proceeding,
we state the following lemmas proving impoprtant 
properties of the set $Changes$ maintained at each flexnode.

\begin{lemma}
\label{lem:changeset:monotonic}
    Let $\st$ and $\st'$ two states in an execution $\EX$ of \optimum{}, such that, $\st$ appears before $\st'$ in $\EX$. If $\flexpr{}\in S^+$ in both states, then $\atT{\flexpr{}.Changes}{\st}\subseteq \atT{\flexpr{}.Changes}{\st'}$.
\end{lemma}

The lemma follows directly from the algorithm as the set $Changes$ in each flexnode is initialized
when it joins and is updated in each dap action by adding more elements. The next lemma 
shows that the size of the participation estimate at each flexnode is always greater than the RLNC code
dimension $n$.

\begin{lemma}
\label{lem:S:size}
    Let $\st$ a state in execution $\EX$ of \optimum{}. If $\flexpr{}\in S^+$ at $\st$,
    then $|\atT{\flexpr{}.S}{\st}|\geq n$.
\end{lemma}

\begin{proof}
    Notice that the value of $S$ variable is updated during the following operations:
    (i) the join operation, (ii) the depart operation, and (iii) the three data access 
    primitives. In all cases $S = \{s:\tup{+,s}\in Changes~\wedge~\tup{-,s}\notin Changes\}$.
    In both join and depart protocols (Algorithms \ref{algo:optimum:join} and \ref{algo:optimum:depart}), the $S$ is computed right after an $SMR.\act{get}()$ 
    operation. Thus, if $\st$ is the state following the update of $S$ during a join or depart operation, then $\atT{\flexpr{}.S}{\st}=S^+$, and by our assumption $|\atT{\flexpr{}.S}{\st}|=|S^+|\geq n$.

    So it remains to examine how the set changes during the dap operations. During the  execution of the primitives, the $Changes$ set is updated with the changes found in the
    received messages. 
    %Let's proceed to examine the membership of the $Changes$ set. The 
    Initially, the $Changes$ at $\flexpr{}$ is set during the join operation 
    of $\flexpr{}$. If the $Changes$ set remains the same then from the previous arguments 
    $|S|=|S^+|\geq n$. 

    Lets assume now that $\st$ is the first state where the set $Changes$ is modified. 
    Let $Diff^+$ and $Diff^-$ denote the new additions and removals added in the $Changes$ set.
    It follows that some flexnode $\flexpr{j}$ discovered those changes from some call of the $SMR.\act{get}()$ in a state $\st'$ that appears before $\st$ in $\EX$. Since, the set $Changes$ does not include $ch\in Diff^+\cup Diff^-$ and since $\st$ is the first state where $Changes$ is modified, then 
    by the {\bf Inclusion} property of the oracle it follows that $\flexpr{j}$ invoked 
    its $SMR.\act{get}()$ after the invocation of the  $SMR.\act{get}()$ operation during the    join operation at $\flexpr{}$. By the same property it also follows that if
    $Changes_j$ is the result of the $SMR.\act{get}()$ operation at $\flexpr{j}$ then 
    $Changes\subseteq Changes_j$. Since $\atT{\flexpr{}.Changes}{\st}= Changes \cup Diff^+\cup Diff^-$, then $\atT{\flexpr{}.Changes}{\st} = Changes_j$. By the constraint of the oracle it must 
    hold that $|S^+_j|\geq n$ and hence the set $\atT{\flexpr{}.S}{\st}$ extracted from $\atT{\flexpr{}.Changes}{\st}$ will be $|\atT{\flexpr{}.S}{\st}|=|S^+_j|\geq n$ completing the proof.
\end{proof}

% \begin{lemma}
%     In any execution $\EX$ of \optimum{}, the estimate $S$ at each flexnode $\flexpr{}$ is growing monotonically. 
% \end{lemma}

% In summary, the main departures from the static environment are the following: 
% (i) \textit{Dynamic Participation}: each flexnode
% $\flexpr{}$ maintains a set $Changes$ recording the flexnode joins and departs
% from the system,
% (ii) \textit{Node Discovery}: using consistent hashing and the set of changes, each flexnode computes the cluster of flexnodes $C$ hosting each object $\obj{}$,
% and (iii) \textit{Read/Write}: a read and write operation in each flexnode 
% is repeated until is executed in a "recent" estimate $S$. In the sequel we 
% show that despite the new elements, the algorithm satisfies the three properties
% A1-A3 of atomicity in each execution. Notice that actions in the algorithm 
% are defined per object. So, for simplicity, we show
% the correctness of the algorithm for a single object, and the same proof 
% applies independently for any object in the system. 

\newcommand{\pestimate}[2]{\atT{#1.S}{#2}}
\newcommand{\pchange}[2]{\atT{#1.Changes}{#2}}

% \myparagraph{Notations and definitions.}
% 	For a flexnode $\flexpr{}$, we use the notation $\atT{\flexpr{}.var}{\state}$ to refer to the value of the state variable $var$, in $\flexpr{}$, at a state $\state$ of an  execution $\EX$. 

    % \begin{definition}[Participation Estimate]
    %     In any given state $\st$ of an execution $\EX$, the variable $\atT{\flexpr{}.S}{\st}$, which we refer to as the \textit{participation estimate} of $\flexpr{}$, 
    %     contains the identifiers of the flexnodes that joined or departed as seen by $\flexpr{}$ by state $\st$. 
    %     % We use the notation $\pestimate{\flexpr{}}{\st}$ as a shorthand to denote the value of the estimate of $\flexpr{}$ in 
    %     % a state $st$.
    % \end{definition}

    % We also need to present some remarks that hold due to the nature 
    % of the chain oracle we use: 

    % \begin{remark}
    %     If $$
    % \end{remark}

\myparagraph{Liveness.}
Let us begin by proving that any read/write/join/depart operations in \optimum{}
terminate in any execution $\EX$, given that no more than $b$,  for $b<\frac{n-k}{3}$,
flexnodes out of $n=|C|$ storing an object $\obj{}$, may arbitrarily fail, where $[n,k]$
the coding parameters for RLNC. We begin with a lemma on the termination or 
write/join/depart operations. the termination of read operations is examined 
separately as the termination of the read also depends on the decodability of the value. 

\begin{lemma}
    Any write/join/depart operation $\op$ invoked by a flexnode $\flexpr{}$ terminates in any execution $\EX$ of \optimum{}, 
    where $b$, $b<\frac{n-k}{3}$, of flexnodes may be Byzantine.
\end{lemma}

\begin{proof}
    Any write/join/depart operation $\op$ determines the set $C$ of the flexnodes to communicate 
    by calling the distance function on their local participation set $S$. Since by 
    Lemma \ref{lem:S:size}, at any state of the execution $|S|\geq n$ for any operation. 
    Thus, by definition, the distance function $\mathfrak{D}_{*}()$ returns $n$ flexnodes.
    If $\op$ is a join, a flexnode $\flexpr{}$ communicates with $n\leq|C|\leq2n$  
    as it captures both the successor and predecessor flexnodes. In any case it waits for 
    $\ceil{\frac{2|C|+1}{3}}$ replies. Since  $b<\frac{n-k}{3}$ then for any number of $k$ 
    $\flexpr{}$ will receive $|C|-b\geq \frac{2n+k}{3}$ replies and thus it will terminate. 
    With similar reasoning any write and depart operations will receive the necessary 
    replies and terminate.
\end{proof}

Although the termination of write/join/depart operation depend solely on the number of 
replies, a read operation terminates successfully if it receives the necessary replies 
and also is able to decode the value 
of the object. 

\begin{lemma}
    Any read operation $\rd$ invoked by a flexnode $\flexpr{}$ terminates in any execution $\EX$ of \optimum{}, 
    where $b$, $b<\frac{n-k}{3}$, of flexnodes may be Byzantine.
\end{lemma}

\begin{proof}[Proof Sketch]
    We need to show that $\rd$ will receive at least $k$ coded elements associated with the same tag $\tg{}$ in order to decode and return the value of the 
    object. We can show this lemma by applying direclty Lemma \ref{lem:join:tag}.
\end{proof}

% \begin{lemma}
%     A read operation $\rd$ on an object $\obj{}$ will fail to decode the value 
%     of $\obj{}$, and thus terminate in an execution $\EX$ of \optimum{}, if
%      $\exists  \mathcal{J},  \mathcal{J}\subseteq \flexSet$, s.t. $\forall j_i\in\mathcal{J}$,
%     $\wrt(\obj{})\bef \act{join}(j_i)$ and $\act{join}(j_i) \bef \rd$, 
%     where $\wrt(\obj{})$ the last write operation on 
%     $\obj{}$, $|\mathcal{J}|=n$, and at most $\frac{n-k}{3}$ joining flexnodes,
%     responsible for $\obj{}$, retrieve the value written by $\wrt(\obj{})$. 
% \end{lemma}

% \begin{proof}
%     We can show that the observation holds by contradiction. Let $\EX$ be an 
%     execution of \optimum{} where a set $\{j_1,\ldots, j_n\}$ join the system 
%     such that $\act{join}(j_i)\bef \act{join}(j_{i+1})$, for $1\leq i \leq n-1$,
%     and $\act{join}(j_1)$ is invoked after the last completed write operation on 
%     $\obj{}$.    
% \end{proof}
    
\myparagraph{Safety.}
To proof the safety (atomicity) of the algorithm we need to show that all 
atomicity properties A1-A3 are satisfied for every read/write operation. 
It is easy to see that if two operations are invoked on the same set of 
flexnodes $C$, atomicity holds by Theorem \ref{thm:single-object:safety}. So we need to examine 
whether the properties hold when changing the set of flexnodes. 

We begin by examining whether the value written on object $\obj{}$ 
by a completed write operation is propagated when new flexnodes join the service. 
First we show that the object $\obj{}$ is propagated successfully after a join 
operation. 

\begin{lemma} [Object Propagation]
\label{lem:join:obj}
    Let $\op$ a join operation by $\flexpr{j}$ in an execution $\EX$ of \optimum{}.
    If  $\st_{add}$ the state of $\op$ after invoking $SMR.add()$, and
    $\atT{S^+}{\st_{add}}$ the 
    participation estimate at the SMR oracle at $\st_{add}$,
    then for any object $\obj{}\in\objSet$, if $\flexpr{j}\in \mathfrak{D}_{suc}(\obj{},\atT{S^+}{\st_{add}})$, then $\obj{}\in \atT{\flexpr{}.D}{\st_{add}}$.
\end{lemma}

\begin{proof}
    Lets assume by contradiction that although $\flexpr{j}\in \mathfrak{D}_{suc}(\obj{},S^+)$, $\flexpr{j}$ did not receive $\obj{}$ and thus, 
    $\obj{}\notin \atT{\flexpr{}.D}{\st_{res}}$. 

    At the invocation of $\op$, $\flexpr{j}$ gets the changes $Changes$ in the network 
    by calling $SMR.get()$ (A\ref{algo:optimum:join}:L\ref{line:join:changes1}), and computes its local participation estimate $S$. The set $S$ is not modified
    during $\op$ so it follows that:
    \[
        \atT{\flexpr{j}.S}{\st_{res}} = \{s:\tup{+,s}\in Changes~\wedge~\tup{-,s}\notin Changes\}
    \]
    Note that $\obj{}$ is 
    inserted in the local set $D$ only if $\flexpr{j}\in \mathfrak{D}_{suc}(\obj{},S)$. In other words $\flexpr{j}$ should be one of the $n$ nodes
    with the closest distance $\distance{\hash{}(o), \hash{}(\flexpr{j})}$.
    If $\flexpr{}$ is the $i^{th}$ closest flexnode to $\obj{}$, for $0<i<n$, 
    then the object is 
    maintained by $n-i$ successors and $i$ predecessors of $\flexpr{j}$ in the 
    logical ring. Let us assume w.l.o.g., that the $\distance{\hash{}(o), \hash{}(\flexpr{j})}= d$. 
    
    Let $\atT{S^\pm}{\st_{res}}$ denote the set of changes at the 
    oracle at state $\st_{res}$. By the \textbf{Inclusion} property of the 
    oracle it follows that $Changes\subseteq\atT{S^\pm}{\st_{res}}$. 
    There are two cases to consider: (i) either $Changes = \atT{S^\pm}{\st_{res}}$,
    or (ii)  $Changes \subset \atT{S^\pm}{\st_{res}}$.

    By the join protocol, $\op$ communicates with a cluster of 
    flexnodes $C=\mathfrak{D}_{suc}(\flexpr{j},S)\cup\mathfrak{D}_{pred}(\flexpr{j},S)$, i.e. with all the flexnodes (successors and predecessors) 
    with distance $n$ from 
    $\flexpr{j}$ (A\ref{algo:optimum:join}:L\ref{line:join:neighbors}).
    Hence, $n\leq|C|\leq2n$. Note that if $\flexpr{j}\in\mathfrak{D}_{suc}(\obj{},S)$ then $\mathfrak{D}_{suc}(\obj{},S)\subseteq C$. Furthermore
    we also assume that no more than $b$ flexnodes from $C$ may fail or depart concurrent 
    with $\op$. Therefore, $|\atT{\flexpr{j}.S}{\st_{res}}\setminus\atT{S^+}{\st_{res}}|\leq b$.
    
    \case{(i)} If $Changes\cup\{\tup{+,\flexpr{j}}\} = \atT{S^\pm}{\st_{res}}$, then $\atT{\flexpr{j}.S}{\st_{res}}\cup\{\flexpr{j}\} = \atT{S^+}{\st_{res}}$.
    So in this case every flexnode $\flexpr{i}$  receiving the join message from $\flexpr{j}$ 
    will also set $S_{\flexpr{i}} = \atT{S^+}{\st_{res}}$. Thus, if $\obj{}\in\flexpr{i}.D$, and since according to our assumption $\flexpr{j}\in \mathfrak{D}_{suc}(\obj{},S^+)$, then $\flexpr{i}$ will include $\obj{}$ in 
    its reply to $\flexpr{j}$. Since no more than $b$ nodes out of the $n$ that host 
    $\obj{}$ may fail, then $\flexpr{j}$ will receive the object from at least
    $2b+1$ replies and $\obj{}\in \atT{\flexpr{}.D}{\st_{res}}$ contradicting
    our assumption.

    \case{(ii)} In this case there are some flexnodes joined or departed the 
    service and thus the oracle set was modified since $\flexpr{j}$ invoked the
    \act{get} operation. This modification affects the set of participating 
    flexnodes at the oracle $\atT{S^+}{\st_{res}}$. If the modification 
    resulted in $\flexpr{j}\notin \suc{\obj{},S^+}$ (e.g., because of flexnode joins) then there is nothing to
    examine. So two cases remain to investigate while $\flexpr{j}\in \suc{\obj{},S^+}$: (a) either $\flexpr{j}\in\mathfrak{D}_{suc}(\obj{},S)$,
    or (b) $\flexpr{j}\notin\mathfrak{D}_{suc}(\obj{},S)$. 
    
    If $\flexpr{j}\in\mathfrak{D}_{suc}(\obj{},S)$ then as shown before
    $\mathfrak{D}_{suc}(\obj{},S)\subseteq C$, and hence same as in 
    case (i) $\flexpr{j}$ will communicate with $n$ flexnodes that
    maintain $\obj{}$ and will receive at least $2b+1$ replies. Therefore,
    will add $\obj{}\in \atT{\flexpr{}.D}{\st_{res}}$ contradicting
    our assumption.

    If $\flexpr{j}\notin\mathfrak{D}_{suc}(\obj{},S)$, then since $\flexpr{j}\in \suc{\obj{},S^+}$ and at most $b$ flexnodes may fail or depart, then it must 
    be the case that $\distance{\hash{}(o), \hash{}(\flexpr{j})}\leq n+b$. Thus,
    at least $n-b$ of the flexnodes that maintain $\obj{}$ are in $C$.
    From this it follows that $\flexpr{j}$ will communicate with at least $2b+1$
    of those flexnodes, and at least $2b+1$ of them will reply assuming that 
    at most $b$ may depart or fail (if $b$ flexnodes departed we should expect all $2b+1$ to reply). In this case $\flexpr{j}$ will receive and add 
    $\obj{}\in \atT{\flexpr{}.D}{\st_{res}}$ contradicting our assumption and 
    completing the proof.
\end{proof}

So by Lemma \ref{lem:join:obj}, we showed that a joining node will be able to 
discover and request any object created in the system before its invocation. 
% As a join operation affects the participation of the service, next we examine
% whether the estimates between two consecutive operations remain consistent. 
Using that lemma we investigate whether a tag is propagated correctly as new 
nodes join the service

\begin{lemma} [Tag Propagation]
\label{lem:join:tag}
Let $\wrt$ be a complete write operation on object $\obj{}$ invoked by $\flexpr{}$ in an execution $\EX$ of \optimum{}.
If $\tg{\wrt}=\atT{\flexpr{}.(\obj{}.\tg{})}{\st_{\wrt}}$ is the tag associated with $v$ at the response 
state $\st_{\wrt}$ of $\wrt$, then for any join operation $\op$ by $\flexpr{j}$ such that $\wrt\bef\op$, if $\st_{\op}$ the response state of $\op$, either $\tg{\wrt}\in \atT{\flexpr{j}.(o.List)}{\st_{\op}}$
or $\exists\tg{\wrt'} \in \atT{\flexpr{j}.(o.List)}{\st_{res}}$, s.t. $\tg{\wrt'}>\tg{\wrt}$
and $\tg{\wrt'}$ was written by a write $\wrt'\bef\op$.
% if $\atT{\flexpr{}'.(\obj{}.\tg{}')}{\st_{res}}$ the maximum tag-encoded pair in $\obj{}.List$ at the completion of the collection operation, then $\atT{\flexpr{}'.(\obj{}.\tg{}')}{\st_{col}}\geq \atT{\flexpr{}.(\obj{}.\tg{})}{\st_{res}}$.
\end{lemma}

\begin{proof}
    We will proof this lemma by induction on the number of join operations that succeed operation $\wrt$.

    As the base case let $\op$ be the only join operation such that $\wrt\bef\op$
    and the participation estimate $S_{\wrt}$ at the response of $\wrt$ is the 
    same as the participation estimate $S_{\op}$ after the $SMR.\act{get}()$
    operation at $\op$. According to the write protocol, $\wrt$ issues a 
    $\daputdata{\obj{}, C_{\wrt}}{\tup{\tg{\wrt}, v}}$ operation on $C_{\wrt}=\mathfrak{D}_{suc}(\obj{},S_{\wrt})=n$ before completing. 
    A quorum $\quo{\wrt}$ of size $\frac{2|C_{\wrt}|+k}{3}$ within $C_{\wrt}$ receives 
    and replies to the \mtype{write} messages of $\wrt$. Every flexnode in 
    $\quo{\wrt}$ adds $\tup{\tg{\wrt},*}$ in its $\obj{}.List$ before replying 
    to $\wrt$. On the other hand, $\op$ will send \mtype{fetch-obj} to a set 
    $C_{\op}$ of flexnodes such that $C_{\wrt}\subseteq C_{\op}$ and will wait
    for a quorum $\quo{\op}$ of $\frac{2|C_{\wrt}|+\beta}{3}$ of flexnodes to reply. 
    Therefore, $|\quo{\wrt}\cap\quo{\op}|\geq b+\beta$, for $b<\frac{|n|-k}{3}$ and 
    $1\leq\beta\leq k$. So, assuming that $b$ of those flexnodes are Byzantine and may reply with outdated values, $\op$ will either discover $\tg{\wrt}$ in $\beta$ lists, 
    or there exists a list $L$, s.t. $\forall\tg{}\in L$, $\tg{}>\tg{\wrt}$. 
    If this is the 
    case, since $\delta$ writes can be concurrently invoked on $\obj{}$ with $\op$ and $L$ contains at most $\delta+1$ tags, then it must be the case that one of those 
    writes, say $\wrt'$, has completed and wrote a tag $\tg{}' > \tg{\wrt}$ before
    the invokation of $\op$; i.e. $\wrt'\bef\op$. Let $\wrt'$ be the 
    completed write with the minimum tag $\tg{}'$ in $L$. If so, $\op$ 
    will discover $\tg{}'$ in more than $\beta$ lists and thus,
    $\tg{}'\in\atT{\flexpr{j}.(o.List)}{\st_{\op}}$. 
    %Since $L$ contains tags all greater than $\tg{\wrt}$ then $\forall\tg{}\in maxTag_{\delta}$ (A\ref{algo:optimum:join}:L\ref{line:join:deltamax}), $\tg{}>\tg{\wrt}$. 
    So either $\tg{\wrt}\in \atT{\flexpr{j}.(o.List)}{\st_{\op}}$ or 
    $\tg{}' \in \atT{\flexpr{j}.(o.List)}{\st_{res}}$, $\tg{}'>\tg{\wrt}$ and $\wrt'\bef\op$.

    For the induction hypothesis, we assume now that this holds after $x$ subsequent joins $\{\op_1,\ldots,\op_x\}$, s.t. $\op_{i}\bef\op_{i+1}$ for $1\leq i< x$, succeed $\wrt$ in $\EX$. Let's now investigate what happens when we extend $\EX$ 
    with the join operation $\op$ from $\flexpr{j}$. As shown in Lemma \ref{lem:join:obj}, during its join operation, $\flexpr{j}$ receives $\obj{}$ from a set of at least $b+1$ flexnodes. We know that
    $\mathfrak{D}_{suc}(\obj{},S_{\op})\subseteq C_{\op}$ are the flexnodes hosting
    object $\obj{}$. Out of those let $J_{\wrt}\subset\mathfrak{D}_{suc}(\obj{},S_{\op})$, , s.t. $|J_{\wrt}|=x$,  be the flexnodes that joined after $\wrt$ and the rest
    $C'_{\wrt}\subset C_{\wrt}$ (maybe empty) the set of flexnodes that received a \mtype{put-data} message from $\wrt$. By the induction hypothesis we know that 
    $\forall \flexpr{}\in J$, $\flexpr{}$ added $\tg{\wrt}$ or a tag $\tg{\wrt'}>\tg{\wrt}$ written by proceeding $\wrt'$ write in their
    local $List$. Note that if $\tg{\wrt'}$ was added by a join operation $\op_i$,
    then $\wrt'\bef\op_i$ and by construction $\wrt'\bef\op$. 
    % Let 
    % \[
    % \tg{max} = \max(\min(\{\tg{\wrt'}: \tg{\wrt'}\in L_i~\wedge~\tg{\wrt'}\text{ written by completed write }\wrt'\}))
    % \]
    % denote the maximum of the minimum tags added in some list $L_i$ of the join operations $\op_i$.
    Let $\tg{max}\geq\tg{\wrt}$ be the maximum tag added in the list of some join $\op_i$ 
    and was written by a write operation $\wrt'\bef\op_i$. Since $\op_i\bef\op$,
    then $\wrt'$ is the last write that precedes any $\op_j$, for $i<j\leq x$.  So any $\op_j$ will
    obtain $\tg{max}$ from more than $\beta$ flexnodes and will add $\tg{max}$ 
    in their list.
    %Before completing $\wrt'$ propagated $\tg{max}$ to a set
    %of $|\quo{\wrt'}|\geq\frac{2|C|+k}{3}$ flexnodes. 
    % Note that $\quo{\wrt'}$ will miss all the flexnodes that in invoked $\op_i$ such that $\wrt'\bef\op_i$.
    So, at least $n-b$ flexnodes in $\mathfrak{D}_{suc}(\obj{},S_{\op})$ will still have $\tg{max}$ in their list despite the new joins. 
    Following similar reasoning as before,  either $\tg{max}\in \atT{\flexpr{j}.(o.List)}{\st_{\op}}$ or 
    $\tg{}' \in \atT{\flexpr{j}.(o.List)}{\st_{res}}$, $\tg{}'>\tg{max}$ and $\wrt'\bef\op$. Since $\tg{max}\geq\tg{\wrt}$ this completes the proof.
    
    % By the algorithm, a tag is removed from the $List$ 
    % only if more than $\delta+1$ higher tags are added in the $List$. Then every newly joined node 
    % will reply with $\tg{\wrt}$ if no higher tag replaced it or $\tg{}\geq \tg{\wrt}$. 
    % Since $|C_{\op}|=|J\cup C'_{\wrt}|=n$ then $\flexpr{j}$ receives more than $b+\beta$ replies 
    % with $\tg{}\geq\tg{\wrt}$ or will observe $\tg{\wrt}$ in at least $\beta$ lists.
    %  So either $\tg{\wrt}\in \atT{\flexpr{j}.(o.List)}{\st_{\op}}$ or 
    % $max(\tg{j}) \in \atT{\flexpr{j}.(o.List)}{\st_{res}}$, $\tg{j}>\tg{\wrt}$ as needed.
\end{proof}

% The following lemma shows the consistency of the estimate in each flexnode. 

% \begin{lemma}
%     In any execution $\EX$ of \optimum{}, if $\state$ and $\state'$  are two states 
%     in $\EX$ such that $\state$ appears before $\state'$ in $\EX$, then for each flexnode $\flexpr{}$ should hold that $\pestimate{\flexpr{}}{\state}\subseteq \pestimate{\flexpr{}}{\state'}$. 
% \end{lemma}
It is important to examine that flexnodes update their $Changes$ set 
to ensure they have a recent estimate of the participation in the service.
The following lemma examines whether the chagnes propagate even when 
some flexnodes are not directly contacted during join/depart operations.

\begin{lemma} [Change Progress]
\label{lem:estimate:consistent}
    Let $\op_1, \op_2$, two complete read or write operations in an execution $\EX$ of \optimum{} invoked by flexnodes $\flexpr{1},\flexpr{2}$ respectively. 
    If $\op_1\bef\op_2$ in $\EX$, and $\pchange{\flexpr{1}}{\st_1}$, $\pchange{\flexpr{2}}{\st_2}$ the participation estimates in the states following the response step of $\op_1$ and the end of phase I in $\op_2$ resp., 
    then it must hold that $\pchange{\flexpr{1}}{\st_1}\subseteq \pchange{\flexpr{2}}{\st_2}$.
\end{lemma}

\begin{proof}[Proof Sketch]
    If $\flexpr{1}=\flexpr{2}$ this result follows from Lemma \ref{lem:changeset:monotonic}.
    
    % We need to investigate the case where $\flexpr{1}\neq\flexpr{2}$.
    % Let us denote by $\atT{\flexpr{1}.Changes}{\st_\}}$
    When the two operations are invoked by different flexnodes (i.e. $\flexpr{1}\neq\flexpr{2}$)  it suffices to make 
    the following observations: (i) any join/depart operation communicates 
    with at least $2b+1$ flexnodes before terminating, and (ii) any read/write
    operation communicates with at least $2b+k$ flexnodes before completging. 
    Therefore, there is a non-empty intersection between those sets, and any 
    server in that intersection both updated its local $Changes$ using the SMR
    oracle, and replied with updates to each read/write operation. Thus,
    even if $\op_2$ starts with an outdated $Changes$ set, it will get 
    updated during the DAP execution and eventually it will hold that 
    $\pchange{\flexpr{1}}{\st_1}\subseteq \pchange{\flexpr{2}}{\st_2}$.
\end{proof}

Below are the lemmas necessary to show Atomicity. 

\begin{lemma} [Write-Write Consistency]
\label{lem:dyn:write:write}
    Let $\wrt_1, \wrt_2$, two complete write operations in an execution 
    $\EX$ of \optimum{} invoked by flexnodes $\flexpr{1},\flexpr{2}$ respectively. 
    If $\wrt_1\bef\wrt_2$ in $\EX$, and $\atT{\flexpr{1}.(o.\tg{\wrt_1})}{\st_1}$, $\atT{\flexpr{2}.(o.\tg{\wrt_2})}{\st_2}$ the tags for $\obj{}$ at the response states
     of $\wrt_1$ and $\wrt_2$, then it must hold that 
    $\atT{\flexpr{1}.(o.\tg{\wrt_1})}{\st_1}< \atT{\flexpr{2}.(o.\tg{\wrt_2})}{\st_2}$.
\end{lemma}

\begin{proof}[Proof Sketch]
 By Lemma \ref{lem:estimate:consistent}, $\wrt_2$ will eventually attempt 
 to execute the $\dagettag{}$ primitive in a cluster of flexnodes from 
 a set $Changes \supseteq \pchange{\flexpr{1}}{\st_1}$. By 
 Lemma \ref{lem:join:tag} the tag written by $\wrt_1$ will be propagated 
 by a join/depart operation $\op$ such that $\wrt_1\bef\op$. So $\wrt_{2}$
 by the same lemma will observe a max tag $\tg{}\geq \atT{\flexpr{1}.(o.\tg{\wrt_1})}{\st_1}$, and hence will increment it and assign a tag with the value to be written 
 $\atT{\flexpr{2}.(o.\tg{\wrt_2})}{\st_2}>\atT{\flexpr{1}.(o.\tg{\wrt_1})}{\st_1}$.
\end{proof}

\begin{lemma} [Write-Read Consistency]
\label{lem:dyn:write:read}
    Let $\wrt_1, \rd_2$, two complete write and read operations resp. in an execution 
    $\EX$ of \optimum{} invoked by flexnodes $\flexpr{1},\flexpr{2}$. 
    If $\wrt_1\bef\rd_2$ in $\EX$, and $\atT{\flexpr{1}.(o.\tg{\wrt_1})}{\st_1}$, $\atT{\flexpr{2}.(o.\tg{\rd_2})}{\st_2}$ the tags for $\obj{}$ at the response states
     of $\wrt_1$ and $\rd_2$, then it must hold that 
    $\atT{\flexpr{1}.(o.\tg{\wrt_1})}{\st_1}\leq \atT{\flexpr{2}.(o.\tg{\rd_2})}{\st_2}$.
\end{lemma}

\begin{proof}[Proof Sketch]
 By Lemma \ref{lem:estimate:consistent}, $\rd_2$ will eventually attempt 
 to execute the $\dagetdata{}$ primitive in a cluster of flexnodes from 
 a set $Changes \supseteq \pchange{\flexpr{1}}{\st_1}$. By 
 Lemma \ref{lem:join:tag} the tag written by $\wrt_1$ will be propagated 
 by a join/depart operation $\op$ such that $\wrt_1\bef\op$. So $\rd_{2}$
 by the same lemma will observe a max tag $\tg{}\geq \atT{\flexpr{1}.(o.\tg{\wrt_1})}{\st_1}$. The addition of each flexnode in the cluster that maintains the object
 $\obj{}$ results also in the removal of a flexnode from that cluster. Notice that the following hold: (i) a tag from a completed write is added in the list of every joined node, (ii) the joined node generates a new encoded element, and (iii) 
 no more than $\delta$ writes can be invoked concurrently with the read, 
 then the read will be able to find this or a higher tag in at least $b+k$
 flexnodes. So it will get at least $k$ \textit{linearly independent} encoded 
 elements and will be able to decode the calue associated with a tag $\atT{\flexpr{2}.(o.\tg{\rd_2})}{\st_2}\geq\atT{\flexpr{1}.(o.\tg{\wrt_1})}{\st_1}$.
\end{proof}

\begin{lemma} [Read-Read Consistency]
\label{lem:dyn:read:read}
    Let $\rd_1, \rd_2$, two complete read operations in an execution 
    $\EX$ of \optimum{} invoked by flexnodes $\flexpr{1},\flexpr{2}$. 
    If $\rd_1\bef\rd_2$ in $\EX$, and $\atT{\flexpr{1}.(o.\tg{\rd_1})}{\st_1}$, $\atT{\flexpr{2}.(o.\tg{\rd_2})}{\st_2}$ the tags for $\obj{}$ at the response states
     of $\rd_1$ and $\rd_2$, then it must hold that 
    $\atT{\flexpr{1}.(o.\tg{\rd_1})}{\st_1}\leq \atT{\flexpr{2}.(o.\tg{\rd_2})}{\st_2}$.
\end{lemma}

\begin{proof}[Proof Sketch]
 Operation $\rd_1$ performs a $\daputdata{*}{\tup{\atT{\flexpr{1}.(o.\tg{\rd_1})}{\st_1},*}}$ before completing. By Lemma \ref{lem:estimate:consistent}, $\rd_2$ will eventually attempt 
 to execute the $\dagetdata{}$ primitive in a cluster of flexnodes from 
 a set $Changes \supseteq \pchange{\flexpr{1}}{\st_1}$. By 
 Lemma \ref{lem:join:tag} the tag $\tup{\atT{\flexpr{1}.(o.\tg{\rd_1})}{\st_1}}$ propagated by $\rd_1$ will be transferred to any newly joined flexnode. 
 Thus with similar reasoning as in the previous lemmas, $\rd_2$ will eventually find 
 and decode the value associated with a tag $\atT{\flexpr{2}.(o.\tg{\rd_2})}{\st_2}\leq \atT{\flexpr{1}.(o.\tg{\rd_1})}{\st_1}$.
\end{proof}

\begin{theorem}
    \optimum{} implements a multi-object, dynamic, atomic R/W distributed shared memory.
\end{theorem}

\begin{proof}[Proof Sketch]
    Using Lemmas \ref{lem:dyn:write:write}, \ref{lem:dyn:write:read}, and 
    \ref{lem:dyn:read:read}, and similar reasoning as in Theorem \ref{thm:single-object:safety}, we can proof that \optimum{} satisfies all three properties 
    A1-A3 of atomicity.
\end{proof}

% The first lemma states that following our joining protocol, any node 
% that joins receives all the objects the node should be responsible for.

% \begin{lemma}
%     In any execution $\EX$ of \optimum, if $\objSet_{c}\subseteq\objSet$ the set of objects stored in the system at the invocation step of 
%     a join operation $\op$ from $q$, and $\objSet_q$ the set of objects stored
%     by $q$ at the response step of $\op$, then $\nexists o\in\objSet_c\setminus\objSet_q$ s.t. $distance(o,q)\leq n$.
% \end{lemma}

% \begin{proof}
%     The proof of this Lemma follows from the algorithm we follow for object 
%     distribution, as well as the definition of the neighboring function $\mathfrak{D}(q,S)$.\nn{[NN:elaborate a bit more.]}
% \end{proof}

% For any two operations that request the value of a virtual node $v_i$
% we need to ensure that the second operation will observe a value more
% recent than the first operation. As the two operations may be invoked 
% on different flexnodes, then we need to examine the replica nodes that 
% each node discovers given its Kademlia tree. So the following Lemma 
% examines the replicas discovered by two consecutive find-nodes operations
% by the Kademlia algorithm.

% \begin{lemma}
%     Let $f_i, f_j\in\flexSet{}$. If node $f_i$ invokes $\suc{S_i,o}$ 
%     operation before $f_j$, and $R_i$ and $R_j$ the set of replica nodes
%     for object $X$ in the two nodes respectively. 
% \end{lemma}

%% file: sec_experiments.tex
Theoretical findings show that \optimum{} implements a consistent, scalable, multi-object, and decentralised shared-memory. We complement our findings with empirical data obtained from experiments conducted on a globally distributed deployment of \optimum{} nodes. Our evaluation focuses on how the latencies of read and write operations scale across different environments.
\subsection{Testbed}
\begin{figure}[t]
    \centering
    \includegraphics[width=\linewidth]{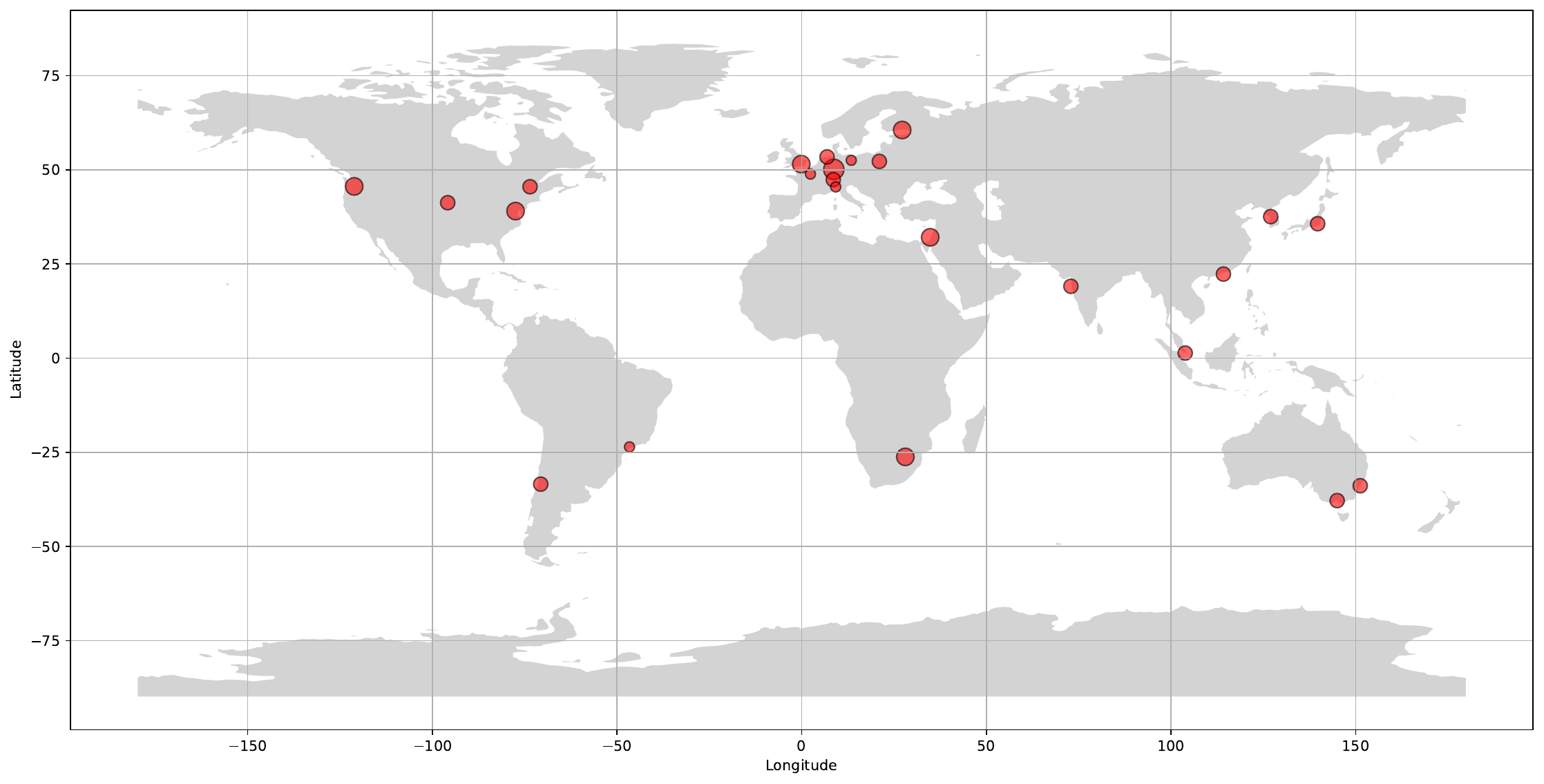}
    \caption{52 \optimum{} nodes used in numerical evaluations}
    \label{fig:distribution_nodes}
\end{figure}

We use a static set of 52 \optimum{} nodes distributed globally across Google data centers, as shown in Fig.~\ref{fig:distribution_nodes}. Each node runs on an e2-highmem-2 instance equipped with 2 vCPUs, 16 GB of RAM, and up to 4 Gbps of egress bandwidth, hosting an \optimum{} node inside a Docker container. The \optimum{} implementation is written in Go and each node uses the in-memory database memdb to locally store the memory objects.

\subsection{Scenarios}
To illustrate the properties of \optimum{}, we employ two MWABD benchmarks: one with full replication, where $n$ equals a subset (sometimes all) of the nodes in the system, and one with a replication cluster of size $n = 5$, where objects are distributed to nodes according to their position on the hashing ring. In contrast, \optimum{} shards each object into $k = 3$ pieces, applies coding via RLNC, and distributes $n = 5$ coded shards across the nodes. Thus, the performance difference between MWABD with full replication and MWABD with a cluster size of $n = 5$ captures the effect of restricting the replication cluster and applying the object distribution, while the difference between MWABD with a restricted cluster and \optimum{} reflects the impact of applying RLNC-based coding. To ensure concurrency, nodes invoke read and write operations every 10 seconds. We evaluate all three algorithms to examine the scalability of the algorithm in four different dimensions:

\input{plots_op_latency_1}

\begin{figure*}[h]
    \centering
    \includegraphics[width=\linewidth, height=4.5cm]{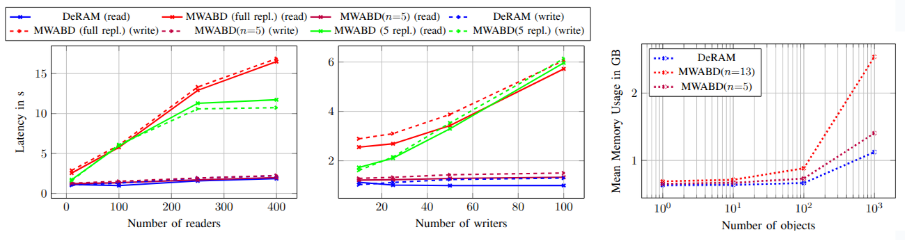}
    \caption{Read and write latencies under varying concurrency in a DSM with 13 nodes and 100 objects. (Left) Varying number of readers (writers fixed at 10). (Middle) Varying number of writers (readers fixed at 10). Mean memory consumption for varying number of objects stored in the DSM. (Right)}
%    \label{fig:distribution_nodes}
\end{figure*}

\begin{enumerate}
    \item {\bf Scalability on Object Size}: We evaluate how read and write latencies scale as the object size increases from $32\text{kB}$ to $16\text{MB}$. A single object is stored across a subsystem of 13 of the 52 nodes, and the system handles 10 read and 3 write requests concurrently.% within a 10-second interval.
    \item {\bf Scalability on Number of Objects}: With the object size fixed at $32\text{kB}$, we assess both latency and storage needs when storing $1$, $10$, $100$, or $1000$ objects in the DSM. As before, we consider a subset of 13 of the 52 nodes, with 10 concurrent readers and 3 concurrent writers.
    \item {\bf Scalability on Number of Participants}: We examine how increasing the number of nodes from 13 to 52 affects read and write latencies under 10 concurrent readers and 3 concurrent writers for a single object. 
    %In this experiment we further evaluate the storage behaviour at individual nodes.
    \item {\bf Scalability on Concurrent Operations}: Using the full 52-node cluster and 100 objects ($32\text{kB}$ each), we measure latencies while varying the number of readers from 10 to 400 and the number of writers from 10 to 100, performing operations over those objects.
    %, all operating on an object of size $32\text{kB}$. 
    We include another benchmark here that applies MWABD with full replication to a subset of 5 out of the 52 nodes.
\end{enumerate}

\subsection{Experimental Results}
Here, we present our results 
%the latencies measured 
in the four scenarios.
%for \optimum{}, MWABD with full replication, and MWABD with a replication cluster of size $n = 5$. 
In summary, clustering and object distribution via the hash ring 
enhances scalability with respect to both the number of participating nodes and the number of concurrent operations,
as latencies remain stable under increasing system load. The use of coding in \optimum{} further reduces read and write latencies, particularly for larger object sizes.

\subsubsection{Object Size}
The left plot in Fig.~\ref{fig:var_object_size_nodes_and_num} shows the read and write latencies of the three storage algorithms for a \textit{single object} ranging from $32\text{kB}$ to $16\text{MB}$. Larger object sizes lead to increased read and write latencies across all algorithms. For every object size, limiting the replication cluster to 5 nodes reduces latencies by up to one third compared with full replication. The benefits of coding become more visible at larger object sizes: for 8 MB and 16 MB objects, \optimum{} latencies are reduced by approximately 25\% and 50\%, respectively. Increasing $k$ may even yield larger benefits here since the individual shards then become smaller.

\subsubsection{Number of Objects}
The plot in the middle of Fig. ~\ref{fig:var_object_size_nodes_and_num} shows the read and write latencies for different numbers of stored objects, each of size $32\text{kB}$. The number of concurrent accesses is held constant at 10 readers and 3 writers. Overall, the results indicate that for none of the three algorithms, latencies increase with the number of objects stored in the DSM. 
However, the memory footprint does scale with increasing number of objects as displayed in Fig. \ref{fig:distribution_nodes}. While the memory footprint scales linearly in the number of stored objects for all three algorithms, MWABD with full replication shows the largest memory footprint out of the three algorithms. For 1000 objects the memory usage for MWABD with full replication is about 1.8 times larger than for MWABD when the hash ring is used and adding coding (DeRAM) reduces the memory footprint further by a factor of 0.8.

\subsubsection{Number of Nodes}
The rightmost plot in Fig. ~\ref{fig:var_object_size_nodes_and_num} illustrates how each algorithm performs as the number of participating nodes increases from the 13-node configuration used in the previous experiments to the full set of 52 nodes. While the latencies of \optimum{} remain unaffected by the growing number of participants, the read and write latencies of MWABD increase linearly with system size. This behavior results from the (per object) restricted replication cluster in \optimum{}, which ensures that the communication load per operation remains constant regardless of the total number of nodes in the system.

\subsubsection{Number of Concurrent Operations}
Finally, we compare the behavior of MWABD with full replication (both on the 52 node and a 5 node network size), MWABD with a cluster-limited replication strategy, and \optimum{} under increasing levels of concurrent read and write activity. 
%We further evaluate a DSM with a small node set (5 nodes) and the corresponding read and write latencies. 
Fig. \ref{fig:distribution_nodes} shows the resulting read and write latencies as the number of readers and writers increases, respectively. MWABD with full replication, both over all the nodes and the smaller cluster of 5 nodes, 
exhibits a linear increase in latency as concurrency grows. On the contrary, the latencies of \optimum{} and MWABD with a restricted replication cluster ($n = 5$) and object distribution remain stable across varying numbers of concurrent readers and writers. Since  MWABD with full replication requires every node to participate in every operation, introduces higher congestion, turning the single cluster to a performance bottleneck and limiting its ability to support high levels of concurrent access. This demonstrates the advantage of localized replication clusters and object distribution.

%% file: plots_op_latency_1.tex
\begin{figure*}[t]
  \centering
  \pgfplotslegendfromname{sharedlegend_1}\vspace{0.6em}
  
  \begin{tikzpicture}
    \begin{groupplot}[
      group style={
        group size=3 by 1,
        horizontal sep=0.8cm,
      },
      width=0.37\textwidth,
      height=4.5cm,
      xmode=log,
      grid=both,
      every axis plot/.append style={line width=1.2pt},
      every mark/.append style={mark options={solid}},
    ]

      % --- (a) Latency vs object size ---
      \nextgroupplot[
        xlabel={Object size in kB},
        ylabel={Latency in s},
        legend to name=sharedlegend_1,
        legend style={font=\small},
        legend columns=3,
      ]

        % Read (solid) + legend entries (ONLY HERE)
        \addplot[mark=x, color=blue]
          table[col sep=comma, x=Size, y expr=\thisrow{deram}/1000]
          {plots/raw_data/read_var_object_size.csv};
        \addlegendentry{DeRAM (read)}

        \addplot[mark=x, color=red]
          table[col sep=comma, x=Size, y expr=\thisrow{mwabd_n_13}/1000]
          {plots/raw_data/read_var_object_size.csv};
        \addlegendentry{MWABD(full repl.) (read)}

        \addplot[mark=x, color=purple]
          table[col sep=comma, x=Size, y expr=\thisrow{mwabd_n_5}/1000]
          {plots/raw_data/read_var_object_size.csv};
        \addlegendentry{MWABD($n{=}5$) (read)}

        % Write (dashed) + legend entries (ONLY HERE)
        \addplot[mark=x, color=blue, dashed]
          table[col sep=comma, x=Size, y expr=\thisrow{deram}/1000]
          {plots/raw_data/write_var_object_size.csv};
        \addlegendentry{DeRAM (write)}

        \addplot[mark=x, color=red, dashed]
          table[col sep=comma, x=Size, y expr=\thisrow{mwabd_n_13}/1000]
          {plots/raw_data/write_var_object_size.csv};
        \addlegendentry{MWABD(full repl.) (write)}

        \addplot[mark=x, color=purple, dashed]
          table[col sep=comma, x=Size, y expr=\thisrow{mwabd_n_5}/1000]
          {plots/raw_data/write_var_object_size.csv};
        \addlegendentry{MWABD($n=5$) (write)}

      % --- (b) Latency vs number of objects ---
      \nextgroupplot[
        xlabel={Number of objects},
      ]

        % Read (solid)
        \addplot[mark=x, color=blue, forget plot]
          table[col sep=comma, x=Size, y expr=\thisrow{deram}/1000]
          {plots/raw_data/read_var_object_num.csv};

        \addplot[mark=x, color=red, forget plot]
          table[col sep=comma, x=Size, y expr=\thisrow{mwabd_n_13}/1000]
          {plots/raw_data/read_var_object_num.csv};

        \addplot[mark=x, color=purple, forget plot]
          table[col sep=comma, x=Size, y expr=\thisrow{mwabd_n_5}/1000]
          {plots/raw_data/read_var_object_num.csv};

        % Write (dashed)
        \addplot[mark=x, color=blue, dashed, forget plot]
          table[col sep=comma, x=Size, y expr=\thisrow{deram}/1000]
          {plots/raw_data/write_var_object_num.csv};

        \addplot[mark=x, color=red, dashed, forget plot]
          table[col sep=comma, x=Size, y expr=\thisrow{mwabd_n_13}/1000]
          {plots/raw_data/write_var_object_num.csv};

        \addplot[mark=x, color=purple, dashed, forget plot]
          table[col sep=comma, x=Size, y expr=\thisrow{mwabd_n_5}/1000]
          {plots/raw_data/write_var_object_num.csv};

      % --- (c) Latency vs number of nodes ---
      \nextgroupplot[
        xlabel={Number of nodes},
        xmode=linear
      ]

        % Read (solid)
        \addplot[mark=x, color=blue, forget plot]
          table[col sep=comma, x=Size, y expr=\thisrow{deram}/1000]
          {plots/raw_data/read_var_nodes.csv};

        \addplot[mark=x, color=red, forget plot]
          table[col sep=comma, x=Size, y expr=\thisrow{mwabd_n_full}/1000]
          {plots/raw_data/read_var_nodes.csv};

        % Write (dashed)
        \addplot[mark=x, color=blue, dashed, forget plot]
          table[col sep=comma, x=Size, y expr=\thisrow{deram}/1000]
          {plots/raw_data/write_var_nodes.csv};

        \addplot[mark=x, color=red, dashed, forget plot]
          table[col sep=comma, x=Size, y expr=\thisrow{mwabd_n_full}/1000]
          {plots/raw_data/write_var_nodes.csv};

    \end{groupplot}
  \end{tikzpicture}

  \caption{Read (solid) and write (dashed) execution latency. (Left) Varying object sizes. (Middle) Varying number of objects stored in the DSM. (Right) Varying number of nodes. In all instances, 10 readers and 3 writers executed requests concurrently in an interval of $10s$}
  \label{fig:var_object_size_nodes_and_num}
\end{figure*}

%% file: sec_conclusion.tex
This paper introduced \optimum{}, a novel decentralized shared memory framework for Web3 that leverages Random Linear Network Coding (RLNC) to enable atomic read/write operations. By integrating RLNC, Optimum effectively mitigates challenges related to node churn, asynchronous communication, and malicious actors while optimizing for latency, storage, and fault tolerance. Its modular architecture ensures high availability and scalability, maintaining safety and liveness across dynamic network conditions.

Future research will focus on tolerating Byzantine readers and writers through the use of signed tags and homomorphic signatures to authenticate value propagation and prevent tag-space depletion. Additionally, we aim to explore dynamic clustering to fully leverage RLNC as a rateless code, allowing for adaptive coding parameters in response to changing networks.